\documentclass[prd,nofootinbib,eqsecnum,superscriptaddress]{revtex4-2}
\usepackage{color}
\usepackage{xcolor}
\usepackage{amsmath}
\usepackage{amssymb}
\usepackage{graphicx}
\usepackage{setspace}

\usepackage[normalem]{ulem}

\usepackage{epsfig}

\usepackage{dcolumn}
\usepackage{bm}
\usepackage[breaklinks=true,colorlinks=true,linkcolor=blue,urlcolor=blue,citecolor=blue,pagebackref=true]{hyperref}

\makeatother

\begin{document}
\title{Computation of $\left\langle \Phi^{2}\right\rangle $
and quantum fluxes at the polar interior of a spinning
black hole}
\author{Noa Zilberman}
\email{nz3745@princeton.edu}
\affiliation{Department of Physics, Technion, Haifa 32000, Israel}
\affiliation{Princeton Gravity Initiative, Princeton University, Princeton NJ 08544, USA}
\author{Marc Casals}
\email{marc.casals@uni-leipzig.de}
\affiliation{Institut f\"ur Theoretische Physik, Universit\"at Leipzig,\\ Br\"uderstra{\ss}e 16, 04103 Leipzig, Germany}
\affiliation{School of Mathematics and Statistics, University College Dublin, Belfield, Dublin 4, D04 V1W8, Ireland}
\affiliation{Centro Brasileiro de Pesquisas F\'isicas (CBPF), Rio de Janeiro, CEP 22290-180, Brazil}
\author{Adam Levi}
\email{leviadam@gmail.com}
\affiliation{Department of Physics, Technion, Haifa 32000, Israel}
\author{Amos Ori}
\email{amos@physics.technion.ac.il}
\affiliation{Department of Physics, Technion, Haifa 32000, Israel}
\author{Adrian C. Ottewill}
\email{adrian.ottewill@ucd.ie}
\affiliation{School of Mathematics and Statistics, University College Dublin, Belfield, Dublin 4, D04 V1W8, Ireland}

\date{\today}
\begin{abstract}
Renormalization of physical quantities for quantum field theories in curved spacetimes can be achieved via the subtraction of counterterms 
in a consistent manner within a regularization scheme such as a point-splitting method.
Pragmatic mode-sum regularization (PMR) is a point-splitting method which is particularly suitable for rotating black hole spacetimes.
We extend and tailor the $t$-splitting variant of PMR specifically for the interior of a Kerr black hole on the axis of rotation, focusing on a minimally-coupled massless scalar field in the physically-motivated Unruh state. The method addresses unique challenges within the black hole interior that do not occur outside. In particular, while the infinite sum over multipolar number $l$ converges in the black hole exterior, it diverges in the interior, necessitating the subtraction of a so-called intermediate divergence which includes introducing an additional “small” split in the direction of the polar angle $\theta$. This procedure is outlined and justified, along with the standard PMR method's subtraction of counterterms mode-by-mode.
We apply this method to calculate the renormalized energy-momentum fluxes $\left\langle T_{uu}\right\rangle _{\text{ren}}^{U}$, $\left\langle T_{vv}\right\rangle _{\text{ren}}^{U}$ (where $u$ and $v$ are the standard Eddington coordinates) and the renormalized field square $\left\langle \Phi^{2}\right\rangle _{\text{ren}}^{U}$ throughout the Kerr black hole interior, spanning from (just off) the event horizon to (just off) the inner horizon. Special emphasis is placed on the vicinity of the inner horizon, where the $t$-splitting results for $\left\langle T_{uu}\right\rangle _{\text{ren}}^{U}$ and $\left\langle T_{vv}\right\rangle _{\text{ren}}^{U}$ asymptote to those obtained directly at the inner horizon using a different method in a previous work.
In an Appendix, we develop an alternative variant of the $t$-splitting PMR method, dubbed the \emph{analytic extension} variant, which does not include the intermediate divergence subtraction. We utilize it to perform independent computations that are used to verify the standard $t$-splitting variant presented in the main text.
\end{abstract}

\maketitle
\section{Introduction}

Within the framework of semiclassical gravity, the gravitational field
is kept classical whereas the matter fields are quantized. Semiclassical
gravity is expected to be a valid framework in the limit that the
physical scales are much larger than the Planck scales and, as such, it has
provided significant results. For example, in black hole (BH) settings,
semiclassical
gravity has led to the pioneering discovery by Hawking \cite{Hawking:1974,Hawking:1975}
that astrophysical BHs emit quantum thermal radiation in their exterior.
In its turn, in the interior region of BHs, recent work within semiclassical
gravity has unveiled an irregularity of the so-called Cauchy
horizon (CH)~\cite{KerrIH:2022,KSCH24} (see~\cite{FluxesIH:2020,Hollands:2020cqg,Hollands:2020prd} in the non-rotating case), which (at least naively\footnote{Semiclassical analyses on fixed Reissner-Nordstr\"om and Kerr metrics (as well as their corresponding de Sitter variants) indicate that the quantum energy-momentum fluxes typically diverge at the CH like $V^{-2}$ (where $V$ is a regular Kruskal coordinate vanishing at the CH) -- which is stronger than the divergence of energy-momentum perturbations in the analogous classical problem. However, when attempting to translate this observation to the near-CH backreaction analysis, one should recall that a semiclassical BH in the Unruh state undergoes evaporation (which is manifested already at the event horizon). This needs to be taken into account when attempting to evolve the Einstein equations from the event horizon towards the CH.}) suggests dominance 
over that due to classical effects~\cite{Dafermos:2017,Ori:1992,Ori:1999,Sbierski:2022cnj,2022PhRvD.106d4060C,PhysRevD.97.104060} (see, e.g.,~\cite{Hiscock:1981,PoissonIsrael:1990,OriMassInflation:1991,luk2017strong,Cardoso:2017soq} in the non-rotating case).

As mentioned, within semiclassical gravity, matter fields are treated
as Quantum Field Theories (QFTs). As is well-known, however, QFTs
suffer from ultraviolet divergences and, hence, the expectation values
of most physical quantities need to be appropriately renormalized. Most importantly,
the \textit{renormalized} expectation value of the stress-energy tensor
(RSET),
$\left\langle T_{\mu\nu}\right\rangle _{\text{ren}}^{\Psi}$\footnote{Typically, a hat is placed over a quantity in order to distinguish its quantum version over its classical version since, mathematically, they are objects of very different types. In order to reduce cluttering, however, we will not make such a distinction. Therefore, in particular, $\Phi$ will equally denote a classical scalar field or its quantum version (which is an operator-valued distribution), and similarly for the stress energy tensor $T_{\mu\nu}$. The distinction between classical and quantum quantities should be clear from the context.}, when  the field is in a quantum state $\Psi$, is the quantity which appears on the right hand side of the
semiclassical Einstein equations:
\begin{equation}
G_{\mu\nu}=8\pi\left\langle T_{\mu\nu}\right\rangle _{\text{ren}}^{\Psi} ,
\end{equation}
where $G_{\mu\nu}$ is the Einstein tensor and we take units where $c=G=1$.
That is, the RSET replaces the classical stress-energy
tensor $T_{\mu\nu}$ in the classical Einstein equations. Ideally,
one would 
evaluate
the RSET 
and the Einstein tensor
in the same spacetime
but that is a very tall order. Thus, typically, one
follows a perturbative approach whereby the RSET is calculated on
a {\it background} spacetime and one would then solve the semiclassical
Einstein equations for the backreacted metric; such a procedure could be carried out iteratively to arbitrary order. In curved spacetimes,
Wald \cite{Wald:1977} established  axioms which a physically-meaningful RSET should satisfy. 

Henceforth we shall focus on the case that the matter
field is a scalar field $\Phi$. In this case, a 
 quantity which is easier to calculate than the RSET but which is also of physical significance is the renormalized field square (or \emph{vacuum polarization}) $\left\langle \Phi^{2}(x)\right\rangle _{\text{ren}}^{\Psi}$: in particular, it is important for spontaneous symmetry breaking (see, e.g.,~\cite{Fawcett} in the context of black holes).

There exist several methods for renormalization
but the one of main interest in this paper involves using the so-called point-splitting regularization
method~\cite{Schwinger:1951,DeWittBook:1965} (see, e.g.,~\cite{BirrellDavies:1984} for a review). Point-splitting-based renormalization methods~\cite{Christensen:1976,Christensen:1978,Wald:1978,BrownOttewill:1986} are particularly useful for calculational purposes and can be used for both the renormalized field square and the RSET, satisfying Wald's physical axioms in the latter case.
The point-splitting method essentially consists of the following.
First, the expectation value at a
spacetime point $x$ of a physical quantity 
which is quadratic
in the quantum field 
(which, mathematically, is an operator-valued distribution) and its derivatives
is temporarily made a bi-tensor\footnote{There is a freedom in the choice of such bi-tensor, while the final physical result is independent of that choice.} by evaluating each
one of the two field factors at a different spacetime point: one factor at $x$
and the other factor at, say, $x'$. 
Such a bi-tensor is then 
regular
as long as the
two spacetime points do not coincide (and are not connected by a null
geodesic~\cite{KayRadzikowskiWald}).
One then subtracts
from this unrenormalized 
bi-tensor 
a 
so-called  counterterm\footnote{The  counterterm is typically expressed as a sum of truly divergent subterms and a finite subterm.}
which is purely-geometrical (and so state-independent).
Finally, one takes the
coincidence limit ($x'\to x$) in the result of such subtraction,
yielding the renormalized expectation value of the quantity of interest.

In the case that the quantity
of interest is the field square 
$\Phi^{2}$ or the stress-energy tensor
$T_{\mu\nu}$, we 
respectively
denote 
the
\textit{un}renormalized bi-tensor by
$\frac{1}{2}G_{\Psi}(x,x')$ 
 or $\left\langle T_{\mu\nu}(x,x')\right\rangle^{\Psi}$,
the  counterterm by
$\frac{1}{2}G^{\text{CT}}(x,x')$ or $T_{\mu\nu}^{\text{CT}}(x,x')$,
and
the point-splitting regularization
procedure then
amounts to
$\left\langle \Phi^{2}(x)\right\rangle _{\text{ren}}^{\Psi}=\frac{1}{2}\lim_{x'\to x}\left(G_{\Psi}(x,x')-G^{\text{CT}}(x,x')\right)$
or $\left\langle T_{\mu\nu}(x)\right\rangle _{\text{ren}}^{\Psi}=\lim_{x'\to x}\left(\left\langle T_{\mu\nu}(x,x')\right\rangle ^{\Psi}-T_{\mu\nu}^{\text{CT}}(x,x')\right)$.
In this paper, for the two-point function $G_{\Psi}(x,x')$ we shall later make the choice of the function $G_{\Psi}^{(1)}(x,x')\equiv\left\langle \left\{ \Phi(x),\Phi(x')\right\} \right\rangle _{\Psi}$ -- also called the Hadamard two-point function (HTPF) -- where $\left\{,\right\}$ denotes anticommutation with respect to $x$ and $x'$. Note, however, that other choices for $G_{\Psi}(x,x')$ are also possible.

An expression for the HTPF in Kerr in terms of modes which are amenable to practical computations was derived in~\cite{Frolov:Thorne} for the exterior of the BH and by us~\cite{HTPF:2022} for the interior. 
From the fact that the
classical stress-energy tensor may be obtained by applying a certain
differential operator quadratic in the field (see Eq.~\eqref{eq:Tab_product} below),
it follows that the RSET may be obtained by applying a related differential
operator to $G_{\Psi}(x,x')-G^{\text{CT}}(x,x')$ and afterwards
taking the coincidence limit.

Unfortunately, from a technical point of view, such a renormalization
procedure is notoriously hard to carry out in practice, at least in
the case of BH background spacetimes. The main reason is that, typically,
one calculates the unrenormalized bi-tensor ($\frac{1}{2}G_{\Psi}(x,x')$ or
$\left\langle T_{\mu\nu}(x,x')\right\rangle ^{\Psi}$ ) via a full
(Fourier and angular) infinite mode decomposition whereas the  counterterms
($\frac{1}{2}G^{\text{CT}}(x,x')$ or $T_{\mu\nu}^{\text{CT}}(x,x')$) are
instead known in terms of geometrical quantities (they are usually
known as an expansion for small geodesic distance between the two
spacetime points\footnote{The geodesic distance between the two spacetime points is unique as
long as the points are `close' enough.} $x$ and $x'$). Only in very special (highly-symmetric) spacetimes  can one {\it analytically} obtain {\it both}
the modes of the unrenormalized quantity and the corresponding
mode sums in closed form; one can then subtract from the closed form expression the  counterterm and finally take the coincidence limit $x'\to x$, thereby performing the entire renormalization procedure analytically. In the other cases, which include all 4-dimensional BH spacetimes, one must resort to a {\it numerical} evaluation of the full mode sum, which can be rather challenging since the convergence of the infinite mode sum slows
down as $x'$ approaches $x$ (and the mode sum diverges in the actual limit
$x'\to x$). 

Therefore, one typically seeks to find a mode decomposition of
the counterterm, so that the renormalization subtraction can be carried
out {\it mode-by-mode}, thus improving the convergence of the mode sum.
For that purpose, it is useful to separate the points $x$ and $x'$
in a coordinate direction which corresponds to a symmetry (in cases where
there is one) of the background spacetime and then re-express the
 counterterm as a mode sum decomposition with respect to the associated coordinate.
Commonly, the background spacetime is stationary and either spherically-symmetric or only axisymmetric, and so, accordingly, one separates the points in
the time direction (so-called $t$-splitting) or corresponding angular
directions (so-called $\theta$- or $\varphi$-splitting, depending
on whether the direction of separation is along the polar angle or
the azimuthal angle, respectively). In the case of $t$-, $\theta$-
or $\varphi$-splitting, the corresponding decomposition of the counterterm is in terms
of, respectively, Fourier frequency $\omega$-modes, spherical/spheroidal $l$-harmonics or azimuthal $m$-modes.
In its turn, the unrenormalized bi-tensor involves all sums: an (infinite) integral over $\omega$, an (infinite) sum over $l$ and a (finite) sum over $m$.

It is worth mentioning that the point-splitting method is expected to fail at the spacetime regions where the Killing vector associated with the direction of symmetry  with respect to which the splitting is carried out has zero norm (since the splitting would be along a null direction, along which the two-point function diverges).
In particular, this would mean that  in a Kerr spacetime $\varphi$-splitting
might fail on the pole and $t$-splitting on the boundary of the ergoregion, which is where the Killing vector $\partial_t$ becomes spacelike. (In particular, at the axis of rotation the ergoregion boundary meets the horizons, leading to the failure of the method there, as we empirically see.) 

If the background is \textit{static} and the quantum state is thermal, $t$-splitting may render the expressions
particularly amenable to computations if one further takes advantage
of a Euclideanization technique, 
whereby the spacetime is made Riemannian via a Wick rotation of the time coordinate.

Another option for performing the renormalization is to calculate {\it differences}
of renormalized expectation values
in two different
states: because the counterterms are state-independent, it is clear
that such difference is equal to the coincidence limit of the difference
between the unrenormalized bi-tensors in the two different states
(e.g., $\left\langle \Phi^{2}(x)\right\rangle _{\text{ren}}^{\Psi_{1}}-\left\langle \Phi^{2}(x)\right\rangle _{\text{ren}}^{\Psi_{2}}=\frac{1}{2}\lim_{x'\to x}\left(G_{\Psi_{1}}(x,x')-G_{\Psi_{2}}(x,x')\right)$
and $\left\langle T_{\mu\nu}(x)\right\rangle _{\text{ren}}^{\Psi_{1}}-\left\langle T_{\mu\nu}(x)\right\rangle _{\text{ren}}^{\Psi_{2}}=\lim_{x'\to x}\left(\left\langle T_{\mu\nu}(x,x')\right\rangle ^{\Psi_{1}}-\left\langle T_{\mu\nu}(x,x')\right\rangle ^{\Psi_{2}}\right)$,
for two states $\Psi_{1}$ and $\Psi_{2}$). Such differences are already regular
and so no actual renormalization needs to be carried out. Such a calculation
is particularly useful if one happens to know through some other means
the value of the renormalized expectation value in some \textit{reference}
state, from which (together with the calculation of the difference
between unrenormalized bi-tensors) the value of the renormalized expectation
value in another state could be thus obtained. We call this the \textit{state
subtraction} method.

Let us from now on focus only on BH spacetimes, for which the main relevant quantum states are the following.
The Unruh state~\cite{Unruh:1976}
describes an astrophysical BH evaporating via the emission of
Hawking radiation and is hence the state of interest in this paper.
In Schwarzschild or Reissner-Nordstr\"om (RN), the Hartle-Hawking (HH) state~\cite{Hartle:Hawking:2011} 
is meant to describe a
BH in thermal equilibrium with its own Hawking radiation and is the only state
for which the Euclideanization procedure turns the Fourier $\omega$-integral into a much more computationally practical discrete sum.
In Kerr,
a proposal for this state was made in~\cite{Frolov:Thorne} but,
unfortunately, off the symmetry axis it is not well-defined for bosons~\cite{Kay:Wald,OttewillWinstanley:2000}, whereas the analogous state for massless fermions  only exists near the event horizon (EH)~\cite{CDNOW}; 
possibly relatedly,
the Euclideanization procedure is 
in principle
not immediately applicable
in Kerr.
Finally, the Boulware state~\cite{Boulware}
is irregular
on the EH and is (only) appropriate to describe the quantum fields
around a star-like object. 

Due to symmetries of the metric and quantum state under consideration, in spherically-symmetric BH spacetimes (e.g. Schwarzschild and RN) the quantities computed (RSET and vacuum polarization) may depend only on the radial coordinate $r$, and in the axially-symmetric (e.g. Kerr) case the dependence may only be on  $r$ and polar angle $\theta$. In particular, a computation at the CH is valid generally at the inner horizon (IH, whose ingoing section is the CH), with some $\theta$-dependence in the rotating case.

Despite the above-mentioned technical difficulties for renormalization
in BH background spacetimes,
significant progress has been made over the years.
Renormalization in QFT in BH spacetimes has a long history starting
in the late 1970's, which we next review classified by method and spacetime, starting with spherically-symmetric BHs and afterwards moving on to rotating BHs.

First, via the method of state subtraction and expected behavior of a reference quantum state in some asymptotic region, the
renormalized expectation values of physical quantities
for the Unruh, Hartle-Hawking and Boulware states
in Schwarzschild spacetime were obtained when
approaching the EH or radial infinity~\cite{Christensen:Fulling,Candelas:1980}. 
Recently, Refs.~\cite{Hollands:2020cqg,Hollands:2020prd,2021PhRvL.127w1301K} 
used the state subtraction method to obtain renormalized expectation values on the CH of  an RN-de Sitter (dS) BH.

Another early calculation of the RSET in a BH spacetime was carried out in~\cite{Fawcett:1983} for the Hartle-Hawking state in Schwarzschild spacetime. This work made use of Euclideanization and then regularization was implemented at the level of the heat kernel representation for the Euclidean Green function, while taking the spacetime coincidence limit in the kernel, although unfortunately this calculation contained a slip unrelated to the renormalization process as noted by Howard~\cite{HowardRSET:1984}. 

Turning to the point-splitting
method, the $t$-splitting variant together with the Euclideanization technique
was used by various authors 
in order to obtain  renormalized expectation values
in static, spherically-symmetric BH spacetimes, namely,
Schwarzschild (e.g.,~\cite{HowardRSET:1984,Ander_His_Sam:1995,PhysRevLett.91.051301,Jensen:Ottewill}), RN (e.g.,~\cite{Ander_His_Sam:1995,PhysRevLett.91.051301}) and
(lukewarm) RNdS (e.g.,~\cite{breen2012hadamard,Winstanley:Young}).
Typically, in order to speed up the convergence of the mode sums, these works use WKB asymptotics for large multipole number $l$ and/or {\it Euclidean} frequency (such WKB asymptotics are not readily generalizable to the case of Lorentzian frequency $\omega$).
It is also worth noting a new variant of the point-splitting method, called the extended coordinate method, which involves splitting in both the angular direction 
and
in the time direction combined with the Euclideanization technique. The extended coordinate method  has been applied
in (4- and higher-dimensional) Schwarzschild~\cite{2016PhRvD..94l5024T,2017PhRvD..96j5020T,taylor2022mode} and in
Schwarzschild-anti-dS~\cite{2018PhRvD..98j5006B}; see~\cite{Freitas:Casals} for renormalization  also via multiple-direction splitting but without  Euclideanization (and so is in principle directly generalizable to Kerr) in the case of Bertotti-Robinson spacetime.

Of most relevance for this paper is the so-called pragmatic mode-sum regularization (PMR) method, which also has the advantage of not using Euclideanization while still using point-splitting.
PMR has been
 developed and used to calculate renormalized expectation values for
a scalar field in the following cases: using $\theta$-splitting, 
  in Schwarzschild~\cite{AAtheta:2016},
  in RNdS~\cite{2021PhRvD.104b5009K} 
  and 
 inside the EH 
 of RN~\cite{FluxesIH:2020,2021PhRvD.104b4066Z,GroupPhiRN:2019};
using $t$-splitting, in Schwarzschild~\cite{AAt:2015,LeviRSET:2017};
separately using $t$-, $\theta$- and $\varphi$-splitting, in
Schwarzschild~\cite{AARSET:2016}.

The literature results mentioned so far were for spherically-symmetric BHs. In
the astrophysically most important case of a stationary, rotating
(Kerr) BH, because of the lack of spherical symmetry, $\theta$-splitting
is not possible and, because of the lack of staticity, the Euclideanization
technique 
is in principle not implementable
either. Thus, progress in Kerr was for a long time made only by calculating
differences of renormalized expectation values of quantities in two
different states. This was some times combined with the expected behavior
of the RSET for one of the states in some specific spacetime region (namely, at infinity
or near a horizon) in order to 
apply the state subtraction method so as to 
gain knowledge about the behavior of the RSET for the other state in that
region: see Refs.~\cite{OttewillWinstanley:2000,duffy2008renormalized,Casals:Ottewill:2005,CDNOW} outside the EH and Ref.~\cite{KerrIH:2022} on the CH (see also Ref.~\cite{McMaken:2024} extending the latter work, and~\cite{KSCH24} on the CH of Kerr-dS).
An exception to that is the axis of symmetry of Kerr, where no rotation is
`felt', and that was used to directly obtain the RSET in 
a `formal' Hartle-Hawking
state on the pole of the EH in Refs.~\cite{Frolov:1982,Frolov:1985spb}. An important technical
breakthrough in renormalization  was achieved using the $t$- and $\varphi$-splitting
variants of PMR in~\cite{LeviEilonOriMeentKerr:2017},
where the authors managed to calculate the RSET outside the EH of
Kerr. This is so far the only time that the calculation
of an RSET has been carried out outside a Kerr BH.

In the current
paper, we present the method and results for an analogous calculation
of certain components of the RSET and the renormalized field square \textit{inside} 
a Kerr BH, all the way from (just off) the EH to (just off) the IH.
We have mentioned our calculation in~\cite{KerrIH:2022} of the RSET energy-flux components on the CH of Kerr using state subtraction (which was done for an array of values of the polar angle $\theta$ and the BH angular momentum). In fact, in~\cite{KerrIH:2022} (see especially its Supplemental Material) we also presented $t$-splitting results for the same RSET components on the pole 
 (i.e., $\theta=0$\footnote{Even though $\theta=\pi$ is also a pole, the value of the scalar field is the same at $\theta=\pi$ as at $\theta=0$, and so we indistinctively refer to ``the pole" or ``the axis of rotation".})
very near -but {\it off}- the IH (as the method we use is inapplicable at exactly the IH itself). The extrapolation of these results onto the IH allowed us to check our result exactly on the IH, which was independently obtained with the state-subtraction method. In particular, this comparison provided a crucial test for the state subtraction procedure used directly at the IH, which was based on a non-conventional reference state.
In this paper we describe in detail the method that we used in~\cite{KerrIH:2022} off the IH, and here we
also use it to obtain new results.

Specifically, the method that we develop here is the $t$-splitting variant of PMR for
the calculation of renormalized quantities on the pole
between (just off) the EH and (just off) the IH of a Kerr spacetime $g_{\alpha\beta}$
for a minimally-coupled massless scalar field in the Unruh state $\left|0\right\rangle _{U}$
(we use throughout a $U$ subscript or superscript 
to indicate that the field is in the Unruh state). The quantities that we give computationally-amenable expressions
for are the renormalized expectation value of the field square (the vacuum polarization), $\left\langle \Phi^{2}\right\rangle _{\text{ren}}^{U}$,
and the energy \textit{flux} components\footnote{The Eddington coordinates are null in spherical symmetry. In the rotating case, they become null on the pole (and on the horizons), which facilitates the interpretation of
$T_{uu}$ and $T_{vv}$  as the energy flux components in the context of this paper.} of the RSET, $\left\langle T_{yy}\right\rangle _{\text{ren}}^{U}$,
where $y\in \{u,v\}$, and $u$ and $v$ are the Eddington coordinates [see
Eq. \eqref{eq:intEddCoor} below]. The significance of these flux components lies in their role in understanding backreaction near the CH, as outlined in Ref.~\cite{KerrIH:2022}.

Broadly speaking, it is more convenient to treat the trace-reversed stress-energy tensor, denoted by $\overline{T}_{\alpha\beta}$, which is related to the original tensor  $T_{\alpha\beta}$ by
\begin{equation}
\overline{T}_{\alpha\beta}\equiv T_{\alpha\beta}-\frac{1}{2}g_{\alpha\beta}T_{\,\,\,\,\mu}^{\mu}\,,\label{eq:trace_rev}
\end{equation}
since it admits the following simple form:
\begin{equation}
\overline{T}_{\alpha\beta}=\Phi_{,\alpha}\Phi_{,\beta}\,.\label{eq:Tab_product}
\end{equation}
It is also worth noting that when taken as sources to the Einstein equation, there is no advantage to using the stress-energy tensor over its trace-reversed counterpart. Indeed, one may work with the alternative version of the Einstein
equation $R_{\alpha\beta}=8\pi\overline{T}_{\alpha\beta}$, where $R_{\alpha\beta}$ is the Ricci tensor, which involves the trace-reversed stress-energy tensor directly. 

However, at the focus of this paper is the pole of Kerr -- where $g_{uu}=g_{vv}=0$ -- hence trace-reversal does not change the flux components. That is, at the pole, the following holds: 
\begin{equation}
\overline{T}_{yy}(r,\theta=0)=T_{yy}(r,\theta=0)\,.
\end{equation}
Hence, in this paper, we treat the flux components directly, and they are given in terms of the field derivatives by  $T_{yy}=\Phi_{,y}\Phi_{,y}$.

As mentioned,
$t$-splitting in BH background spacetimes involves a separation of the spacetime points in the $t$ direction 
(which is a symmetry of the background) 
and a decomposition of the unrenormalized bitensor, and so also of the corresponding renormalized quantity, that involves two infinite sums in the multipolar number $l$ and frequency $\omega$, as well as a finite sum in the azimuthal number $m$.
We note that the interior of 
the BH reveals some distinct features which give rise to specific technical challenges
which do not appear in the exterior (as in the calculation in \cite{LeviEilonOriMeentKerr:2017}). In particular, expressions for renormalized
quantities inside the BH contain the mentioned double infinite sums such that the
innermost, multipolar $l$-sum diverges. We refer to this as the intermediate divergence (ID)
problem and we show how to deal with it (namely, by including a `small'
split in the polar angle direction, on the top of the split in the
$t$ direction).

We then use the $t$-splitting PMR method in order to derive the results
on the pole for the renormalized energy fluxes $\left\langle T_{yy}\right\rangle _{\text{ren}}^{U}$ 
on approaching the IH 
that were already shown in Ref.~\cite{KerrIH:2022} (agreeing with the state-subtraction results computed directly at the IH therein), 
as well as to obtain new results. The new results on the pole are these energy fluxes all the way between the two horizons (see Figs.~\ref{Fig:a0p8_general}--\ref{Fig:near_EH_both})
as well as the renormalized field square $\left\langle \Phi^{2}\right\rangle _{\text{ren}}^{U}$ (see Figs.~\ref{Fig:a0p8_general-phi}--\ref{Fig:a0p9_near_IH-phi}).
Apart from the IH vicinity, we also focus on the EH vicinity at the pole, where we obtain numerical support for regularity of the Unruh state there (reflected in the vanishing
of $\left\langle T_{uu}\right\rangle _{\text{ren}}^{U}$ as $(r-r_+)^{2}$ in the $r\to r_+$ limit), which is a property that has not yet been rigorously proven in the case of Kerr.
In Table \ref{Tbl:numValues} 
we provide a summary of the numerical values of various
quantities of physical interest (such as the fluxes and the field square) at the pole of the horizons for the values of the Kerr BH angular momentum that we considered in this paper.

The rest of this paper is organized as follows. In Sec.~\ref{sec:Preliminaries.}
we introduce Kerr spacetime, the scalar field and its Eddington modes, the Unruh state
as well as a conserved quantity. In Sec.~\ref{sec:The-t-splitting-procedure}, the $t$-splitting procedure is extended and customized to the Kerr interior (at the pole, $\theta=0$), accommodating for the complexities arising there and describing the required extra steps in the procedure. In Sec.~\ref{sec:Numerical-results} we present the aforementioned numerical results, and we end with a discussion in Sec.~\ref{sec:discussion}.
The Appendices complement the rest of the paper, and are as follows:
Appendix~\ref{App:Large-l-asymptotics} gives the  asymptotic
expressions for large multipole number $l$ (and $m=0$) for the interior radial function $\psi_{\omega l}^{\text{int}}$
and the exterior reflection coefficient $\rho_{\omega l}^{\text{up}}$, which
are then used in Sec.~\ref{subsec:Intermediate-blindspots}; Appendix
\ref{App:The-intermediate-divergence} illustrates and justifies our treatment of the intermediate divergence problem arising in the sum over $l$; Appendix~\ref{App:Numerical-implementation}
focuses on the numerical methods implemented in the current work; finally,
Appendix~\ref{App:The-analytic-extension} offers an alternative
approach to the computation, which we call the \emph{analytic extension} variant of t-splitting,
following computations of several quantities which may
be compared with their standard $t$-splitting counterparts.
Supplementing this paper is a {\rm Mathematica} notebook which includes the PMR t-splitting counterterms for the field square and for the full stress-energy tensor, given at a general spacetime point in a Kerr spacetime. (The results for the full stress-energy tensor are given in Boyer-Lindquist coordinates, and are then translated to the flux components in coordinates  $(u,v,\theta,\phi)$ at the pole.)

We use 
metric signature $(-+++)$ and units where $c=G=1$.

\section{Preliminaries\label{sec:Preliminaries.}}

In this section, we introduce various issues at the basis of this paper: the background Kerr
BH spacetime, the wave equation satisfied by the scalar field propagating
on Kerr spacetime, computationally-convenient modes of the
scalar field (namely, the interior and exterior Eddington modes), and a brief presentation of the Unruh state.
We finish the section by noting the existence of a conserved quantity and its interpretation in the Unruh state.

\subsection{The Kerr metric\label{subsec:preliminary:Kerr_metric}}

We begin with the Kerr metric, a solution to the vacuum Einstein equations
describing a spinning BH of mass $M$ and angular momentum $J$, given
in Boyer-Lindquist coordinates $\left(t,r,\theta,\varphi\right)$
by the line element 

\begin{equation}
\text{d}s^{2}=-\left(1-\frac{2Mr}{\rho^{2}}\right)\text{d}t^{2}+\frac{\rho^{2}}{\Delta}\text{d}r^{2}+\rho^{2}\text{d}\theta^{2}+\left(r^{2}+a^{2}+\frac{2Mra^{2}}{\rho^{2}}\sin^{2}\theta\right)\sin^{2}\theta\text{d}\varphi^{2}-\frac{4Mra}{\rho^{2}}\sin^{2}\theta\text{d}\varphi\text{d}t,\label{eq:Kerr_metric}
\end{equation}
where $a\equiv J/M$ and
\begin{align*}
\rho^{2} & \equiv r^{2}+a^{2}\cos^{2}\theta\,,\\
\Delta & \equiv r^{2}-2Mr+a^{2}\,.
\end{align*}
In this paper we only consider the sub-extremal case, that is, in
which the BH parameters satisfy $\left|a\right|/M<1$.

\begin{figure}[h!]
\centering \includegraphics{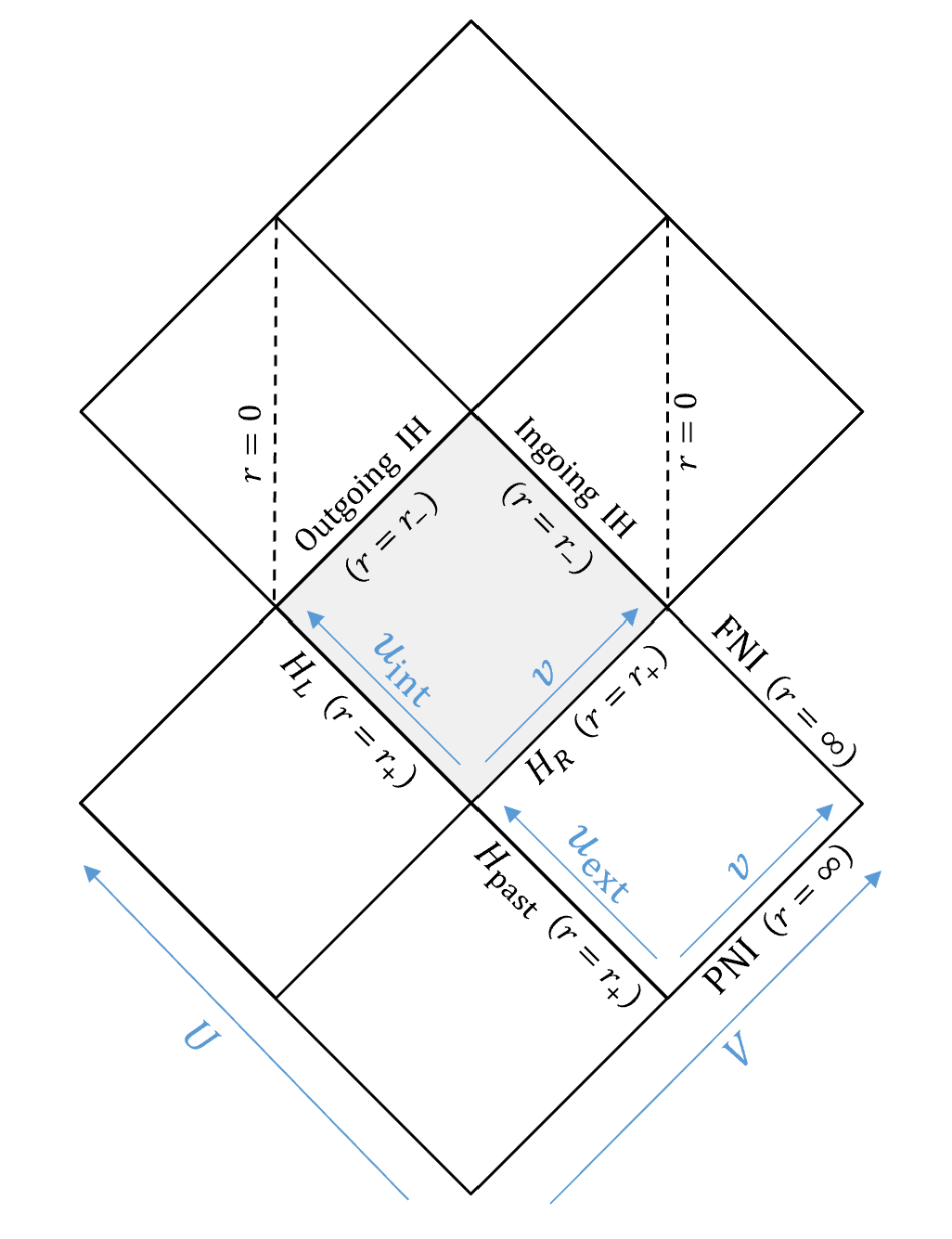}\caption{A portion of the Penrose diagram of a sub-extremal Kerr BH. The so-called \emph{past} and \emph{future} null infinities are denoted by PNI and FNI, respectively. The three relevant arms corresponding to  $r=r_+$  are denoted by $H_\text{past}$ (the \emph{past horizon}),  $H_R$ (the EH)  and $H_L$ the \emph{left horizon}.
The two relevant arms of the IH hypersurface are marked as well. The shaded region (where
$r_{-}<r<r_{+}$) is what we refer to as the BH interior, which is
the main focus of this paper. Three sets of coordinates {[}the interior
Eddington \eqref{eq:intEddCoor}, the exterior Eddington \eqref{eq:extEddCoor}
and the Kruskal coordinates \eqref{eq:extKrusCoor} and \eqref{eq:intKrusCoor}]
are portrayed as blue arrows on which the corresponding coordinate ranges from $-\infty$ to $\infty$. The vertical dashed lines are the $r=0$ ring singularities, which are only present at $\theta=\pi/2$.}
\label{Fig:Penrose}
\end{figure}

We note that $\Delta$ given above may be written as
\begin{equation}
\Delta=\left(r-r_{+}\right)\left(r-r_{-}\right)\label{eq:Delta}
\end{equation}
where the roots
\begin{equation}
r_{\pm}=M\pm\sqrt{M^{2}-a^{2}}\label{eq:horizons}
\end{equation}
mark the locations of the EH (at $r=r_{+}$) and the
IH (at $r=r_{-}$). We dub\footnote{Strictly, one needs two patches of Boyer-Lindquist coordinates in order to cover both the interior and the exterior regions, since these coordinates are irregular on the horizons; see Eqs.~\eqref{eq:extKrusCoor}, \eqref{eq:intKrusCoor} and \eqref{eq:phi_pm} for the coordinates $\{U,V,\theta,\varphi_+ \}$, which are regular across the  EH.} the region $r>r_{+}$ the
BH \emph{exterior} (or ``outside the BH''), and the region $r_{-}<r<r_{+}$
-- the BH \emph{interior} (or ``inside the BH'' -- corresponding
to the shaded region in Fig.~\ref{Fig:Penrose}). 

Both $r=r_{+}$ and $r=r_{-}$ correspond to null causal surfaces.
Regarding the IH, however, we point out that in the (physically-realistic)
case of gravitational collapse, only the \emph{ingoing} section (the
right one of the two segments denoted ``IH'' in Fig.~\ref{Fig:Penrose})
maintains the causal role of a Cauchy horizon, being the boundary
of the domain of predictability for an initial data surface reaching spacelike infinity (in the external universe).\footnote{In the case of an eternal BH, the relevant initial data surface reaches spacelike infinity in both external universes (including the parallel universe), and then both sections of the IH are Cauchy horizons.}

The surface gravity parameter $\kappa_{\pm}$ corresponding to the
horizon at $r_{\pm}$ is given by
\begin{equation}
\kappa_{\pm}\equiv\frac{r_{+}-r_{-}}{2\left(r_{\pm}^{2}+a^{2}\right)}\,.\label{eq:kappa_pm}
\end{equation}

We introduce the ``tortoise coordinate'' $r_{*}$ defined through
$dr/dr_{*}=\Delta/\left(r^{2}+a^{2}\right)$. In this paper, we pick
a constant of integration such that
\begin{equation}
r_{*}=r+\frac{1}{2\kappa_{+}}\log\left(\frac{\left|r-r_{+}\right|}{r_{+}-r_{-}}\right)-\frac{1}{2\kappa_{-}}\log\left(\frac{\left|r-r_{-}\right|}{r_{+}-r_{-}}\right)\,.\label{eq:rstarKerr}
\end{equation}
Note that $r_{*}$ diverges at both horizons; in particular, $r_{*}\to-\infty$
at $r=r_{+}$ and $r_{*}\to\infty$ at $r=r_{-}$.

The future-directed \emph{Eddington} \emph{coordinates}, $u$ and
$v$, may be defined in the BH exterior by
\begin{equation}
u_{\text{ext}}\equiv t-r_{*},\,\,\,v\equiv t+r_{*}\label{eq:extEddCoor}
\end{equation}
and in the BH interior by
\begin{equation}
u_{\text{int}}\equiv r_{*}-t,\,\,\,v\equiv r_{*}+t\,.\label{eq:intEddCoor}
\end{equation}
The Eddington coordinates are null at the pole of Kerr (where $g^{uu}=g^{vv}=0$) and off the pole at both $r=r_{+}$ and $r=r_{-}$ (but not elsewhere).\footnote{Although $u$ and $v$ are not necessarily regular at the horizons,
they are null there in the sense that they are co-directed with the
corresponding Kruskal coordinates (which are regular and null at the
corresponding horizon).}

While $v$ parameterizes the EH, the interior and exterior $u$ coordinates
diverge there (see Fig.~\ref{Fig:Penrose}). This motivates defining
the \emph{Kruskal} \emph{coordinates} $U$ and $V$, which remain
regular across the EH. In the BH exterior they are defined by

\begin{equation}
U\left(u_{\text{ext}}\right)\equiv-\frac{1}{\kappa_{+}}\exp\left(-\kappa_{+}u_{\text{ext}}\right),\,\,\,V\left(v\right)\equiv\frac{1}{\kappa_{+}}\exp\left(\kappa_{+}v\right)\,,\label{eq:extKrusCoor}
\end{equation}
and in the BH interior by
\begin{equation}
U\left(u_{\text{int}}\right)\equiv\frac{1}{\kappa_{+}}\exp\left(\kappa_{+}u_{\text{int}}\right),\,\,\,V\left(v\right)\equiv\frac{1}{\kappa_{+}}\exp\left(\kappa_{+}v\right)\,.\label{eq:intKrusCoor}
\end{equation}
The $V\left(v\right)$ coordinate is the same in the interior and
exterior regions. Regarding $U\left(u\right)$, the interior $U\left(u_{\text{int}}\right)$
is a smooth (and in fact analytic) continuation of the exterior $U\left(u_{\text{ext}}\right)$. 

An analogous set of Kruskal coordinates may be defined to expose the
regularity of the metric at the IH, but these IH coordinates are not
required in this paper. 

Finally, we note that in Kerr, all free-falling observers share the
same asymptotic value of $d\varphi/dt$ on approaching $r\to r_{\pm}$,
\begin{equation}
\Omega_{\pm}\equiv\frac{a}{2Mr_{\pm}}\,.\label{eq:Omega_pm}
\end{equation}
Since the $t$ coordinate diverges at both horizons (as in the spherically
symmetric case), the above fact implies that $\varphi$ also diverges
there (unlike in the spherically symmetric case). Hence, we use $\Omega_{\pm}$ to define an azimuthal coordinate
which stays regular at the horizon at $r_{\pm}$:
\begin{equation}
\varphi_{\pm}\equiv \varphi-\Omega_{\pm}t\,.\label{eq:phi_pm}
\end{equation}

\subsection{The wave equation and its separation\label{subsec:preliminary:wave_equation_separation}}

We consider a massless, minimally-coupled  scalar field $\Phi$,
obeying the wave equation
\begin{equation}
\square\Phi=0, \label{eq:KG}
\end{equation}
where $\square$ is the covariant d'Alembertian.

Due to separability of this equation on a Kerr background, we decompose
the field into $\left(\omega lm\right)$ modes
\begin{equation}
\Phi_{\omega lm}\left(t,r,\theta,\varphi\right)=\text{const}\cdot\frac{\psi_{\omega lm}\left(r\right)}{\sqrt{r^{2}+a^{2}}}e^{-i\omega t}Z_{lm}^{\omega}\left(\theta,\varphi\right)\,.\label{eq:decomKerr}
\end{equation}
The angular functions $Z_{lm}^{\omega}\left(\theta,\varphi\right)$
are the \emph{spheroidal harmonics}, given by
\begin{equation}
Z_{lm}^{\omega}\left(\theta,\varphi\right)=\frac{1}{\sqrt{2\pi}}S_{lm}^{\omega}\left(\theta\right)e^{im\varphi}\,,\label{eq:spheroidalHar}
\end{equation}
where $S_{lm}^{\omega}\left(\theta\right)$ is the (real) \emph{spheroidal
wave function} (see Ref.~\cite{BertiCardosoCasals:2006} and references
therein),
satisfying the eigenvalue equation:

\begin{equation}
\frac{1}{\sin\theta}\frac{d}{d\theta}\left(\sin\theta\frac{dS_{lm}^{\omega}\left(\theta\right)}{d\theta}\right)+\left(a^{2}\omega^{2}\cos^{2}\theta-\frac{m^{2}}{\sin^{2}\theta}+E_{lm}\left(a\omega\right)\right)S_{lm}^{\omega}\left(\theta\right)=0,\label{eq:angKerr}
\end{equation}
with $E_{lm}\left(a\omega\right)$ the corresponding eigenvalue, obtained
by requiring regularity at $\theta=0,\pi$.
We normalize the spheroidal wave functions as in Eq.~(2.10) Ref.~\cite{BertiCardosoCasals:2006}, namely:
\begin{equation}\label{eq:norm S}
\int_0^\pi \left[S_{lm}^{\omega}(\theta)\right]^2  \sin\theta d\theta =1.
\end{equation}
Of particular practical relevance for this paper is the fact that the angular function is zero at the poles if $m\neq 0$, i.e.,  $S_{lm}^{\omega}\left(\theta=0,\pi\right)\propto \delta_m^0$.

The so-called \emph{radial function }$\psi_{\omega lm}\left(r\right)$
solves a simple scattering-like equation which we term \emph{the radial
equation}:

\begin{equation}
\frac{d^{2}\psi_{\omega lm}}{dr_{*}^{2}}+V_{\omega lm}\left(r\right)\psi_{\omega lm}=0,\label{eq:radKerr}
\end{equation}
with the effective potential

\begin{equation}
V_{\omega lm}\left(r\right)\equiv\frac{K_{\omega m}^{2}\left(r\right)-\lambda_{lm}\left(a\omega\right)\Delta}{\left(r^{2}+a^{2}\right)^{2}}-G^{2}\left(r\right)-\frac{dG\left(r\right)}{dr_{*}}\label{eq:KerrPot}
\end{equation}
where
\begin{equation}
K_{\omega m}\left(r\right)\equiv\left(r^{2}+a^{2}\right)\omega-am,\;\;\;\lambda_{lm}\left(a\omega\right)\equiv E_{lm}\left(a\omega\right)-2am\omega+a^{2}\omega^{2},\;\;\;G\left(r\right)\equiv\frac{r\Delta}{\left(r^{2}+a^{2}\right)^{2}}\,.\label{eq:KerrPotFun}
\end{equation}

Note that $V_{\omega lm}$ involves the frequency $\omega$ in a non-trivial
way via the angular eigenvalue. Carrying Eq.~\eqref{eq:KerrPot}
to the asymptotic domains outside and inside the BH, $r\to\infty$,
$r\to r_{+}$ and $r\to r_{-}$, we obtain three different limiting
values:

\begin{equation}
V_{\omega lm}\simeq\begin{cases}
\omega^{2}, & r\to\infty\,\,\,\,\,\,\,\,\left(r_{*}\to\infty\right),\\
\omega_{+}^{2}, & r\to r_{+}\,\,\,\,\,\,\,\,\left(r_{*}\to-\infty\right),\\
\omega_{-}^{2}, & r\to r_{-}\,\,\,\,\,\,\,\,\left(r_{*}\to\infty\right),
\end{cases}\label{eq:asymPot}
\end{equation}
where we define 
\begin{equation}
\omega_{\pm}\equiv\omega-m\Omega_{\pm}\,.\label{eq:omega_pm}
\end{equation}
For that reason (and due to the potential being short-range), the
asymptotic behavior of solutions to Eq.~\eqref{eq:radKerr} is generally
of the form $e^{\pm i\omega r_{*}}$ as $r\to\infty$, $e^{\pm i\omega_{+}r_{*}}$
as $r\to r_{+}$, and $e^{\pm i\omega_{-}r_{*}}$ as $r\to r_{-}$,
corresponding to free waves in these $r_{*}\to\pm\infty$ domains.

\subsection{The Eddington modes\label{subsec:preliminary:The-Eddington-modes}}

The Eddington modes are solutions to Eq.~\eqref{eq:KG} which conform
with the general separated form of Eq.~\eqref{eq:decomKerr} and which
admit initial data that uniformly oscillate with either $u$ or $v$
along the relevant initial null hypersurface. Their decomposition
allows for a convenient numerical computation of these modes, requiring
one to merely solve the ODE given in Eq.~\eqref{eq:intEddCoor} for
the radial function (along with the standard computation of the spheroidal
harmonics).

We briefly introduce two families of Eddington modes (each consisting
of an ingoing set and an outgoing set): the exterior Eddington modes
which are defined in the BH exterior ($r_{+}<r<\infty$) and the interior
Eddington modes which are defined in the BH interior ($r_{-}<r<r_{+}$).
For a more detailed introduction of the various modes, see Sec. III
in Ref.~\cite{HTPF:2022}.

\paragraph{The exterior Eddington modes\label{subsec:preliminary:extEdd}}

We begin by defining two sets of exterior radial functions {[}solutions
to Eq.~\eqref{eq:radKerr} in the BH exterior{]}, ``in'' functions
denoted $\psi_{\omega lm}^{\text{in}}$ and ``up'' functions denoted
$\psi_{\omega lm}^{\text{up}}$, which are determined by the boundary
conditions:

\begin{equation}
\psi_{\omega lm}^{\text{in}}\left(r\right)\simeq\begin{cases}
\tau_{\omega lm}^{\text{in}}e^{-i\omega_{+}r_{*}} & r_{*}\to-\infty\\
e^{-i\omega r_{*}}+\rho_{\omega lm}^{\text{in}}e^{i\omega r_{*}} & r_{*}\to\infty
\end{cases}\,,\label{eq:psiIN_asym}
\end{equation}
\begin{equation}
\psi_{\omega lm}^{\text{up}}\left(r\right)\simeq\begin{cases}
e^{i\omega_{+}r_{*}}+\rho_{\omega lm}^{\text{up}}e^{-i\omega_{+}r_{*}} & r_{*}\to-\infty\\
\tau_{\omega lm}^{\text{up}}e^{i\omega r_{*}} & r_{*}\to\infty
\end{cases}\,.\label{eq:psiUP_asym}
\end{equation}
The coefficients $\tau_{\omega lm}^{\Lambda}$ and $\rho_{\omega lm}^{\Lambda}$
(with $\Lambda$ either ``in'' or ``up'') are respectively the
transmission and reflection coefficients, and may be determined numerically.

The ``in'' and ``up'' exterior Eddington modes, respectively denoted
by $f_{\omega lm}^{\text{in}}$ and $f_{\omega lm}^{\text{up}}$,
are then defined in terms of $\psi_{\omega lm}^{\text{in}}$ and $\psi_{\omega lm}^{\text{up}}$
 as
\begin{align}
f_{\omega lm}^{\text{in}}\left(x\right) & =\frac{1}{\sqrt{4\pi\left|\omega\right|\left(r^{2}+a^{2}\right)}}Z_{lm}^{\omega}\left(\theta,\varphi\right)e^{-i\omega t}\psi_{\omega lm}^{\text{in}}\left(r\right)\,,\label{eq:EddModesIn}\\
f_{\omega lm}^{\text{up}}\left(x\right) & =\frac{1}{\sqrt{4\pi\left|\omega_{+}\right|\left(r^{2}+a^{2}\right)}}Z_{lm}^{\omega}\left(\theta,\varphi\right)e^{-i\omega t}\psi_{\omega lm}^{\text{up}}\left(r\right)\,,\label{eq:EddModesUp} 
\end{align}
where $x$ denotes a spacetime point. The prefactors are determined
such that the modes are normalized to unity (with respect to the Klein-Gordon
inner product; see e.g. Ref.~\cite{HTPF:2022}).

\paragraph{The interior Eddington modes\label{subsec:preliminary:intEdd}}

Since the interior Eddington modes are defined exclusively in the
BH interior, where $r$ and $t$ switch roles as temporal and spatial
coordinates, defining the two spanning sets of modes will only require
 a single  (internal) radial function, denoted $\psi_{\omega lm}^{\text{int}}$,
defined as a solution of Eq.~\eqref{eq:radKerr} 
equipped with the initial condition at $r\to r_+$,
\begin{equation}
\psi_{\omega lm}^{\text{int}}\simeq e^{-i\omega_{+}r_{*}},\,\,\,\,\,r\to r_{+}\,.\label{eq:psi_int_BC}
\end{equation}

The ``right'' and ``left'' interior Eddington modes, respectively
denoted $f_{\omega lm}^{R}$ and $f_{\omega lm}^{L}$, are then defined
using $\psi_{\omega lm}^{\text{int}}$ and its complex conjugate as
\begin{align}
f_{\omega lm}^{R}\left(x\right) & =\frac{1}{\sqrt{4\pi\left|\omega_{+}\right|\left(r^{2}+a^{2}\right)}}Z_{lm}^{\omega}\left(\theta,\varphi\right)e^{-i\omega t}\psi_{\omega lm}^{\text{int}}\left(r\right)\,,\label{eq:EddModesRL}\\
f_{\omega lm}^{L}\left(x\right) & =\frac{1}{\sqrt{4\pi\left|\omega_{+}\right|\left(r^{2}+a^{2}\right)}}Z_{lm}^{\omega}\left(\theta,\varphi\right)e^{-i\omega t}\psi_{\omega lm}^{\text{int}*}\left(r\right)\,,\nonumber 
\end{align}
where, again, the prefactors ensure Klein-Gordon normalization.

\subsection{Field quantization and Unruh state\label{subsec:Field-quantization-Unruh}}

In this subsection we sketch the quantization of the field and the
definition of our quantum state of interest, namely the Unruh state.
For details, we refer the reader to Secs.~III and IV in Ref.~\cite{HTPF:2022}. The
scalar field $\Phi$ is typically quantized by expanding it in terms
of a choice of modes and then promoting the coefficients in the mode
expansion to (creation and annihilation) operators. In the previous
subsection we introduce the Eddington modes since they are convenient
for practical calculations. However, the Unruh state is instead more
naturally defined in terms of some other modes, the so-called Unruh
modes, which consist of the union of the following two subsets of
modes. 
The first subset, which merges with the Eddington $f^\text{in}_{\omega lm}$ in the BH exterior, is defined as having no upward excitations coming from $\{U\in\mathbb{R},V=0\}$ (i.e., from the union of $H_L$ and $H_\text{past}$, see Fig.~\ref{Fig:Penrose})  and having positive frequencies with respect to the Eddington coordinate $v$ along  $\{U\to -\infty, V>0\}$ (i.e., along past null infinity, see Fig.~\ref{Fig:Penrose}); the second subset
is defined by being positive frequency with respect to the Kruskal
coordinate $U$ along $H_L\cup H_\text{past}$ and having no upward excitations from past null infinity. The Unruh state
\cite{Unruh:1976} is then defined as the quantum state which is annihilated
by the annihilation operator coefficients when expanding the quantum
field in Unruh modes. The Unruh state is the state of relevance for
astrophysical BHs since it models a BH evaporating via the emission
of quantum, thermal (Hawking) radiation.

Regularity of the Unruh state at the EH is anticipated for decades.
We note, however, that it has not yet been actually proven in the case of a scalar field in
Kerr, but our numerical results here (see Fig.~\ref{Fig:near_EH_both})
provide -- for the first time, to the best of our knowledge -- numerical
support for it.\footnote{As mentioned in the Introduction, in Ref.~\cite{KerrIH:2022} we calculated the renormalized flux components in the Unruh state on the CH by the method of state subtraction. We note that the reference state we used is similarly expected to be regular on the CH, which  is also supported by numerical calculations, but no actual rigorous proof is known.}
Furthermore, regularity of the Unruh state on the EH and up to, but excluding, the CH, has been proven in~\cite{hafner2024hadamard} for massless fermions in Kerr (and in~\cite{2023AnHP...24.2401K} for scalars in Kerr-dS for sufficiently small angular momentum of the BH). 

We denote by $\left\langle T_{\mu\nu}(x)\right\rangle _{\text{ren}}^{U}$
the RSET at the spacetime point $x$ when the field is in the Unruh
state. As mentioned in the Introduction, in this paper we are interested
in the calculation of, specifically, the energy \textit{flux} components
of the RSET in the Unruh state, $\left\langle T_{yy}(x)\right\rangle _{\text{ren}}^{U}$,
where $y=u,v$, as well as the Unruh-state renormalized field square $\left\langle \Phi^2(x)\right\rangle _{\text{ren}}^{U}$.

\subsection{The conserved quantity\label{subsec:preliminary:fluxes} }

As reflected from Eq.~\eqref{eq:trace_rev} along with the fact that
$g_{uu}=g_{vv}$ {[}in coordinates $\left(u,v,\theta,\tilde{\varphi}\right)$
where $\tilde{\varphi}$ may be $\varphi$, $\varphi_{-}$ or $\varphi_{+}${]},
the difference between the two flux components $T_{uu}$ and $T_{vv}$ (at any angle $\theta$)
equals its trace-reversed counterpart: 
\begin{equation}
\overline{T}_{uu}\left(r,\theta\right)-\overline{T}_{vv}\left(r,\theta\right)=T_{uu}\left(r,\theta\right)-T_{vv}\left(r,\theta\right)\,.\label{eq:Tuu-Tvv(bar=00003Dstan)}
\end{equation}
Energy-momentum conservation, along with stationarity of the background 
and of the quantum state, implies $r$-independence of this quantity (now applied to the RSET) times
$r^{2}+a^{2}$ (related to an area element), that is 
\[
\frac{d}{dr}\left[\left(r^{2}+a^{2}\right)\left(\left\langle T_{uu}\left(r,\theta\right)\right\rangle _{\text{ren}}-\left\langle T_{vv}\left(r,\theta\right)\right\rangle _{\text{ren}}\right)\right]=0\,.
\]
We accordingly define the $r$-independent (yet $\theta$-dependent)
quantity $\mathcal{F}\left(\theta\right)$, which we sometimes dub ``the conserved quantity'':
\begin{equation}
\mathcal{F}\left(\theta\right)\equiv\left(r^{2}+a^{2}\right)\left(\left\langle T_{uu}\left(r,\theta\right)\right\rangle _{\text{ren}}-\left\langle T_{vv}\left(r,\theta\right)\right\rangle _{\text{ren}}\right)\,.\label{eq:conserved_quantity}
\end{equation}

In the Unruh state, carrying $\mathcal{F}\left(\theta\right)$
to infinity (where only the outflux $\langle T_{uu}\rangle _{\text{ren}}$ exists) shows it coincides with the Hawking energy outflux (per
unit solid angle in the $\theta$ direction). The corresponding mode-sum
expression (see Eq.~(B42) in Ref.~\cite{HTPF:2022}) is 
\begin{equation}
\mathcal{F}\left(\theta\right)=\frac{\hbar}{8\pi^{2}}\sum_{l=0}^\infty\sum_{m=-l}^l\int_{0}^{\infty}\left[S_{lm}^{\omega}\left(\theta\right)\right]^{2}\omega\left[\coth\left(\frac{\pi\omega_{+}}{\kappa_{+}}\right)-1\right]\left(1-\left|\rho_{\omega lm}^{\text{up}}\right|^{2}\right)\text{d}\omega\,.\label{eq:F(theta)}
\end{equation}
Evidently, the RHS only depends on scattering in the BH exterior (via
$\rho_{\omega lm}^{\text{up}}$) -- but is independent of the radial
function (and of $r$).

In this paper, we mainly focus on the axis of rotation of Kerr, i.e. $\theta=0$. Plugging  $\theta=0$ in Eq.~\eqref{eq:F(theta)}, only $m=0$ contributes to the $m$-sum (as mentioned, due to $S_{lm}^{\omega}\left(\theta=0\right)\propto \delta_m^0$). Hence, the Hawking energy outflux per unit solid angle in the polar direction is

\begin{equation}
\mathcal{F}_0\equiv\mathcal{F}(\theta=0)=\frac{\hbar}{8\pi^{2}}\sum_{l=0}^\infty\int_{0}^{\infty}\left[S_{\omega l}(0)\right]^{2}\omega\left[\coth\left(\frac{\pi\omega}{\kappa_{+}}\right)-1\right]\left(1-\left|\rho_{\omega l}^{\text{up}}\right|^{2}\right)\text{d}\omega\,.\label{eq:F(0)}
\end{equation}
where $S_{\omega l}$ and $\rho^\text{up}_{\omega l}$ are, respectively, $S^\omega_{lm}$ and $\rho^\text{up}_{\omega lm}$ restricted to $m=0$, as defined later in Eq.~\eqref{eq:m=0 quantities}.

\section{The PMR $t$-splitting procedure inside Kerr\label{sec:The-t-splitting-procedure}}

The $t$-splitting method generally involves splitting the point of interest in the $t$ direction, and utilizing the symmetry of a $t$-independent  background to decompose the known counterterms into frequency modes and perform the regularization procedure on a frequency mode-by-mode basis.
This method has been used by, e.g., \cite{Candelas:1980,Candelas_Jensen:1986,Ander_His_Sam:1995} for computations outside
and inside static spherical BHs. However, the methods involved are generally inapplicable in Kerr (off the axis of symmetry).

\begin{figure}[h!]
\centering \includegraphics[scale=0.3]{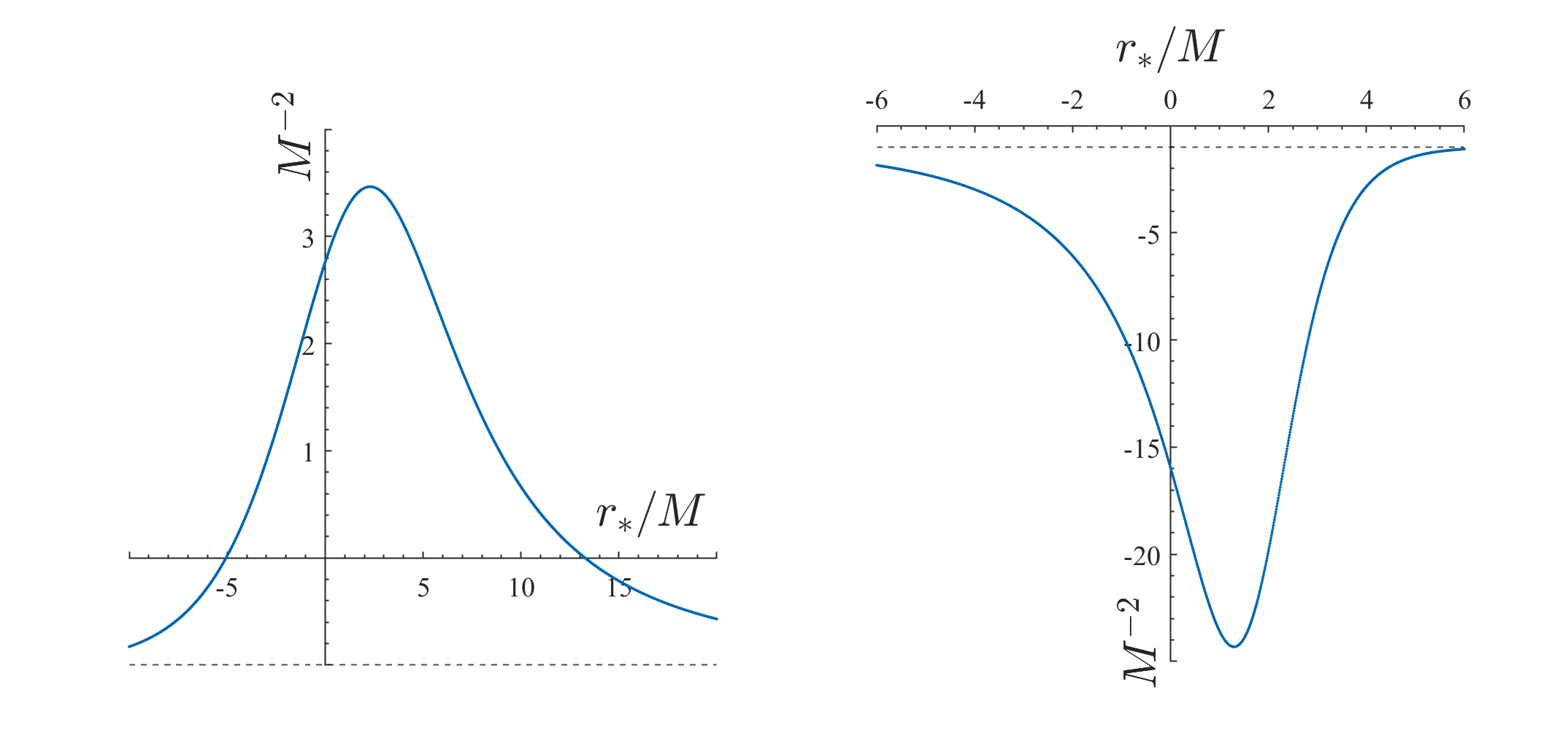} \caption{From the form of the radial equation \eqref{eq:radKerr}, the quantity
admitting the usual role of a potential is\emph{ minus }$V_{\omega lm}$,
the so-called effective potential given in Eq.~\eqref{eq:KerrPot}.
Here we portray $-V_{\omega lm}$ as a function of $r_{*}$ given
in Eq.~\eqref{eq:rstarKerr}, for the mode $\omega=1/M$, $l=10$
and $m=0$ (this choice of $m$ accords with the pole, and the choice
of $l$ is aimed to portray the typical large-$l$ behavior). The horizontal
dashed line corresponds to the asymptotic value of $-V_{\omega lm}$
for that mode, that is $-\omega^{2}$, at all asymptotic domains $r_*\to \pm\infty$ [see Eq.~\eqref{eq:asymPot} with $m=0$]. \emph{Left}: $-V_{\omega lm}$
for the mentioned mode at the BH exterior, acting as a potential barrier.
Here, the EH (respectively spatial infinity) is located at $r_{*}\to-\infty$ (respectively $r_{*}\to\infty$). 
\emph{Right}: $-V_{\omega lm}$ for the mentioned
mode at the BH interior, acting as a potential well.
Here, the EH (IH) is located at $r_{*}\to-\infty$ ($r_{*}\to\infty$).
}
\label{Fig:potential}
\end{figure}

In recent years, the $t$-splitting variant of the PMR method  was
developed \cite{AAt:2015,LeviRSET:2017}, allowing practical computations on \emph{stationary} backgrounds. It has been since implemented outside a
rotating BH \cite{LeviEilonOriMeentKerr:2017}, as well as outside
and inside a spherical (charged or neutral) BH \cite{AAt:2015,LeviRSET:2017}. In this section, we describe the implementation
of the $t$-splitting variant of PMR for computing $\left\langle \Phi^{2}\right\rangle _{\text{ren}}^{U}$
and the renormalized fluxes, $\left\langle T_{uu}\right\rangle _{\text{ren}}^{U}$
and $\left\langle T_{vv}\right\rangle _{\text{ren}}^{U}$,
in the \emph{interior} (up to, but excluding, the IH) of a Kerr BH, at the \emph{pole} (that
is, $r_{-}<r<r_{+}$ and $\theta=0$). There are two notable differences
between the BH exterior and interior, which are then reflected in
some aspects of the corresponding $t$-splitting scheme: (\emph{i})
Unlike its exterior behavior, for large values of $l$ the effective
potential \eqref{eq:KerrPot} inside the BH acts as a potential \emph{well}
rather than a potential \emph{barrier} (see Fig.~\ref{Fig:potential},
which allows appreciating this difference visually for a selected
mode). Then, while outside the BH the field modes decay exponentially
in $l$ for fixed $\omega$ (making the numerical implementation very
efficient), in the interior we encounter a \emph{diverging} mode-sum
(see Eq.~\eqref{eq:Pren_out}; this problem will be discussed in what follows and in Appendix~\ref{App:The-intermediate-divergence}).
(\emph{ii}) Outside the BH there exist  null geodesics connecting the points $x$ and $x'$ involved in the splitting,
which introduce an oscillatory behavior of the $\omega$-integrand
(see Ref.~\cite{LeviRSET:2017}). This is not the case inside, where
$t$ changes its nature and becomes spacelike. (See, however, the analytic extension variant in Appendix~\ref{App:The-analytic-extension}.)

We now present a schematic overview of the $t$-splitting procedure, specializing in its PMR implementation 
inside a Kerr BH, at the pole. For that matter, we denote the quantity
of interest (either $\left\langle \Phi^{2}\right\rangle^U $ or the
fluxes $\left\langle T_{yy}\right\rangle^U $) generally by $P$, whose individual mode contribution is generally denoted by $E_{\omega lm}$. The latter 
is comprised of an angular dependence (being the squared spheroidal
wavefunction, $\left[S_{lm}^{\omega}\left(\theta\right)\right]^{2}$)
times a function of $r$ {[}composed of the internal radial function
$\psi_{\omega lm}^{\text{int}}$ \eqref{eq:psi_int_BC} and its derivative
$d\psi_{\omega lm}^{\text{int}}/dr_{*}${]}. We shall use the term `\emph{bare}' expression for an expectation value $P$ 
(or, loosely, just `bare quantity')
when such expectation value has not yet been renormalized, in other words, 
a bare expression is the 
formal expression\footnote{Henceforth, the expressions for all bare quantities are understood
to be merely formal expressions.} for the corresponding unrenormalized bi-tensor evaluated at coincidence ($x=x'$). Since this paper's focus
is on the pole, only $m=0$ has a non-vanishing contribution to the
sum over $m$ (since, as indicated earlier, $S_{l\left(m\neq0\right)}^{\omega}\left(0\right)=0$).
Thus we may remove the $m$ index, and write the bare 
mode-sum for a quantity $P$  at the pole as 
\begin{equation}
P_{\text{bare}}\left(x\right)=\int_{0}^{\infty}\left(\sum_{l=0}^{\infty}E_{\omega l}\left(x\right)\right)\text{d}\omega\,,\label{eq:Pbare-1}
\end{equation}
where $ E_{\omega l}$ is $E_{\omega lm}$ reduced to $m=0$.

We would like to regularize this sum using $t$-splitting, in which
we introduce a split in the $t$ direction, denoted $\varepsilon\equiv t'-t$.
Outside the BH \cite{LeviRSET:2017}, the renormalized quantity $P_{\text{ren}}$
is given by
\begin{equation}
P_{\text{ren}}\left(x\right)=\lim_{\varepsilon\to0}\left[\int_{0}^{\infty}\cos\left(\omega\varepsilon\right)\left(\sum_{l=0}^{\infty}E_{\omega l}\left(x\right)\right)\text{d}\omega-C\left(\varepsilon,x\right)\right],\label{eq:Pren_out}
\end{equation}
where $C\left(\varepsilon,x\right)$ is the  counterterm (specifically,
$G^{\text{CT}}(x,x')/2$ or $T_{\mu\nu}^{\text{CT}}(x,x')$ mentioned
in the Introduction in the cases of, respectively, $\left\langle \Phi^{2}\right\rangle $
or the fluxes) \cite{DeWittBook:1965,Christensen:1978} translated
to be given in terms of $\varepsilon$ as described in Refs.~\cite{AAt:2015,LeviRSET:2017}. However, the sum $\sum_{l=0}^{\infty}E_{\omega l}$
fails to converge inside the BH: In fact, we find  that at large
$l$ the sequence $E_{\omega l}$ behaves as
\begin{equation}
E_{\omega l}^{\text{div}}\equiv c_{0}+c_{1}\omega^{2}+c_{2}l(l+1)\label{eq:Ewl_div}
\end{equation}
with an additional  $O\left(1/l\left(l+1\right)\right)$ piece,
where the coefficients $c_{0},c_{1}$ and $c_{2}$ are  independent
of $l$ and $\omega$ (but are functions of  position). We establish
this form (analytically for the leading order, being $c_{0}$ for
the field square and $c_{2}l\left(l+1\right)$
for the fluxes,
and empirically for the remaining terms) in Sec.~\ref{subsec:Intermediate-blindspots}
and the figures within.  In particular, it is crucial that the $O\left(l^{0}\right)$
term, $c_{0}+c_{1}\omega^{2}$, presents the \emph{exact} dependence
on $\omega$, not just a leading order in an expansion in $\omega$.

The diverging piece $E_{\omega l}^{\text{div}}$ constitutes what
we shall call the \emph{intermediate-divergence }(ID) problem, and
we hereafter refer to $E_{\omega l}^{\text{div}}$ as the ID\footnote{We note that, in Refs.~\cite{HowardRSET:1984,Anderson:1990,Ander_His_Sam:1995}, divergences in the $l$-sum for fixed frequency where also observed in the Euclidean Green function for points separated along the time direction outside a Schwarzschild black hole. These were referred to as ``superficial" divergences and they were removed by subtracting Dirac-$\delta$  distributions  (and derivatives of them).
We also note that, on the other hand,  still outside a BH but in the Lorentzian case and either in Schwarzschild or Kerr, no such IDs are present~\cite{AAt:2015,LeviEilonOriMeentKerr:2017}.}. The
treatment of this delicate technical issue is presented in Appendix
\ref{App:The-intermediate-divergence}, and requires introducing
an additional ``small'' split in the $\theta$ direction (namely,
a split that is taken to zero before closing the split in the $t$
direction).  At this point, we only mention that the ID {[}which,
crucially, has the form given in Eq.~\eqref{eq:Ewl_div}{]} may be
simply subtracted -- postponing the justification of this subtraction
to the mentioned Appendix.

Then, our renormalized quantity is
\begin{equation}
P_{\text{ren}}\left(x\right)=\lim_{\varepsilon\to0}\left[\int_{0}^{\infty}\cos\left(\omega\varepsilon\right)E^{\text{basic}}\left(\omega,x\right)\text{d}\omega-C\left(\varepsilon,x\right)\right]\,,\label{eq:P_ren}
\end{equation}
where 
\begin{equation}
E^{\text{basic}}\left(\omega,x\right)\equiv\sum_{l=0}^{\infty}\left[E_{\omega l}\left(x\right)-E_{\omega l}^{\text{div}}\left(x\right)\right],\label{eq:E_basic_def}
\end{equation}
which we shall refer to as the \emph{basic integrand function}.

Finally, we next rewrite Eq.~\eqref{eq:P_ren} in its PMR form. 
First, $C\left(\varepsilon,x\right)$ is expanded for small $\varepsilon$. The $O(\varepsilon^0)$ term in the expansion is denoted by $e\left(x\right)$. The rest of the expansion of $C\left(\varepsilon,x\right)$ is Fourier-decomposed and the Fourier modes are denoted by $E^{\text{sing}}\left(\omega,x\right)$.
This $\omega$-dependent PMR counterterm, $E^{\text{sing}}\left(\omega,x\right)$, may be subtracted from
the integrand in $\omega$ prior to integration  over $\omega$, and the so-called \emph{finite counterterm} $e\left(x\right)$ remains to be subtracted after integration. This way, the entire computation
is done at coincidence. The final expression for the PMR renormalized
quantity at a point $x$ inside the BH is then
\begin{equation}
P_{\text{ren}}\left(x\right)=\int_{0}^{\infty}\left[E^{\text{basic}}\left(\omega,x\right)-E^{\text{sing}}\left(\omega,x\right)\right]\text{d}\omega-e\left(x\right)\,.\label{eq:P_ren_PMR}
\end{equation}

  This form, including the construction of the PMR counterterms $E^\text{sing}(\omega,x)$ and $e(x)$, is  established in
Refs.~\cite{AAt:2015,LeviRSET:2017} for a stationary BH exterior,
and extended here to the interior.

The various components ($E_{\omega l}\left(x\right)$, $E_{\omega l}^{\text{div}}\left(x\right)$,
and the counterterms $E^{\text{sing}}\left(\omega,x\right)$ and $e\left(x\right)$)
are given explicitly (in our case of the polar Kerr interior), for
both the field square and the fluxes in the Unruh state, in the following
sections: For the individual bare mode contribution $E_{\omega l}$,
see Sec.~\ref{subsec:The-bare-mode}; the ID (the diverging piece
$E_{\omega l}^{\text{div}}$) is discussed in Sec.~\ref{subsec:Intermediate-blindspots},
with the leading order in $l$ analytically worked out; the PMR
counterterms $E^{\text{sing}}\left(\omega,x\right)$ and $e\left(x\right)$
are given in Sec.~\ref{subsec:Counterterms}. The information provided
in these three subsections, along with Eq.~\eqref{eq:P_ren_PMR},
comprises the $t$-splitting PMR recipe for the computation of $\left\langle \Phi^{2}\right\rangle _{\text{ren}}^{U}$
and $\left\langle T_{yy}\right\rangle _{\text{ren}}^{U}$
at the pole inside a Kerr BH.

\subsection{The bare mode contribution\label{subsec:The-bare-mode}}

In what follows, we write the individual mode contribution to the
bare mode-sum expression of the quantities of interest -- $\left\langle \Phi^{2}\right\rangle ^{U}$
and $\left\langle T_{yy}\right\rangle ^{U}$ -- for a
general $r$ value inside a Kerr BH, at the pole ($\theta=0$). The
presented results follow immediately from computations done in Ref.~\cite{HTPF:2022}, as we hereafter describe.

\subsubsection{$\left\langle \Phi^{2}\right\rangle _{\text{bare}}^{U}$\label{subsec:phi2_bare}}

We begin with the HTPF in the Unruh state at a Kerr BH interior, given
in Eqs.~(B8) and (B9) in Ref.~\cite{HTPF:2022} (along with Eq.~(6.38) therein)
as
\begin{equation}
G_{U}^{\left(1\right)}\left(x,x'\right)=\hbar\int_{0}^{\infty}\left[\sum_{l=0}^{\infty}\sum_{m=-l}^{l}G_{\omega lm}\left(x,x'\right)\right]\text{d}\omega\,,\label{eq:Gmodesum}
\end{equation}
where 
\begin{align}
 & G_{\omega lm}\left(x,x'\right)=\nonumber \\
 & \frac{\left|\omega_{+}\right|}{\omega_{+}}\left[\coth\left(\frac{\pi\omega_{+}}{\kappa_{+}}\right)\left(\left\{ f_{\omega lm}^{L}\left(x\right),f_{\omega lm}^{L*}\left(x'\right)\right\} +\left|\rho_{\omega lm}^{\text{up}}\right|^{2}\left\{ f_{\omega lm}^{R}\left(x\right),f_{\omega lm}^{R*}\left(x'\right)\right\} \right)\right.\label{eq:Gwlm}\\
 & \left.+2\text{cosech}\left(\frac{\pi\omega_{+}}{\kappa_{+}}\right)\Re\left(\rho_{\omega lm}^{\text{up}}\left\{ f_{\omega lm}^{R}\left(x\right),f_{\omega lm}^{L*}\left(x'\right)\right\} \right)+\frac{\omega_{+}}{\omega}\left|\tau_{\omega lm}^{\text{in}}\right|^{2}\left\{ f_{\omega lm}^{R}\left(x\right),f_{\omega lm}^{R*}\left(x'\right)\right\} \right]\,.\nonumber 
\end{align}
From the HTPF one may easily obtain a mode-sum expression for the
bare (i.e., unrenormalized) expectation value of the field square
in the Unruh state,  
\[
\left\langle \Phi^{2}\left(x\right)\right\rangle _{\text{bare}}^{U}=\frac{\hbar}{2}\int_{0}^{\infty}\left(\sum_{l=0}^{\infty}\sum_{m=-l}^{l}\left[\lim_{x'\to x}G_{\omega lm}\left(x,x'\right)\right]\right)\text{d}\omega\,.
\]
The square brackets yield an expression whose $\theta$-dependence
is factored out as $\left[S_{lm}^{\omega}\left(\theta\right)\right]^{2}$,
which gives rise to a major simplification at the pole: since $S_{lm}^{\omega}\left(\theta=0\right)\propto\delta_{0m}$,
we are left only with the $m=0$ contribution to the sum over $m$.
To ease notation from this point on, we denote $m=0$ quantities by
simply removing the $m$ index. For example, we denote
\begin{equation}\label{eq:m=0 quantities}
S_{\omega l}\left(\theta\right)\equiv S_{l\left(m=0\right)}^{\omega}\left(\theta\right),\,\,\,\,\rho_{\omega l}^{\text{up}}\equiv\rho_{\omega l\left(m=0\right)}^{\text{up}},\,\,\,\,\,\psi_{\omega l}^{\text{int}}\left(r\right)\equiv\psi_{\omega l\left(m=0\right)}^{\text{int}}\left(r\right)
\end{equation}
\[
V_{\omega l}\left(r\right)\equiv V_{\omega l\left(m=0\right)}\left(r\right),\,\,\,\lambda_{l}\left(a\omega\right)\equiv\lambda_{l\left(m=0\right)}\left(a\omega\right),\,\,\,\text{etc.}
\]
Plugging the mode functions given in Eq.~\eqref{eq:EddModesRL} into
Eq.~\eqref{eq:Gwlm} and taking $m=0$ followed by the coincidence
limit and $\theta=0$, one obtains the bare mode-sum $\left\langle \Phi^{2}\right\rangle _{\text{bare}}^{U}$
at the pole in the interior of a Kerr BH 
\begin{equation}
\left\langle \Phi^{2}\right\rangle _{\text{bare}}^{U}=\int_{0}^{\infty}\left(\sum_{l=0}^{\infty}E_{\omega l}\right)\text{d}\omega\,,\label{eq:phi2_bare}
\end{equation}
where\footnote{Here we write $\left\langle \Phi^{2}\right\rangle _{\text{bare}}^{U}$
as in Eq.~\eqref{eq:P_ren_PMR}, choosing $E_{\omega l}$ to denote
the individual mode contribution. For $\left\langle T_{yy}\right\rangle _{\text{bare}}^{U}$
which follows next, we choose a different notation as the analog of
$E_{\omega l}$ (to be explained in Sec. (\ref{subsec:bare_Tyy})).
The same note applies to $E_{\omega l}^{\text{div}}$ in what follows.}
\begin{align}
E_{\omega l} & =\hbar\frac{\left[S_{\omega l}\left(0\right)\right]^{2}}{8\pi^{2}\omega\left(r^{2}+a^{2}\right)}\times\nonumber \\
 & \left[\coth\left(\frac{\pi\omega}{\kappa_{+}}\right)\left|\psi_{\omega l}^{\text{int}}\right|^{2}\left(1+\left|\rho_{\omega l}^{\text{up}}\right|^{2}\right)+2\,\text{cosech}\left(\frac{\pi\omega}{\kappa_{+}}\right)\Re\left(\rho_{\omega l}^{\text{up}}\left(\psi_{\omega l}^{\text{int}}\right)^{2}\right)+\left(1-\left|\rho_{\omega l}^{\text{up}}\right|^{2}\right)\left|\psi_{\omega l}^{\text{int}}\right|^{2}\right]\,.\label{eq:Ewl_phi2}
\end{align}

\subsubsection{$\left\langle T_{uu}\right\rangle _{\text{bare}}^{U}$
\text{and} $\left\langle T_{vv}\right\rangle _{\text{bare}}^{U}$\label{subsec:bare_Tyy}}

As in Eq.~(B10) in Ref.~\cite{HTPF:2022}, taking $\alpha\beta$ to
be $yy$, (where, again, $y$ denotes either $u$ or $v$), we begin
with the following ``bare'' expression for the trace-reversed flux components at a general $\theta$, with the azimuthal coordinate taken as  $\tilde{\varphi}$ which may be either $\varphi$, $\varphi_{+}$ or $\varphi_{-}$:\footnote{To conform with the $t$-splitting procedure described above, we change
the order of summation and integration appearing in Eq.~(B10) in Ref.
\cite{HTPF:2022} to have the integration over $\omega$ performed
last. (Clearly, performing such an exchange there should not matter
since this bare quantity diverges in any case.) 
The procedure used here has been constructed and justified for this
specific order of summation and integration.}
\[
\left\langle \overline{T}_{yy}\right\rangle _{\text{bare}}^{U}=\int_{0}^{\infty}\left(\sum_{l=0}^{\infty}\sum_{m=-l}^{l}\overline{T}_{yy\left(\omega lm\right)}\right)\text{d}\omega\,,
\]
with the individual mode contribution:
\[
\overline{T}_{yy\left(\omega lm\right)}\equiv\frac{\hbar}{2}\lim_{x'\to x}\left[G_{\omega lm}\left(x,x'\right)_{,yy'}\right]\,.
\]
In Appendix B of Ref.~\cite{HTPF:2022} we obtain an expansion of
the latter in powers of $\Delta$ {[}given in Eq.~\eqref{eq:Delta}{]}:
\[
\overline{T}_{yy\left(\omega lm\right)}=\overline{T}_{yy\left(\omega lm\right)}^{\mathcal{A}}+\overline{T}_{\left(\omega lm\right)}^{\mathcal{B}}\Delta+\overline{T}_{\left(\omega lm\right)}^{\mathcal{C}}\Delta^{2}\,,
\]
with the coefficients $\overline{T}_{yy\left(\omega lm\right)}^{\mathcal{A}}$,
$\overline{T}_{\left(\omega lm\right)}^{\mathcal{B}}$ and $\overline{T}_{\left(\omega lm\right)}^{\mathcal{C}}$ given in Eqs.~(B35)-(B38) therein, including a dependence on the choice of azimuthal coordinate  $\tilde{\varphi}$.

Focusing now on $\theta=0$ (which is hereafter implied in the notation), a few simplifications occur:

(\emph{i}) As mentioned earlier, since $S_{lm}^{\omega}\left(\theta=0\right)\propto\delta_{0m}$, only $m=0$ remains in the sum over $m$.

(\emph{ii}) The metric components $g_{yy}$ are identically $0$, hence trace-reversal (through Eq.~\eqref{eq:trace_rev}) does not change the flux components, i.e. $T_{yy}=\overline{T}_{yy}$.

(\emph{iii}) The flux components $T_{yy}$ are the same whether the azimuthal coordinate is $\varphi$, $\varphi_+$ or $\varphi_-$.

We may thus write the bare Unruh mode contribution to the fluxes $\left\langle T_{yy}\right\rangle_{\text{bare}}^{U}$ {[}in
the form of Eq.~\eqref{eq:Pbare-1}, with $T_{yy\left(\omega l\right)}$
taking the role of $E_{\omega l}${]}
\begin{equation}
\left\langle T_{yy}\right\rangle _{\text{bare}}^{U}=\int_{0}^{\infty}\left(\sum_{l=0}^{\infty}T_{yy\left(\omega l\right)}\right)\text{d}\omega\,,\label{eq:Tyy_bare}
\end{equation}
where $T_{yy\left(\omega l\right)}$ is the $m=0$, $\theta=0$  version of $\overline{T}_{yy\left(\omega l m\right)} $ of Ref.~\cite{HTPF:2022}, 
and similarly admits the following expansion in $\Delta$:
\begin{equation}
T_{yy\left(\omega l\right)}=T_{yy\left(\omega l\right)}^{\mathcal{A}}+T_{\left(\omega l\right)}^{\mathcal{B}}\Delta+T_{\left(\omega l\right)}^{\mathcal{C}}\Delta^{2}\,.\label{eq:Tyy_wl_expanded}
\end{equation}
Taking $m=0$ and $\theta=0$ in Eqs.~(B35)--(B38) in Ref.~\cite{HTPF:2022}, 
\begin{align}
 & T_{uu\left(\omega l\right)}^{\mathcal{A}}=\hbar\frac{\left[S_{\omega l}\left(0\right)\right]^{2}}{32\pi^{2}\omega\left(r^{2}+a^{2}\right)}\label{eq:TAuu}\\
 & \left(\coth\left(\frac{\pi\omega}{\kappa_{+}}\right)\left[\left|\psi_{\omega l,r_{*}}^{\text{int}}\right|^{2}+\omega^{2}\left|\psi_{\omega l}^{\text{int}}\right|^{2}+2\omega^{2}+\left|\rho_{\omega l}^{\text{up}}\right|^{2}\left(\left|\psi_{\omega l,r_{*}}^{\text{int}}\right|^{2}+\omega^{2}\left|\psi_{\omega l}^{\text{int}}\right|^{2}-2\omega^{2}\right)\right]\right.\nonumber \\
 & \left.+2\,\text{cosech}\left(\frac{\pi\omega}{\kappa_{+}}\right)\Re\left(\rho_{\omega l}^{\text{up}}\left[\left(\psi_{\omega l,r_{*}}^{\text{int}}\right)^{2}+\omega^{2}\left(\psi_{\omega l}^{\text{int}}\right)^{2}\right]\right)+\left(1-\left|\rho_{\omega l}^{\text{up}}\right|^{2}\right)\left(\left|\psi_{\omega l,r_{*}}^{\text{int}}\right|^{2}+\omega^{2}\left|\psi_{\omega l}^{\text{int}}\right|^{2}-2\omega^{2}\right)\right)\,,\nonumber 
\end{align}
\begin{align}
 & T_{vv\left(\omega l\right)}^{\mathcal{A}}=\hbar\frac{\left[S_{\omega l}\left(0\right)\right]^{2}}{32\pi^{2}\omega\left(r^{2}+a^{2}\right)}\label{eq:TAvv}\\
 & \left(\coth\left(\frac{\pi\omega}{\kappa_{+}}\right)\left[\left|\psi_{\omega l,r_{*}}^{\text{int}}\right|^{2}+\omega^{2}\left|\psi_{\omega l}^{\text{int}}\right|^{2}-2\omega^{2}+\left|\rho_{\omega l}^{\text{up}}\right|^{2}\left(\left|\psi_{\omega l,r_{*}}^{\text{int}}\right|^{2}+\omega^{2}\left|\psi_{\omega l}^{\text{int}}\right|^{2}+2\omega^{2}\right)\right]\right.\nonumber \\
 & \left.+2\,\text{cosech}\left(\frac{\pi\omega}{\kappa_{+}}\right)\Re\left(\rho_{\omega l}^{\text{up}}\left[\left(\psi_{\omega l,r_{*}}^{\text{int}}\right)^{2}+\omega^{2}\left(\psi_{\omega l}^{\text{int}}\right)^{2}\right]\right)+\left(1-\left|\rho_{\omega l}^{\text{up}}\right|^{2}\right)\left(\left|\psi_{\omega l,r_{*}}^{\text{int}}\right|^{2}+\omega^{2}\left|\psi_{\omega l}^{\text{int}}\right|^{2}+2\omega^{2}\right)\right)\,,\nonumber 
\end{align}
 
\begin{align}
T_{\left(\omega l\right)}^{\mathcal{B}} & =-\hbar\frac{\left[S_{\omega l}\left(0\right)\right]^{2}r}{16\pi^{2}\omega\left(r^{2}+a^{2}\right)^{3}}\left(\coth\left(\frac{\pi\omega}{\kappa_{+}}\right)\Re\left(\psi_{\omega l}^{\text{int}}\psi_{\omega l,r_{*}}^{\text{int}*}\right)\left(1+\left|\rho_{\omega l}^{\text{up}}\right|^{2}\right)\right.\label{eq:TB}\\
 & \left.+2\text{cosech}\left(\frac{\pi\omega}{\kappa_{+}}\right)\Re\left(\rho_{\omega l}^{\text{up}}\psi_{\omega l}^{\text{int}}\psi_{\omega l,r_{*}}^{\text{int}}\right)+\left(1-\left|\rho_{\omega l}^{\text{up}}\right|^{2}\right)\Re\left(\psi_{\omega l}^{\text{int}}\psi_{\omega l,r_{*}}^{\text{int}*}\right)\right)\,,\nonumber 
\end{align}

\begin{align}
 & T_{\left(\omega l\right)}^{\mathcal{C}}=\hbar\frac{\left[S_{\omega l}\left(0\right)\right]^{2}r^{2}}{32\pi^{2}\omega\left(r^{2}+a^{2}\right)^{5}}\label{eq:TC}\\
 & \left(\coth\left(\frac{\pi\omega}{\kappa_{+}}\right)\left|\psi_{\omega l}^{\text{int}}\right|^{2}\left(1+\left|\rho_{\omega l}^{\text{up}}\right|^{2}\right)+2\,\text{cosech}\left(\frac{\pi\omega}{\kappa_{+}}\right)\Re\left(\rho_{\omega l}^{\text{up}}\left(\psi_{\omega l}^{\text{int}}\right)^{2}\right)+\left(1-\left|\rho_{\omega l}^{\text{up}}\right|^{2}\right)\left|\psi_{\omega l}^{\text{int}}\right|^{2}\right)\,.\nonumber 
\end{align}

Combining Eqs.~\eqref{eq:TAuu}-\eqref{eq:TC} with Eqs.~\eqref{eq:Tyy_bare} and \eqref{eq:Tyy_wl_expanded},
taking either $y=u$ or $v$, one obtains the bare mode-sum expression
for the fluxes $\left\langle T_{uu}\right\rangle _{\text{bare}}^{U}$
and $\left\langle T_{vv}\right\rangle _{\text{bare}}^{U}$
at the pole inside a Kerr BH.

\subsection{Intermediate divergence\label{subsec:Intermediate-blindspots}}

As mentioned at the beginning of Sec.~\ref{sec:The-t-splitting-procedure},
the ID captures the large-$l$ diverging behavior of the individual
mode contribution, and is generally (for our quantities of interest,
$\left\langle \Phi^{2}\right\rangle $ and $\left\langle T_{yy}\right\rangle $)
of the form given in Eq.~\eqref{eq:Ewl_div}. In this subsection,
we aim to justify this form as well as analytically compute the large-$l$
leading order ID coefficient for both $\left\langle \Phi^{2}\right\rangle $
and $\left\langle T_{yy}\right\rangle $ (being $c_{0}$
for $\left\langle \Phi^{2}\right\rangle $ and $c_{2}$ for $\left\langle T_{yy}\right\rangle $).

\subsubsection{The ID of $\left\langle \Phi^{2}\right\rangle $\label{subsec:IBS_phi2}}

We begin with Eq.~\eqref{eq:Ewl_phi2}, and wish to take its
large-$l$ regime ($l\gg1$). To this end, we denote
\begin{equation}
\tilde{l}\equiv l+\frac{1}{2}\,,\label{eq:l_tilde}
\end{equation}
and note that, at large $l$ (see e.g. Eq.~(67) in Ref.~\cite{RokhlinXiao:2007}),
\begin{equation}
S_{\omega l}\left(0\right)\simeq\sqrt{\tilde{l}}\,.\label{eq:S_large_l}
\end{equation}
In addition,  large-$l$ $m=0$ modes outside the BH undergo total
reflection, namely (see Eq.~\eqref{eq:rho_largeL})
\begin{equation}\label{eq:|rho|_largeL}
\left|\rho_{\omega l}^{\text{up}}\right|\simeq1\,.
\end{equation}
Applying these two very simple facts to Eq.~\eqref{eq:Ewl_phi2},
we are left with the intermediate expression at large $l$:
\begin{equation}
E_{\omega l}\simeq\hbar\frac{\tilde{l}}{4\pi^{2}\omega\left(r^{2}+a^{2}\right)}\left(\coth\left(\frac{\pi\omega}{\kappa_{+}}\right)\left|\psi_{\omega l}^{\text{int}}\right|^{2}+\text{cosech}\left(\frac{\pi\omega}{\kappa_{+}}\right)\Re\left[\rho_{\omega l}^{\text{up}}\left(\psi_{\omega l}^{\text{int}}\right)^{2}\right]\right)\,.\label{eq:phi2_IBS_intermediate}
\end{equation}

Putting together the large-$l$ WKB form of the interior radial
function given in Eq.~\eqref{eq:psiWKB}, and the large-$l$ reflection
coefficient $\rho_{\omega l}^{\text{up}}$ given in Eq.~\eqref{eq:rho_largeL},
one finds the following leading order large-$l$ expressions: 

\begin{align}
\frac{\left|\psi_{\omega l}^{\text{int}}\right|^{2}}{\omega\left(r^{2}+a^{2}\right)} & \simeq\frac{1}{\tilde{l}\sqrt{\left(r_{+}-r\right)\left(r-r_{-}\right)}}\left(\coth\left(\frac{\pi\omega}{\kappa_{+}}\right)+\left(-1\right)^{l}\text{cosech}\left(\frac{\pi\omega}{\kappa_{+}}\right)\cos\left[2\tilde{l}g\left(r\right)\right]\right),\label{eq:intermediate1}
\end{align}
\begin{equation}
\frac{\Re\left(\rho_{\omega l}^{\text{up}}\left(\psi_{\omega l}^{\text{int}}\right)^{2}\right)}{\omega\left(r^{2}+a^{2}\right)}\simeq\frac{-1}{\tilde{l}\sqrt{\left(r_{+}-r\right)\left(r-r_{-}\right)}}\left(\text{cosech}\left(\frac{\pi\omega}{\kappa_{+}}\right)+\left(-1\right)^{l}\coth\left(\frac{\pi\omega}{\kappa_{+}}\right)\cos\left[2\tilde{l}g\left(r\right)\right]\right),\label{eq:intermediate2}
\end{equation}
where $g\left(r\right)$ is given in Eq.~\eqref{eq:g(r)} below (in
fact, the exact form of $g\left(r\right)$ does not matter here).
Plugging these into Eq.~\eqref{eq:phi2_IBS_intermediate} (and recalling
$\coth^{2}x-\text{cosech}^{2}x=1$), we easily obtain the desired
large-$l$ plateau (for fixed $\omega$) of the sequence,
\[
E_{\omega l}\simeq\frac{\hbar}{4\pi^{2}\sqrt{\left(r_{+}-r\right)\left(r-r_{-}\right)}}\,,
\]
which is evidently independent of $\omega$ and $l$. That is, casting
into the general form of Eq.~\eqref{eq:Ewl_div}, we  analytically obtain for the
ID of $\left\langle \Phi^{2}\right\rangle $ (the full ID for the field square corresponds to just the leading-order asymptotics),
\begin{equation}
E_{\omega l}^{\text{div}}=c_{0}=\frac{\hbar}{4\pi^{2}\sqrt{\left(r_{+}-r\right)\left(r-r_{-}\right)}}\,,\label{eq:Ewl_div_phi2}
\end{equation}
and 
\begin{equation}
c_{1}=c_{2}=0\,.\label{eq:c1c2_zero}
\end{equation}

\begin{figure}[h!]
\centering \includegraphics[scale=0.3]{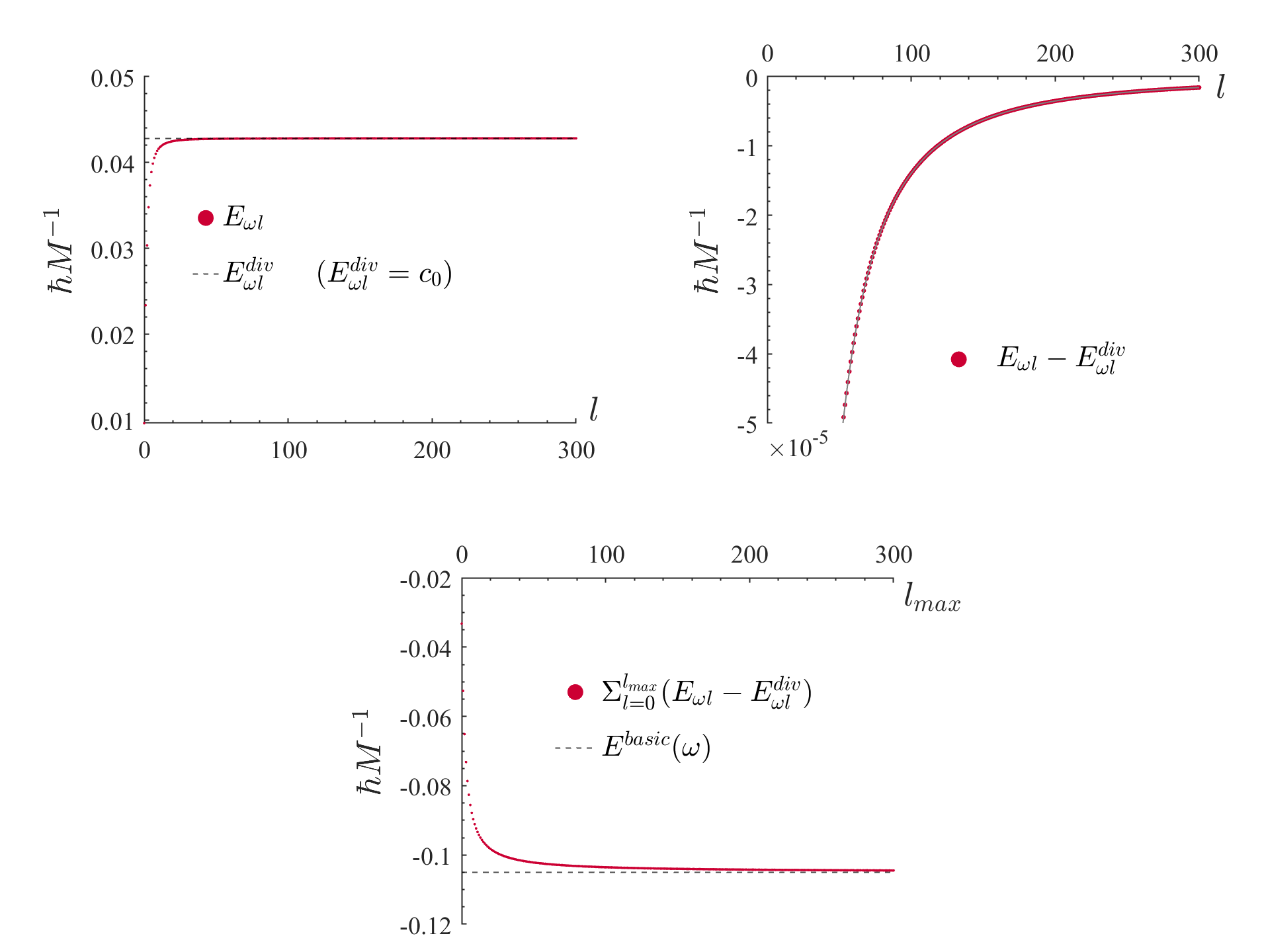}
\caption{The ID subtraction procedure is demonstrated for $\Phi^{2}$
for the specific case of $\omega=1/M$, at the point $r/M=0.9$ inside
an $a/M=0.8$ BH. \emph{Top left:} the red points
are the numerically-computed bare mode contribution, $E_{\omega l}$.
The dashed horizontal line presents the analytic value for $E_{\omega l}^{\text{div}}$
which is the constant $c_{0}$ given in Eq.~\eqref{eq:Ewl_div_phi2}.
\emph{Top right:} the difference $E_{\omega l}-E_{\omega l}^{\text{div}}$
as a function of $l$, which allows appreciating the convergence of
the sequence $E_{\omega l}$ to its large-$l$ analytically-computed
plateau value. The remaining large-$l$ behavior is numerically found to be $O\left(1/l(l+1)\right)$, as demonstrated by the $1/l(l+1)$ fit represented by a thin black line. The plot is trimmed vertically at $-5\times10^{-5}\hbar M^{-1}$,
for scale purposes. \emph{Bottom:} the red points
are the partial sums of $E_{\omega l}-E_{\omega l}^{\text{div}}$
over $l$ up to $l_{\text{max}}$, as a function of the latter. The
dashed horizontal line, corresponding to $l_{\text{max}}\to\infty$,
constitutes the resultant (basic) integrand value $E^{\text{basic}}\left(\omega=1/M\right)$.}
\label{Fig:phi_vs_l_4tiles}
\end{figure}

The correspondence of the large-$l$ behavior of $E_{\omega l}$
and the constant $E_{\omega l}^{\text{div}}$ computed here, as well
as the convergence of the $l$-sum of $E_{\omega l}-E_{\omega l}^{\text{div}}$,
is demonstrated in Fig.~\ref{Fig:phi_vs_l_4tiles} for a fixed typical
$\omega$, taken here to be $\omega=1/M$. The top left panel shows
both quantities $E_{\omega l}$ and $E_{\omega l}^{\text{div}}$ as
a function of $l$, and the top right panel shows their difference
$E_{\omega l}-E_{\omega l}^{\text{div}}$ (which is numerically found to behave like $1/l\left(l+1\right)$, as illustrated by the attached fit).
 The partial sums $\sum_{l=0}^{l_{\text{max}}}\left(E_{\omega l}-E_{\omega l}^{\text{div}}\right)$,
 displayed in the bottom  panel of the figure, demonstrate the
resulting convergence as  $l_{\text{max}}\to\infty$, yielding 
the (basic) integrand value at that $\omega$, $E^{\text{basic}}\left(\omega\right)$
defined in Eq.~\eqref{eq:E_basic_def} {[}this is done through a fit
as described in Appendix~\ref{subsec:numerical_methods}{]}. This
$E^{\text{basic}}$ value, represented by the dashed horizontal line,
is then highlighted in the left panel of Fig.~\ref{Fig:phi_integrand_2tiles}
by a bold red point at $\omega M=1$.\footnote{All figures in this section, Figs.~\ref{Fig:phi_vs_l_4tiles}-\ref{Fig:Tuu_integrand_2tiles},
correspond to the specific case $r/M=0.9$ -- which is a typical
point between the horizons -- inside an $a/M=0.8$ BH. They are 
aimed to illustrate the regularization procedure. The general picture
emerging from these figures is typical to the case $a/M=0.9$ as
well, and to generic $r$ values -- provided that $r$ is not too
close to the horizons. However, as the horizons are approached, it
becomes increasingly difficult to perform the procedure demonstrated
here (as the singular piece $E^{\text{sing}}\left(\omega\right)$
diverges at the horizons, see Sec.~\ref{subsec:Counterterms}).
In addition, although we did not explicitly check this, we expect
a similar picture to emerge also at other $a/M$ values (as long as
$a/M$ is not too close to $0$ or $1$). 

{[}In Figs.~\ref{Fig:phi_vs_l_4tiles} and \ref{Fig:Tuu_vs_l_4tiles}
we demonstrated the large-$l$ behavior for a specific $\omega$ value,
being $\omega=1/M$, but this large-$l$ behavior is shared by all
fixed $\omega$ values{]}.\label{fn:typical_case_figures}}

\subsubsection{The ID of $\left\langle T_{uu}\right\rangle $ and $\left\langle T_{vv}\right\rangle $\label{subsec:IBS_Tyy}}

The ID for $\left\langle T_{yy}\right\rangle $, which
we denote by $T_{yy\left(\omega l\right)}^{\text{div}}$, has been claimed above to take the form given in Eq.~\eqref{eq:Ewl_div},
with generally non-vanishing coefficients $c_{0},c_{1}$ and $c_{2}$
{[}unlike $\left\langle \Phi^{2}\right\rangle $ for which $c_{1},c_{2}=0$,
see Eq.~\eqref{eq:c1c2_zero}{]}. In the current subsection
we justify this form in two parts: first, analytically computing the
large-$l$ leading order coefficient $c_{2}$; and second, numerically
establishing the $O\left(l^{0}\right)$ behavior being $c_{0}+c_{1}\omega^{2}$,
as well as the convergence of the sum over $l$  following the ID
subtraction {[}as in Eq.~\eqref{eq:E_basic_def}{]}.

We begin with the individual mode contribution $T_{yy\left(\omega l\right)}$
given in Eqs.~\eqref{eq:Tyy_wl_expanded}-\eqref{eq:TC}, depending
in particular on the radial function $\psi_{\omega l}^{\text{int}}$.
The latter evolves according to the radial equation \eqref{eq:radKerr},
governed by the effective potential $V_{\omega l}$ obtained by setting $m=0$ in Eq.~\eqref{eq:KerrPot}. Notably, the dependence of $V_{\omega l}$ on $l$
comes entirely from  the ($m=0$) spheroidal-eigenvalue term $\propto\lambda_{l}\left(a\omega\right)$,
whose asymptotic behavior at large $l$ takes the form (see Theorem
10 in Ref.~\cite{RokhlinXiao:2007})
\begin{equation}
\lambda_{l}(a\omega)=l\left(l+1\right)+\frac{1}{2}a^{2}\omega^{2}+\hat{\lambda}_{ l}(a\omega)\label{eq:lambda-1}
\end{equation}
with the remainder term $\hat{\lambda}_{ l}(a\omega)$ vanishing at $l\to\infty$
(like $1/l\left(l+1\right)$). We then obtain the following simple form
for the effective potential holding asymptotically at large-$l$,
\begin{equation}
V_{\omega l}\simeq-\frac{l\left(l+1\right)\Delta}{\left(r^{2}+a^{2}\right)^{2}}\,,\label{eq:pot_large_l}
\end{equation}
with an $O\left(l^{0}\right)$ remainder. Then, the radial equation
\eqref{eq:radKerr} admits the large-$l$ form
\begin{equation}
\frac{d^{2}\psi_{\omega l}}{dr_{*}^{2}}\simeq\frac{l\left(l+1\right)\Delta}{\left(r^{2}+a^{2}\right)^{2}}\psi_{\omega l}\,.\label{eq:d2psi_large_l}
\end{equation}
 We may thus expect the large-$l$ behavior of the bare flux expression
to be some power of $l\left(l+1\right)$ (which will be confirmed
below).  

Recall that $E_{\omega l}$, the mode contribution to $\left\langle \Phi^{2}\right\rangle _{\text{bare}}^{U}$
given in Eq.~\eqref{eq:Ewl_phi2}, yielded an $O\left(l^{0}\right)$
ID $E_{\omega l}^{\text{div}}$ (given in Eq.~\eqref{eq:Ewl_div_phi2}).
However,  from Eq.~\eqref{eq:Tab_product}, when switching from
$\left\langle \Phi^{2}\right\rangle $ to (trace-reversed)  RSET components we
encounter (a product of) two differentiations of the mode functions.
 These two differentiations may potentially lead to an ``amplification''
of the leading order by a factor of $l\left(l+1\right)$. Specifically,
such amplification may emerge from derivatives with respect to $r_{*}$.
{[}In principle such amplifications may also arise from derivatives
with respect to $\theta$ and $\varphi$, but these derivatives do
not exist for the (trace-reversed) energy flux components; and derivatives with respect
to $t$  merely lead to additional $\omega$ factors (rather than
$l(l+1)$).{]} We shall hence concentrate on contributions involving
$\psi_{\omega l,r_{*}}^{\text{int}}$, which indeed involves amplification
by a factor of $\tilde{l}$ compared to $\psi_{\omega l}^{\text{int}}$
(as one can see, e.g., from its WKB form -- see Eq.~\eqref{eq:psi_pm}
or \eqref{eq:psiWKB}, which includes factors of $e^{\pm i\tilde{l}g\left(r\right)}$).

We now look more closely at the expression for $T_{yy\left(\omega l\right)}$
as given in Eqs.~\eqref{eq:Tyy_wl_expanded}-\eqref{eq:TC}. It is the
sum of three different terms: $T_{yy\left(\omega l\right)}^{\mathcal{A}}$,
$T_{\left(\omega l\right)}^{\mathcal{B}}\Delta$, and $T_{\left(\omega l\right)}^{\mathcal{C}}\Delta^{2}$.
Evidently, $T_{yy\left(\omega l\right)}$ depends on the
radial function $\psi_{\omega l}^{\text{int}}$ and its first derivative
$\psi_{\omega l,r_{*}}^{\text{int}}$ \emph{quadratically}, either through
(\emph{i}) $\left|\psi_{\omega l}^{\text{int}}\right|^{2}$
or $\left(\psi_{\omega l}^{\text{int}}\right)^{2}$ (appearing in
$T_{yy\left(\omega l\right)}^{\mathcal{A}}$ and $T_{\left(\omega l\right)}^{\mathcal{C}}$),
(\emph{ii}) combinations involving one derivative, $\psi_{\omega l}^{\text{int}}\psi_{\omega l,r_{*}}^{\text{int}*}$
or $\psi_{\omega l}^{\text{int}}\psi_{\omega l,r_{*}}^{\text{int}}$
(appearing in $T_{\left(\omega l\right)}^{\mathcal{B}}$
only), or (\emph{iii}) combinations of two first-order derivatives,
$\left|\psi_{\omega l,r_{*}}^{\text{int}}\right|^{2}$ or $\left(\psi_{\omega l,r_{*}}^{\text{int}}\right)^{2}$
(appearing in $T_{yy\left(\omega l\right)}^{\mathcal{A}}$
only). Based on the above, the highest order in $l$ that can potentially
occur  may emerge only from these latter terms, quadratic in $\psi_{\omega l,r_{*}}^{\text{int}}$
{[}namely $\left|\psi_{\omega l,r_{*}}^{\text{int}}\right|^{2}$ or
$\left(\psi_{\omega l,r_{*}}^{\text{int}}\right)^{2}${]}, which are
anticipated to contribute at order  $l\left(l+1\right)$. (In particular,
$T_{\left(\omega l\right)}^{\mathcal{B}}$ and $T_{\left(\omega l\right)}^{\mathcal{C}}$
do not contribute at this level.) We shall now see concretely that
this is indeed the case.

To proceed, we focus on $T_{yy\left(\omega l\right)}^{\mathcal{A}}$
from Eq.~\eqref{eq:TAuu} or \eqref{eq:TAvv}, keeping only the terms
quadratic in the derivatives $\psi_{\omega l,r_{*}}^{\text{int}}$
(conveniently, these terms are shared by both the $uu$ and $vv$
components, as expected from $T_{uu}-T_{vv}$ being regular). This
leaves us with the potential $\propto l\left(l+1\right)$ contribution \footnote{In the asymptotic expression in Eq.~\eqref{eq:Twl(psi'^2)}, as well as in other ones below (including Eq.~\eqref{eq:  T1_and_T2}) within this subsection, we only claim the {\it leading-order} term in $l(l+1)$  to be correct.
}
to $T_{yy\left(\omega l\right)}$: 
\begin{align}
 & T_{yy\left(\omega l\right)}\simeq\hbar\frac{\left[S_{\omega l}\left(0\right)\right]^{2}}{32\pi^{2}\omega\left(r^{2}+a^{2}\right)}\label{eq:Twl(psi'^2)}\\
 & \left[\coth\left(\frac{\pi\omega}{\kappa_{+}}\right)\left|\psi_{\omega l,r_{*}}^{\text{int}}\right|^{2}\left(1+\left|\rho_{\omega l}^{\text{up}}\right|^{2}\right)+2\,\text{cosech}\left(\frac{\pi\omega}{\kappa_{+}}\right)\Re\left[\rho_{\omega l}^{\text{up}}\left(\psi_{\omega l,r_{*}}^{\text{int}}\right)^{2}\right]+\left(1-\left|\rho_{\omega l}^{\text{up}}\right|^{2}\right)\left|\psi_{\omega l,r_{*}}^{\text{int}}\right|^{2}\right]\,.
 \nonumber
\end{align}

We may now analyze the $O\left(l\left(l+1\right)\right)$ content
of the above terms quadratic in $\psi_{\omega l,r_{*}}^{\text{int}}$,
by using Eq.~\eqref{eq:d2psi_large_l}, from which we obtain 
\[
\frac{1}{2}\left[\left(\psi_{\omega l}^{\text{int}}\right)^{2}\right]_{,r_{*}r_{*}}=\left(\psi_{\omega l,r_{*}}^{\text{int}}\right)^{2}+\psi_{\omega l}^{\text{int}}\psi_{\omega l,r_{*}r_{*}}^{\text{int}}\simeq\left(\psi_{\omega l,r_{*}}^{\text{int}}\right)^{2}+\frac{l\left(l+1\right)\Delta}{\left(r^{2}+a^{2}\right)^{2}}\left(\psi_{\omega l}^{\text{int}}\right)^{2}
\]
yielding 
\[
\left(\psi_{\omega l,r_{*}}^{\text{int}}\right)^{2}\simeq\frac{1}{2}\left[\left(\psi_{\omega l}^{\text{int}}\right)^{2}\right]_{,r_{*}r_{*}}-\frac{l\left(l+1\right)\Delta}{\left(r^{2}+a^{2}\right)^{2}}\left(\psi_{\omega l}^{\text{int}}\right)^{2}\,.
\]
Similarly, using Eq.~\eqref{eq:d2psi_large_l} and its conjugate,
we obtain
\[
\left|\psi_{\omega l,r_{*}}^{\text{int}}\right|^{2}\simeq\frac{1}{2}\left[\left|\psi_{\omega l}^{\text{int}}\right|^{2}\right]_{,r_{*}r_{*}}-\frac{l\left(l+1\right)\Delta}{\left(r^{2}+a^{2}\right)^{2}}\left|\psi_{\omega l}^{\text{int}}\right|^{2}\,.
\]
Using these two last relations, we may now rewrite Eq.~\eqref{eq:Twl(psi'^2)}
(again, 
only claiming its leading order in $l(l+1)$ to be correct)
as
\begin{equation}
T_{yy\left(\omega l\right)}\simeq T_{yy\left(\omega l\right)}^{\left(1\right)}l\left(l+1\right)+T_{yy\left(\omega l\right)}^{\left(2\right)}\label{eq:  T1_and_T2}
\end{equation}
where
\begin{align}
 & T_{yy\left(\omega l\right)}^{\left(1\right)}=-\hbar\frac{\left[S_{\omega l}\left(0\right)\right]^{2}}{32\pi^{2}\omega\left(r^{2}+a^{2}\right)}\frac{\Delta}{\left(r^{2}+a^{2}\right)^{2}}\label{eq:Twl1}\\
 & \left[\coth\left(\frac{\pi\omega}{\kappa_{+}}\right)\left|\psi_{\omega l}^{\text{int}}\right|^{2}\left(1+\left|\rho_{\omega l}^{\text{up}}\right|^{2}\right)+2\,\text{cosech}\left(\frac{\pi\omega}{\kappa_{+}}\right)\Re\left[\rho_{\omega l}^{\text{up}}\left(\psi_{\omega l}^{\text{int}}\right)^{2}\right]+\left(1-\left|\rho_{\omega l}^{\text{up}}\right|^{2}\right)\left|\psi_{\omega l}^{\text{int}}\right|^{2}\right]\nonumber 
\end{align}
and
\begin{align}
 & T_{yy\left(\omega l\right)}^{\left(2\right)}=\hbar\frac{\left[S_{\omega l}\left(0\right)\right]^{2}}{64\pi^{2}\omega\left(r^{2}+a^{2}\right)}\nonumber \\
 & \left[\coth\left(\frac{\pi\omega}{\kappa_{+}}\right)\left(\left|\psi_{\omega l}^{\text{int}}\right|^{2}\right)_{,r_{*}r_{*}}\left(1+\left|\rho_{\omega l}^{\text{up}}\right|^{2}\right)+2\,\text{cosech}\left(\frac{\pi\omega}{\kappa_{+}}\right)\Re\left(\rho_{\omega l}^{\text{up}}\left[\left(\psi_{\omega l}^{\text{int}}\right)^{2}\right]_{,r_{*}r_{*}}\right)+\left(1-\left|\rho_{\omega l}^{\text{up}}\right|^{2}\right)\left(\left|\psi_{\omega l}^{\text{int}}\right|^{2}\right)_{,r_{*}r_{*}}\right]\,.\label{eq:Twl2}
\end{align}

Focusing first on $T_{yy\left(\omega l\right)}^{\left(1\right)}$,
one may recognize it is proportional to $E_{\omega l}$, the mode
contribution of $\left\langle \Phi^{2}\right\rangle _{\text{bare}}^{U}$
given in Eq.~\eqref{eq:Ewl_phi2}:
\begin{equation}
T_{yy\left(\omega l\right)}^{\left(1\right)}=-\frac{\Delta}{4\left(r^{2}+a^{2}\right)^{2}}E_{\omega l}\,.\label{eq:  T1_vs_E}
\end{equation}
Next, a comparison of Eqs.~\eqref{eq:Ewl_phi2} and \eqref{eq:Twl2}
reveals that $T_{yy\left(\omega l\right)}^{\left(2\right)}$
is related to $E_{\omega l}$ via 
\begin{equation}
T_{yy\left(\omega l\right)}^{\left(2\right)}=\frac{1}{8\left(r^{2}+a^{2}\right)}\,\frac{d^{2}}{dr_{*}^{2}}\left[\left(r^{2}+a^{2}\right)E_{\omega l}\left(r\right)\right]\,.\label{eq:  T2_vs_E}
\end{equation}
Taking the large-$l$ limit in the last two equations -- recalling
that the $l\to\infty$ limit of $E_{\omega l}$ is $E_{\omega l}^{\text{div}}$
given in Eq.~\eqref{eq:Ewl_div_phi2} -- we obtain
\begin{equation}
\lim_{l\to\infty}T_{yy\left(\omega l\right)}^{\left(1\right)}=-\frac{\Delta}{4\left(r^{2}+a^{2}\right)^{2}}\,E_{\omega l}^{\text{div}}\,\label{eq:  T1_vs_Ediv}
\end{equation}
and
\begin{equation}
\lim_{l\to\infty}T_{yy\left(\omega l\right)}^{\left(2\right)}=\frac{1}{8\left(r^{2}+a^{2}\right)}\,\frac{d^{2}}{dr_{*}^{2}}\left[\left(r^{2}+a^{2}\right)E_{\omega l}^{\text{div}}\left(r\right)\right]\,.\label{eq:  T2_vs_Ediv}
\end{equation}
(Recall that $E_{\omega l}^{\text{div}}$ depends on $r$ but is independent
of $l$.) We can therefore rewrite the large-$l$ asymptotic behavior
of Eq.~\eqref{eq:  T1_and_T2} as
\begin{equation}
T_{yy\left(\omega l\right)}\simeq T_{yy\left(\omega l\right)}^{\left(1\right)}l\left(l+1\right)\simeq\left[\frac{-\Delta}{4\left(r^{2}+a^{2}\right)^{2}}\,E_{\omega l}^{\text{div}}\right]\,l\left(l+1\right)\,.\label{eq:  Tyy_large_l}
\end{equation}
Comparing this result to our ``canonical'' large-$l$ form given
in Eq.~\eqref{eq:Ewl_div} we find that
\begin{equation}
c_{2}=\frac{-\Delta}{4\left(r^{2}+a^{2}\right)^{2}}E_{\omega l}^{\text{div}}\,.\label{eq:c2(Ediv)}
\end{equation}

Plugging in the explicit form of $E_{\omega l}^{\text{div}}$ found
in Eq.~\eqref{eq:Ewl_div_phi2} (as well as the explicit form of $\Delta$),
we obtain
\begin{equation}
c_{2}=\hbar\frac{\sqrt{\left(r_{+}-r\right)\left(r-r_{-}\right)}}{16\pi^{2}\left(r^{2}+a^{2}\right)^{2}}\,.\label{eq:c2(r)}
\end{equation}

This settles the large-$l$ leading order $O\left(l\left(l+1\right)\right)$
contribution to the ID. 

\begin{figure}[h!]
\centering \includegraphics[scale=0.3]{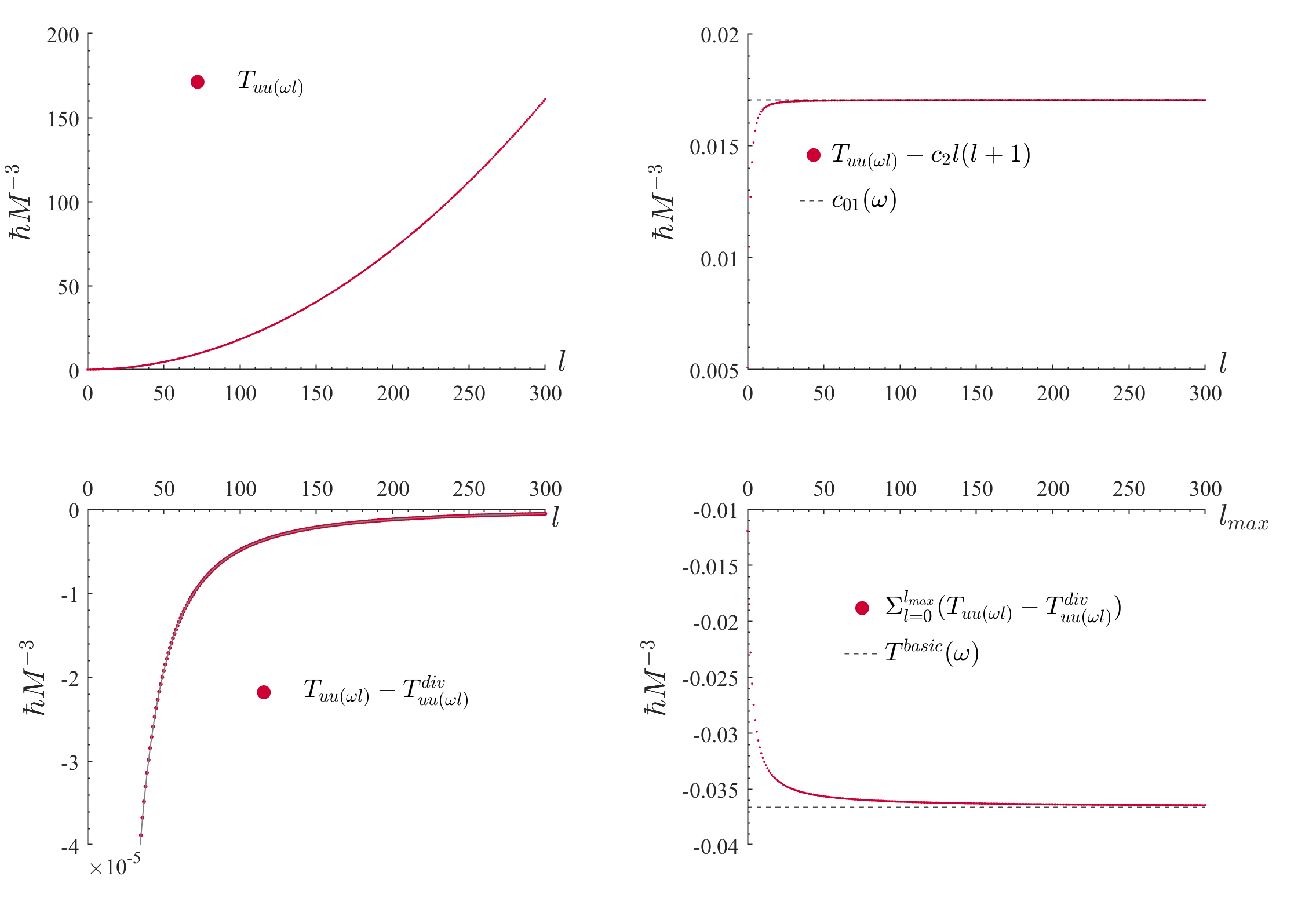}\caption{The ID subtraction procedure for $T_{uu}$ as
a function of $l$ is demonstrated for the specific case of $\omega=1/M$,
at the point $r/M=0.9$ inside an $a/M=0.8$ BH. \emph{Top
left:} the numerically-computed bare mode contribution, $T_{uu\left(\omega l\right)}$
(given in Sec.~\ref{subsec:bare_Tyy}), for the mentioned fixed $\omega$. Its behavior is asymptotically
parabolic in $l$, as determined by the large-$l$ leading order of $T_{uu\left(\omega l\right)}^{\text{div}}$ {[}given in Eq.~\eqref{eq:Tyy_div_c01,c2}{]}. \emph{Top right:}
the difference between $T_{uu\left(\omega l\right)}$ and its large-$l$ analytically-computed $\propto l\left(l+1\right)$
leading order, namely $T_{uu\left(\omega l\right)}-c_{2}l\left(l+1\right)$,
as a function of $l$. At large $l$ this leaves a constant in $l$, denoted $c_{01}\left(\omega\right)$ and defined in Eq.~\eqref{eq:c01_flux_def},
which is numerically extracted and portrayed here by a dashed horizontal line. The behavior of $c_{01}\left(\omega\right)$
as a function of $\omega$ is explored in Fig.~\ref{Fig:Tuu_vs_w_2tiles}. \emph{Bottom left:} subtracting this large-$l$
plateau value $c_{01}\left(\omega\right)$ from the quantity presented (by the red points) in the previous panel, one remains with the difference
$T_{uu\left(\omega l\right)}-T_{uu\left(\omega l\right)}^{\text{div}}$ (recall $T_{uu\left(\omega l\right)}^{\text{div}}=c_{2}l\left(l+1\right)+c_{01}\left(\omega\right)$)
which is portrayed here as a function of $l$. The plot is trimmed vertically at $-4\times10^{-5}\hbar M^{-1}$ for scale purposes, allowing to appreciate the remaining large-$l$ converging behavior. This large-$l$ behavior is numerically
found to be $O\left(1/l(l+1)\right)$, and the thin black line represents the corresponding $1/l\left(l+1\right)$ fit. \emph{Bottom right:} the red points are the partial sums of $T_{uu\left(\omega l\right)}-T_{uu\left(\omega l\right)}^{\text{div}}$
over $l$ up to $l_{\text{max}}$, displayed as a function of the latter. The dashed horizontal line corresponds to the $l_{\text{max}}\to\infty$
limit, and hence constitutes the basic integrand value $T_{uu}^{\text{basic}}\left(\omega=1/M\right)$
defined in Eq.~\eqref{eq:Tyy_basic_def}.}
\label{Fig:Tuu_vs_l_4tiles}
\end{figure}

Numerically exploring the next-to-leading order at large-$l$, we find that it is $\propto l^0$, as demonstrated in the top right panel of Fig.~\ref{Fig:Tuu_vs_l_4tiles}. We denote this $l$-independent term by $c_{01}\left(\omega\right)$, and write the divergent piece\footnote{We see numerically that the next order after $l^0$ is $1/l(l+1)$ (and is, in particular, convergent in a sum over $l$), as illustrated at the bottom left panel of Fig.~\ref{Fig:Tuu_vs_l_4tiles}.}
$T_{yy\left(\omega l\right)}^{\text{div}}$ as\footnote{Even though $c_{2}$ and $c_{01}$ are functions of $r$, we do not
make that dependence explicit here.}
\begin{align}
T_{yy\left(\omega l\right)}^{\text{div}} & =c_{2}l\left(l+1\right)+c_{01}\left(\omega\right)\,,\label{eq:Tyy_div_c01,c2}
\end{align}
with $c_{2}$ as given in Eq.~\eqref{eq:c2(r)}.  This  $c_{01}\left(\omega\right)$ term may thus be defined as
\begin{equation}
c_{01}\left(\omega\right)\equiv\lim_{l\to\infty}\left[T_{yy\left(\omega l\right)}-c_{2}l\left(l+1\right)\right]\,.\label{eq:c01_flux_def}
\end{equation}
A numerical exploration shows that this limit indeed exists, and furthermore, it takes exactly the form
\begin{equation}
c_{01}\left(\omega\right)=c_{0}+c_{1}\omega^{2}\,\label{eq:c01_parabolic}
\end{equation}
with $c_0$ and $c_1$ independent of $\omega$ and $l$ (as we shortly show numerically). After establishing this form, it indeed becomes justified (as mentioned above, and as demonstrated -- partly numerically, partly analytically -- in detail in
Appendix~\ref{App:The-intermediate-divergence}) to subtract $T_{yy\left(\omega l\right)}^{\text{div}}$
from the bare mode contribution $T_{yy\left(\omega l\right)}$.
The convergent sequence resulting from this subtraction is consequently summed over $l$ to yield the
basic integrand function, denoted by $T_{yy}^{\text{basic}}\left(\omega\right)$:
\begin{equation}
T_{yy}^{\text{basic}}\left(\omega\right)\equiv\sum_{l=0}^{\infty}\left(T_{yy\left(\omega l\right)}-T_{yy\left(\omega l\right)}^{\text{div}}\right)\,.\label{eq:Tyy_basic_def}
\end{equation}

This multiple-step procedure, resulting in the basic integrand value
$T_{yy}^{\text{basic}}$ for a given $\omega$, is illustrated
in Fig.~\ref{Fig:Tuu_vs_l_4tiles} for the case $y=u$. In the top
left panel we present the bare mode contribution $T_{uu\left(\omega l\right)}$
as a sequence in $l$ for a fixed $\omega$ value (taken here to be
$\omega=1/M$, as for the $\left\langle \Phi^{2}\right\rangle $ case
in Fig.~\ref{Fig:phi_vs_l_4tiles}). In the top right panel, we subtract
the analytically-computed leading order of the ID, $c_{2}l\left(l+1\right)$
with $c_{2}$ as given in Eq.~\eqref{eq:c2(r)}, and are left with
the large-$l$ plateau value $c_{01}\left(\omega\right)$\footnote{In particular, this plateau shows that the limit $\lim_{l\to\infty}\left[T_{uu\left(\omega l\right)}-c_{2}l\left(l+1\right)\right]$,
defining $c_{01}\left(\omega\right)$, indeed exists (namely, there
are no intervening, weaker, large-$l$ divergences).
}.  In the bottom left panel we further subtract this $c_{01}\left(\omega\right)$
term (obtained numerically) and are left with a sequence $T_{uu\left(\omega l\right)}-T_{uu\left(\omega l\right)}^{\text{div}}$, which is numerically found to behave as $1/l(l+1)$
(as demonstrated by a fit). The corresponding convergence of the $l$-sum of this sequence is demonstrated in the bottom right panel,
which portrays the partial sums $\sum_{l=0}^{l_{\text{max}}}\left(T_{uu\left(\omega l\right)}-T_{uu\left(\omega l\right)}^{\text{div}}\right)$ and their approach to $T_{uu}^{\text{basic}}\left(\omega\right)$
as $l_{\text{max}}\to\infty$. This $T_{uu}^{\text{basic}}$ value {[}which is extracted through a fit as described in Appendix
\ref{subsec:numerical_methods}{]}, represented by the dashed horizontal
line, is then highlighted in the left panel of Fig.~\ref{Fig:Tuu_integrand_2tiles}
by a bold red point at $\omega M=1$.\footnote{In practice, in the numerical implementation of the computation we
extract $T_{yy}^{\text{basic}}\left(\omega\right)$ and
$c_{01}\left(\omega\right)$ simultaneously using a slightly different
(but mathematically equivalent) method -- see Appendix~\ref{subsec:numerical_methods}.}

We now turn to empirically establish the form \eqref{eq:c01_parabolic}
of the $O\left(l^{0}\right)$ piece of the ID, $c_{01}\left(\omega\right)$.
Our claim is that this quantity admits the \emph{exact} parabolic
form $c_{0}+c_{1}\omega^{2}$. This is numerically demonstrated in the left panel
of Fig.~\ref{Fig:Tuu_vs_w_2tiles}, by attaching a fit $c_{0}+c_{1}\omega^{2}$
to the numerically extracted (see Appendix~\ref{subsec:numerical_methods})
$c_{01}\left(\omega\right)$  term, presented  as a function of
$\omega$. The accuracy of this fit is demonstrated in the right
panel, which portrays the difference between $c_{01}\left(\omega\right)$
and its $c_{0}+c_{1}\omega^{2}$ fit -- indicating (together with
the left panel) a relative difference smaller than   $10^{-94}$
in the case depicted here. (This was achievable due to the fact that
we had a typical precision of hundreds of figures -- at least $250$ -- for the raw
material $\rho^\text{up}_{\omega l}$,
$\psi_{\omega l}^{\text{int}}$ and $\psi_{\omega l,r}^{\text{int}}$,
see Appendix~\ref{subsec:Computation-of-raw-data}.) 
This striking agreement (presumably limited only by numerical  precision)
provides solid empirical evidence for the validity of Eq.~\eqref{eq:c01_parabolic}.

In fact, we empirically-numerically found a closed expression for $c_1$, being\footnote{Notably, this expression for $c_1$ may also be written as $c_1=\left(1/4\right)\left(1-v_\omega/2\right)E_{\omega l}^\text{div}$ where $E_{\omega l}^\text{div}$ is given in Eq.~\eqref{eq:Ewl_div_phi2} and the function $v_\omega$ is the prefactor of $-\omega^2$ in the subleading large-$l$ form of the effective potential $V_{\omega l}$, which is $v_\omega=a^2\Delta\left[2\left(a^2+r^2\right)^2\right]^{-1}-1$. Playing around these relations is in fact how we arrived at Eq.~\eqref{eq:c1}}
\begin{equation}
c_1=\frac{\hbar}{8}\left(3-\frac{a^2\Delta}{2\left(a^2+r^2\right)^2}\right)\frac{1}{4\pi^2\sqrt{\left(r_+-r\right)\left(r-r_-\right)}}\,.
\label{eq:c1}
\end{equation}
[Note, however, that the values of neither $c_0$ nor $c_1$ are actually used in our numerical computation (see \ref{subsec:numerical_methods}).]

It should also be noted that while Figs.~\ref{Fig:Tuu_vs_l_4tiles} and \ref{Fig:Tuu_vs_w_2tiles}
(as well as Fig.~\ref{Fig:Tuu_integrand_2tiles} to follow) focus specifically
on $T_{uu}$, a  similar behavior is seen for $T_{vv}$
(see also footnote \ref{fn:typical_case_figures}). 

\begin{figure}[h!]
\centering \includegraphics[scale=0.25]{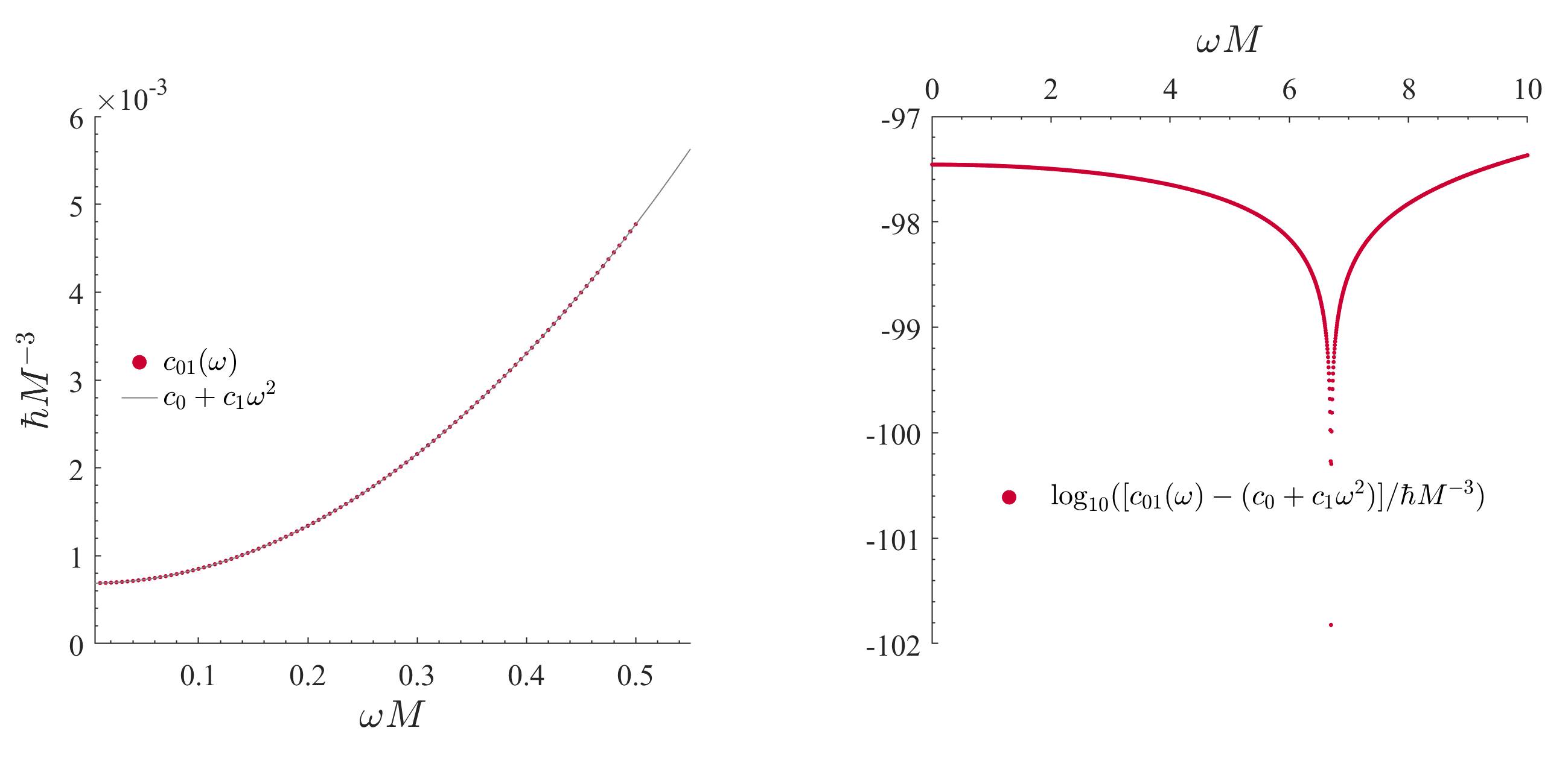}
\caption{For the same parameters as in Fig.~\ref{Fig:Tuu_vs_l_4tiles} ($r/M=0.9$,
$a/M=0.8$), this figure demonstrates the robustness of the parabolic
behavior of the large-$l$ plateau term $c_{01}\left(\omega\right)$
{[}defined in Eq.~\eqref{eq:c01_flux_def}{]}.\emph{ Left}: the red
points present $c_{01}\left(\omega\right)$ as a function of $\omega M$,
extracted numerically per $\omega$ (as demonstrated in the top right
panel of Fig.~\ref{Fig:Tuu_vs_l_4tiles}).
The  thin black line is a fit of the form $c_{0}+c_{1}\omega^{2}$
(where $c_{0}$ and $c_{1}$ are extracted to a precision of roughly
$\sim100$ figures). The plot is horizontally terminated at $\omega=1/2M$
for visual purposes (but this behavior was verified up to $\omega=10/M$,
as seen in the right panel). \emph{Right}: The
(decimal logarithm of the) difference between the numerically extracted
$c_{01}\left(\omega\right)$ and its parabolic fit $c_{0}+c_{1}\omega^{2}$.
}
\label{Fig:Tuu_vs_w_2tiles}
\end{figure}

\subsection{The $t$-splitting PMR counterterms\label{subsec:Counterterms}}

In the previous subsection we analyzed the 
ID for the vacuum polarization and the fluxes, namely
$E_{\omega l}^{\text{div}}$ and $T_{yy\left(\omega l\right)}^{\text{div}}$, respectively. In this subsection we present the PMR counterterms. That is, the `sing' modes $E^{\text{sing}}$, as well as their finite counterpart $e$,
derived from the counterterms $C\left(\varepsilon,x\right)$. In the notation of the beginning of this section, used to  illustrate the method for the generic quantity $P$, the symbols $E^{\text{sing}}$ and $e$ were used generically both for the fluxes and the vacuum polarization. Here, similarly to the above subsection, we use the $E^{\text{sing}}$ and $e$ notation to correspond specifically to the vacuum polarization, and  $T_{yy}^{\text{sing}}$ and $e_{yy}$ for the fluxes.
We shall first present the PMR counterterms for the vacuum polarization and then for the fluxes.

\subsubsection{The $t$-splitting PMR counterterms for $\left\langle \Phi^{2}\right\rangle$ \label{subsec:CT_phi2}}

We first write Eq.~\eqref{eq:P_ren_PMR} for $\left\langle \Phi^{2}\right\rangle $:
\begin{equation}
\left\langle \Phi^{2}\left(x\right)\right\rangle _{\text{ren}}^{U}=\int_{0}^{\infty}\left[E^{\text{basic}}\left(\omega,x\right)-E^{\text{sing}}\left(\omega,x\right)\right]\text{d}\omega-e\left(x\right)\,.\label{eq:Phi^2_ren_PMR}
\end{equation}

The $t$-splitting PMR  counterterms for $\left\langle \Phi^{2}\right\rangle $,
denoted $E^{\text{sing}}\left(\omega,x\right)$ and $e\left(x\right)$,
are obtained as described in Ref.~\cite{AAt:2015} -- namely, by
expanding  
the DeWitt-Schwinger counterterm (see Ref.~\cite{DeWittBook:1965})
in the point separation $\varepsilon=t'-t$ up to order $\varepsilon^0$ included, and then Fourier transforming
this expansion into frequency space. See the Supplemental Material for the results of this computation at a general spacetime point in Kerr. In our case (at the pole, where the dependence on the spacetime point reduces to a dependence
on $r$), this yields simply\footnote{Note that the notations used here, which were motivated by a uniform
treatment of $\left\langle \Phi^{2}\right\rangle $ and $\left\langle T_{yy}\right\rangle $
in Eq.~\eqref{eq:P_ren_PMR}, slightly differ from those used in Ref.~\cite{AAt:2015}.
Comparing Eq.~\eqref{eq:Phi^2_ren_PMR} with Eqs.~(3.15)-(3.16) therein,
our $E^{\text{sing}}$ replaces $\hbar F_{\text{sing}}$ and $e\left(x\right)$
replaces $\hbar d\left(x\right)$. In addition, since we are at the
BH interior (where an ID $E_{\omega l}^{\text{div}}$ exists), $\hbar F\left(\omega,x\right)$
is replaced by $E^{\text{basic}}\left(\omega,x\right)$ which is defined
by $\sum_{l=0}^{\infty}\left[E_{\omega l}\left(x\right)-E_{\omega l}^{\text{div}}\left(x\right)\right]$
(with $E_{\omega l}$ and $E_{\omega l}^{\text{div}}$ given respectively
in Eqs.~\eqref{eq:psi_int_BC} and \eqref{eq:Ewl_div_phi2}). 

Moreover, comparing Eq.~\eqref{eq:Esing_phi^2} with Eq.~(3.17) in
Ref.~\cite{AAt:2015}, we see that writing the latter for the case considered here (polar Kerr) would come with the
prefactors $a\left(r,\theta,\varphi\right)=-\left(a^{2}+r^{2}\right)/4\pi^{2}\Delta$
and $c\left(r,\theta,\varphi\right)=0$. That is, while generally
there is an additional $1/\left(\omega+\mu\right)$ term in $E^{\text{sing}}$
(where $\mu$ is a scale ambiguity), in our case this term is not
present. }

\begin{equation}
E^{\text{sing}}\left(\omega,r\right)=\hbar\frac{a^{2}+r^{2}}{4\pi^{2}\Delta}\omega\label{eq:Esing_phi^2}
\end{equation}
and the finite counterterm
\begin{equation}
e\left(r\right)=\frac{\hbar}{48\pi^{2}}\frac{M^{2}}{\Delta}\frac{\left(a^{2}-r^{2}\right)^{2}}{\left(a^{2}+r^{2}\right)^{3}}\,.\label{eq:finite_ct_phi^2}
\end{equation}

\begin{figure}[h!]
\centering \includegraphics[scale=0.25]{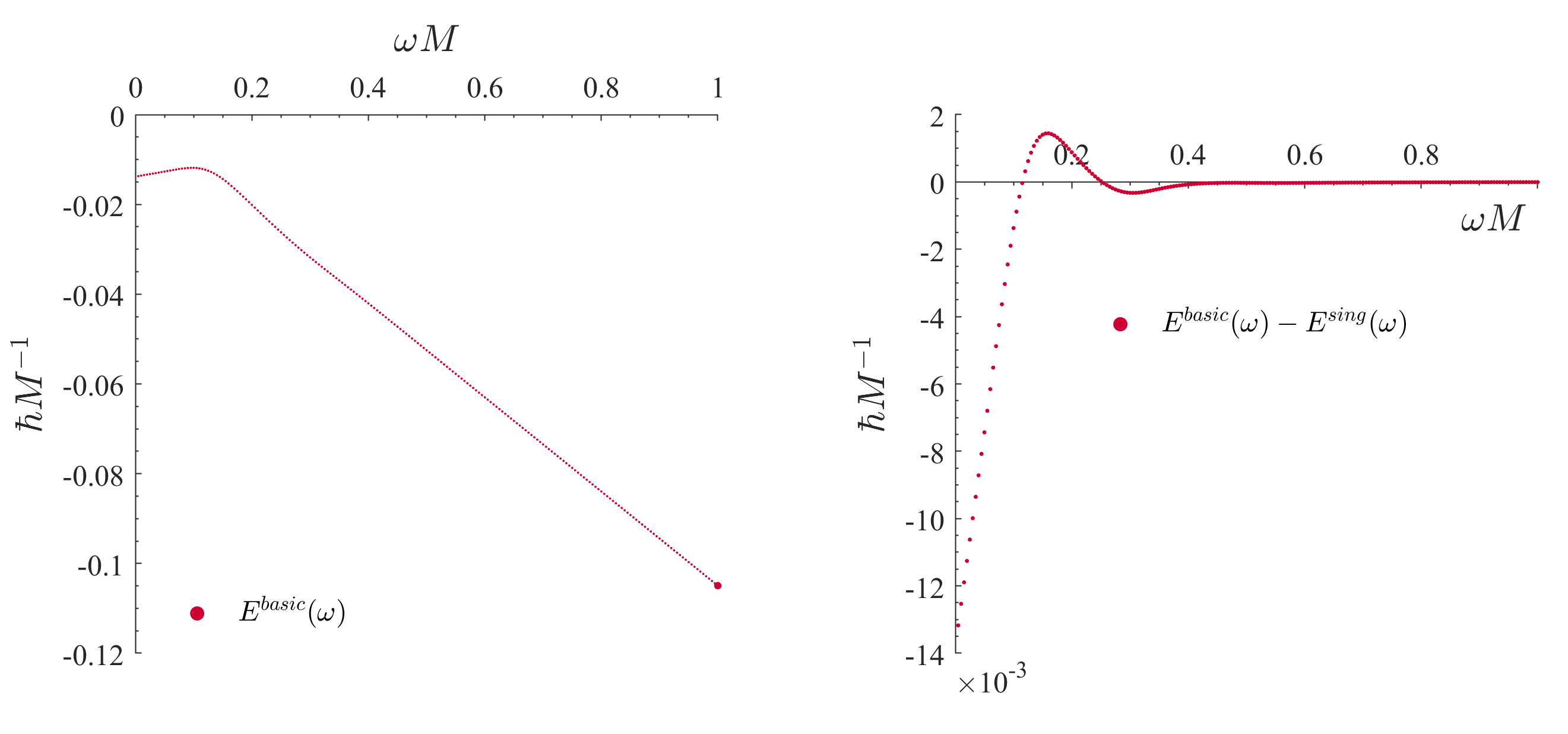}
\caption{\emph{Left:} for the same parameters as in Fig.~\ref{Fig:phi_vs_l_4tiles}
($r/M=0.9$, $a/M=0.8$), the basic integrand $E^{\text{basic}}\left(\omega\right)$
is portrayed as a function of $\omega M$. $E^{\text{basic}}\left(\omega\right)$
is obtained  as demonstrated in the bottom  panel of Fig.~\ref{Fig:phi_vs_l_4tiles},
and the bold red point appearing here at $\omega M=1$ corresponds
to the $E^{\text{basic}}$ value obtained therein for this specific
$\omega$ value. The large-$\omega$ linear behavior of $E^{\text{basic}}\left(\omega\right)$
in $\omega$ is inherited from the singular piece $E^{\text{sing}}\left(\omega\right)$.
\emph{Right: }the difference $E^{\text{basic}}\left(\omega\right)-E^{\text{sing}}\left(\omega\right)$,
as a function of \textcolor{black}{$\omega M$, being the quantity
to be integrated over $\omega$ {[}see Eq.~\eqref{eq:P_ren_PMR}{]}.
}(The plots here are presented up to $\omega=1/M$, for increased
visibility of the features at small $\omega$, but in practice we
carried the integration up to $\omega=10/M$).}
\label{Fig:phi_integrand_2tiles}
\end{figure}

Fig.~\ref{Fig:phi_integrand_2tiles} allows to appreciate the 
part of the regularization given in Eq.~\eqref{eq:Phi^2_ren_PMR},
in which the singular piece $E^{\text{sing}}\left(\omega,r\right)$
is subtracted from the basic integrand and then the remainder is integrated
over $\omega$. The basic integrand 
$E^{\text{basic}}\left(\omega,x\right)$,
extracted numerically per $\omega$ (as demonstrated in the bottom
 panel of Fig.~\ref{Fig:phi_vs_l_4tiles}), is presented in the
left panel of Fig.~\ref{Fig:phi_integrand_2tiles} as a function of
$\omega$. Being dominated at large $\omega$ by its singular piece
$E^{\text{sing}}\left(\omega\right)$ {[}given in Eq.~\eqref{eq:Esing_phi^2}{]},
it diverges linearly in $\omega$. The right panel  shows the regular
difference $E^{\text{basic}}\left(\omega\right)-E^{\text{sing}}\left(\omega\right)$
as a function of $\omega$. This regularized integrand is then integrated
over in order to obtain the final result for $\left\langle \Phi^{2}\right\rangle _{\text{ren}}^{U}$,
after subtracting the final counterterm $e\left(r\right)$, as prescribed
in Eq.~\eqref{eq:Phi^2_ren_PMR}. 

\subsubsection{The $t$-splitting PMR counterterms for $\left\langle T_{uu}\right\rangle $
and $\left\langle T_{vv}\right\rangle $\label{subsec:CT_fluxes}}

We write Eq.~\eqref{eq:P_ren_PMR} for the flux component $\left\langle T_{yy}\right\rangle $
as

\begin{equation}
\left\langle T_{yy}\left(x\right)\right\rangle _{\text{ren}}^{U}=\int_{0}^{\infty}\left(T_{yy}^{\text{basic}}\left(\omega,x\right)-T_{yy}^{\text{sing}}\left(\omega,x\right)\right)\text{d}\omega-e_{yy}\left(x\right)\,,\label{eq:Tyy_ren_PMR}
\end{equation}
where $T_{yy}^{\text{basic}}\left(\omega,x\right)$ is
given in Eq.~\eqref{eq:Tyy_basic_def}, and we denote by $T_{yy}^{\text{sing}}\left(\omega,x\right)$
and $e_{yy}\left(x\right)$ the $\left\langle T_{yy}\right\rangle $-versions
of the aforementioned PMR counterterms $E^{\text{sing}}\left(\omega,x\right)$ and
$e\left(x\right)$, respectively.

One may obtain the $t$-splitting PMR counterterms for the RSET
as described in Ref.~\cite{LeviRSET:2017} -- which includes translating the Christensen
counterterms given in Ref.~\cite{Christensen:1976} to be expressed
in terms of the point separation $\varepsilon$ (expanded to order $\varepsilon^0$ included), then Fourier decomposing them
to obtain an expansion 
in $\omega$. The results of this computation at a general spacetime point in a Kerr background are given for the full stress-energy tensor in Boyer-Lindquist coordinates in the Supplemental Material.

The flux components [in coordinates $(u,v,\theta,\phi)$] at the pole are then  obtained (also given in the Supplemental Material), yielding (the dependence on $x$ reduces to a dependence on $r$)\footnote{By comparing to Ref.~\cite{LeviRSET:2017} {[}see, in particular,
Eqs.~(3.8-3.10) therein{]}, our $T_{yy}^{\text{sing}}\left(\omega,x\right)$
stands for $\hbar F_{yy}^{\text{Sing}}\left(\omega,x\right)$ and
the integrand function $F_{yy}\left(\omega,x\right)$ is replaced
by $T_{yy}^{\text{basic}}\left(\omega,x\right)$ defined
in Eq.~\eqref{eq:Tyy_basic_def} {[}therein, $T_{yy\left(\omega l\right)}$
is the bare mode contribution given in Sec. (\ref{subsec:bare_Tyy})
and $T_{yy\left(\omega l\right)}^{\text{div}}$ is the
ID given in Eq.~\eqref{eq:Tyy_div_c01,c2}{]}. Casting Eq.~\eqref{eq:Tsing_PMR} into the form of Eq.~(3.9) in Ref.
\cite{LeviRSET:2017}, the prefactors in the case considered here (polar Kerr) would be $a_{yy}\left(r\right)=\left(a^{2}+r^{2}\right)/\left(2\pi^{2}\Delta\right)\,$,
$b_{yy}\left(r\right)=M\left(a^{2}\left(2M-3r\right)+r^{3}\right)/\left(24\pi^{2}\left(a^{2}+r^{2}\right)^{2}\Delta\right)$
and $c_{yy}=d_{yy}=0$. That is, while generally the singular part
$T_{yy}^{\text{sing}}\left(\omega,r\right)$ may also include
$\ln\omega$  and $1/\left(\omega+\mu e^{-\gamma}\right)$ terms
(with $\mu$ a scale ambiguity and $\gamma$ Euler's constant), these
terms are absent in our case.}
\begin{equation}
T_{yy}^{\text{sing}}\left(\omega,r\right)=\frac{\hbar}{6}\frac{a^{2}+r^{2}}{2\pi^{2}\Delta}\omega^{3}+\hbar M\frac{\left(a^{2}\left(2M-3r\right)+r^{3}\right)}{24\pi^{2}\left(a^{2}+r^{2}\right)^{2}\Delta}\omega\,,\label{eq:Tsing_PMR}
\end{equation}
and the finite counterterm
\begin{align}
 & e_{yy}\left(r\right)=\label{eq:finite_ct_Tyy}\\
 & \frac{\hbar M^{2}}{1440\pi^{2}\left(a^{2}+r^{2}\right)^{7}\Delta}\left[18a^{10}+a^{8}\left(-11M^{2}+15Mr-216r^{2}\right)+r^{8}\left(-81M^{2}+99Mr-32r^{2}\right)\right.\nonumber \\
 & \left.+2a^{4}r^{4}\left(-334M^{2}+55Mr+68r^{2}\right)+2a^{2}r^{6}\left(297M^{2}-298Mr+73r^{2}\right)-2a^{6}r^{2}\left(29M^{2}-410Mr+138r^{2}\right)\right]\,.\nonumber 
\end{align}
Note that the $t$-splitting PMR counterterms, both for $\left\langle \Phi^{2}\right\rangle $
{[}Eqs.~\eqref{eq:Esing_phi^2} and \eqref{eq:finite_ct_phi^2}{]} and $\left\langle T_{yy}\right\rangle $
{[}Eqs.~\eqref{eq:Tsing_PMR} and \eqref{eq:finite_ct_Tyy}{]} are proportional
to $1/\Delta$, hence diverge as $1/\delta r_{\pm}$
 at the horizons, where
$$\delta r_{\pm}\equiv
 \frac{\left|r-r_{\pm}\right|}{M}$$
is a (dimensionless) radial coordinate distance to the corresponding
horizon, see Eq.~\eqref{eq:Delta}{]}. This implies that $t$-splitting
will be increasingly difficult to apply as the horizons are approached.

\begin{figure}[h!]
 \includegraphics[scale=0.22]{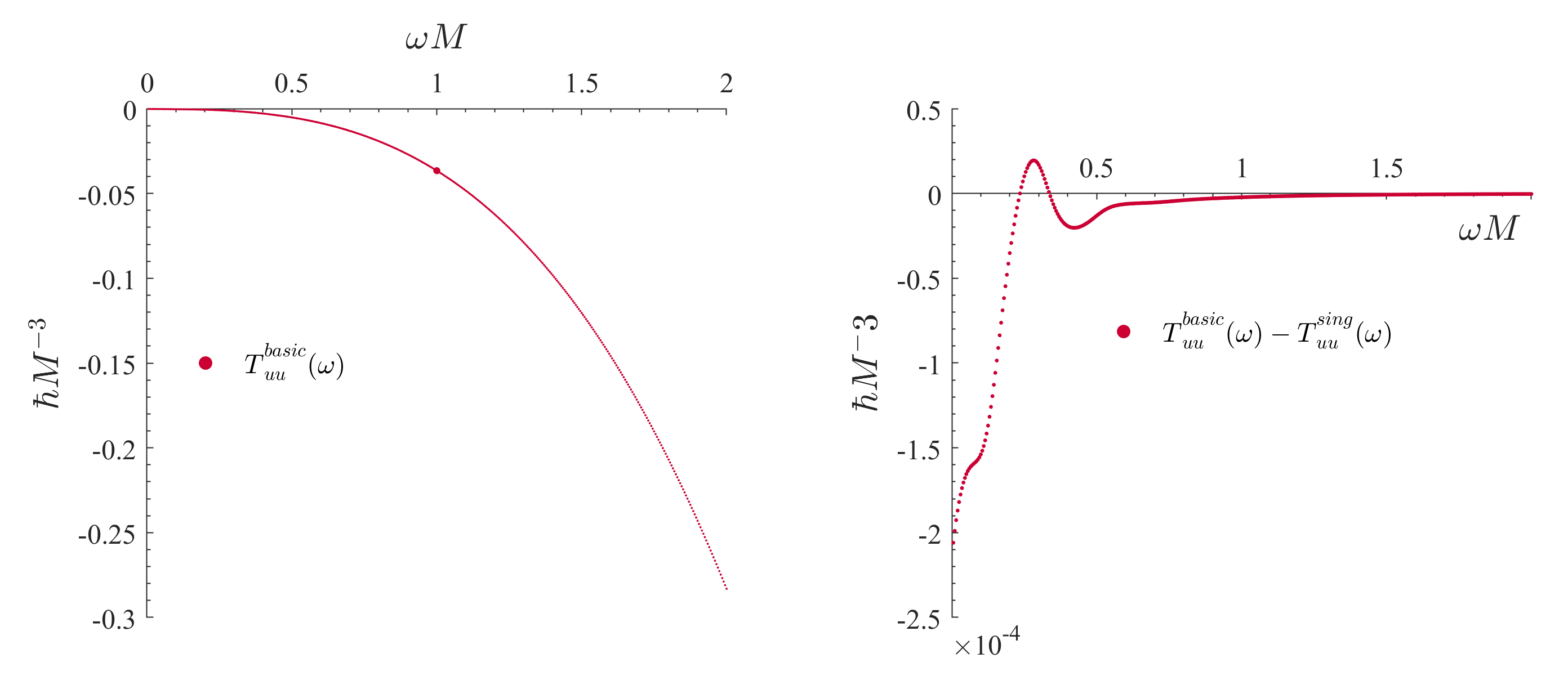}
\caption{\emph{Left:} for the same parameters as in Fig.~\ref{Fig:Tuu_vs_l_4tiles}
($r/M=0.9$, $a/M=0.8$), the basic integrand $T_{uu}^{\text{basic}}\left(\omega\right)$
is portrayed as a function of $\omega M$. $T_{uu}^{\text{basic}}\left(\omega\right)$
is obtained  as demonstrated in the bottom right panel of Fig.~\ref{Fig:Tuu_vs_l_4tiles},
and the bold red point appearing here at $\omega M=1$ corresponds
to the $T_{uu}^{\text{basic}}$ value obtained therein
for this specific $\omega$ value. The $\omega^{3}$ behavior seen
here is inherited from the large-$\omega$ leading order of the singular
piece $T_{uu}^{\text{sing}}\left(\omega\right)$, given
in Eq.~\eqref{eq:Tsing_PMR}. \emph{Right: }the difference $T_{uu}^{\text{basic}}\left(\omega\right)-T_{uu}^{\text{sing}}\left(\omega\right)$,
as a function of $\omega M$\textcolor{blue}{, }\textcolor{black}{being
the quantity to be integrated over {[}see Eq.~\eqref{eq:P_ren_PMR}{]}}.
(The plots here are presented up to $\omega=2/M$, for increased
visibility of the features at small $\omega$, but in practice we
carried the integration up to $\omega=10/M$.)}
\label{Fig:Tuu_integrand_2tiles}
\end{figure}

Fig.~\ref{Fig:Tuu_integrand_2tiles} allows one to appreciate the final
part of the regularization for $\left\langle T_{uu}\right\rangle _{\text{ren}}^{U}$,
given in Eq.~\eqref{eq:Tyy_ren_PMR}, in which the singular piece
is subtracted from the basic integrand and then integrated over $\omega$.
The basic integrand $T_{uu}^{\text{basic}}\left(\omega\right)$,
extracted numerically per $\omega$ (as demonstrated in the bottom
right panel of Fig.~\ref{Fig:Tuu_vs_l_4tiles} and described in Appendix~\ref{subsec:numerical_methods}), is presented in the left panel
of Fig.~\ref{Fig:Tuu_integrand_2tiles} as a function of $\omega$.
Being dominated at large $\omega$ by its singular piece $T_{uu}^{\text{sing}}\left(\omega\right)$
{[}given in Eq.~\eqref{eq:Tsing_PMR}{]}, it diverges like $\omega^{3}$.
The right panel  shows the regular difference $T_{uu}^{\text{basic}}\left(\omega\right)-T_{uu}^{\text{sing}}\left(\omega\right)$
as a function of $\omega$. This regularized integrand (which numerically
seems to decay at large $\omega$ as $1/\omega^{3}$) is then integrated
to obtain the final result for
$\left\langle T_{uu}\right\rangle _{\text{ren}}^{U}\left(r\right)$ at the pole,
after subtracting the corresponding finite counterterm $e_{uu}\left(r\right)$,
as prescribed in Eq.~\eqref{eq:Tyy_ren_PMR}. 

\section{Numerical results\label{sec:Numerical-results}}

Using the $t$-splitting method described in the previous section,
we may now compute $\left\langle \Phi^{2}\right\rangle _{\text{ren}}^{U}$
and the fluxes $\left\langle T_{uu}\right\rangle _{\text{ren}}^{U}$
and $\left\langle T_{vv}\right\rangle _{\text{ren}}^{U}$
at the pole inside a Kerr BH, observing the range between the EH and
the IH. The main computational methods and various numerical details
are postponed to Appendix~\ref{App:Numerical-implementation}.

Since different  $a/M$ values may yield qualitatively-different
behaviors, we choose to work here with two different values: $a/M=0.8$
and $a/M=0.9$.\footnote{In particular, it turns out that an $a/M=0.8$ BH has positive IH
flux values at the pole, whereas in the $a/M=0.9$ case the polar
IH flux values are negative.} For both $a/M$ values we compute $\left\langle \Phi^{2}\right\rangle _{\text{ren}}^{U}$
and the fluxes $\left\langle T_{uu}\right\rangle _{\text{ren}}^{U}$
and $\left\langle T_{vv}\right\rangle _{\text{ren}}^{U}$
as a function of $r$ between the horizons, as well as pay special
attention to the horizon vicinities. 

Notably, the $t$-splitting method cannot be implemented directly
at the horizons (in particular, the corresponding counterterms diverge
there as $1/\Delta$, see Sec.~\ref{subsec:Counterterms}), and
approaching them is increasingly difficult. This limits our ability
to directly explore the very close vicinity of the horizons using
$t$-splitting, and in what follows we typically approach the horizons
up to a distance in $r$ of $10^{-5}M$ or $10^{-4}M$ (with precision dropping
rapidly beyond that). Nevertheless, this vicinity suffices to obtain
the asymptotic behavior near the horizons for the fluxes (see Secs.
\ref{subsec:fluxes_EH-vicinity} and \ref{subsec:fluxes_IH-vicinity}).
(However, for $\left\langle \Phi^{2}\right\rangle _{\text{ren}}^{U}$ 
near the IH this vicinity does not suffice to expose the final asymptotic behavior, as we discuss in Sec.~\ref{subsec:Plots-for-Phi^2}). 

Some ``anchors'' are present at the horizons: First, we have the
numerically computed values of $\left\langle T_{uu}\right\rangle _{\text{ren}}^{U}$
and $\left\langle T_{vv}\right\rangle _{\text{ren}}^{U}$
at the IH from the state-subtraction computation (see Ref.~\cite{KerrIH:2022}).
In addition, from the expected regularity of the Unruh
state at the EH (see Sec.~\ref{subsec:Field-quantization-Unruh}), we know that $\left\langle T_{uu}\right\rangle _{\text{ren}}^{U}=0$
there {[}seen by transforming to the regular $U$ Kruskal coordinate
at the EH, Eqs.~\eqref{eq:extKrusCoor} and \eqref{eq:intKrusCoor}{]}.
This fact allows obtaining the (negative) value of $\left\langle T_{vv}\right\rangle _{\text{ren}}^{U}$
at the EH {[}utilizing the conserved quantity $\left(r^{2}+a^{2}\right)\left(\left\langle T_{uu}\right\rangle _{\text{ren}}^{U}-\left\langle T_{vv}\right\rangle _{\text{ren}}^{U}\right)$,
which may be computed independently{]}. Regularity also implies a
$\delta r_{+}^{2}$
behavior of $\left\langle T_{uu}\right\rangle _{\text{ren}}^{U}$
on approaching the EH, and a Taylor series in $\delta r_{+}$ for
$\left\langle T_{vv}\right\rangle _{\text{ren}}^{U}$.
For $\left\langle \Phi^{2}\right\rangle_{\text{ren}}^{U}$ in the Unruh state we have no a priori knowledge, but an analytic result exists at the EH
in a different quantum state, a `formal' HH state (defined only at the pole) ~\cite{Frolov:1982}, as will be
briefly discussed at the end of this section (see Sec.~\ref{subsec:Frolov-state}).

For a verification of our results at a general $r$ value inside
the BH, it is beneficial to compare them against an independent method
of computation. For that purpose, in addition to the procedure
used here, we have developed an alternative method (the \emph{analytic
extension }method), which is another variant of the $t$-splitting PMR method inside
the BH. This method is described in Appendix~\ref{App:The-analytic-extension} 
(the difference between the two variants lies in the manner in which the
extension from the exterior to the interior is carried out). We numerically
implemented this alternative procedure for the computation of $\left\langle \Phi^{2}\right\rangle _{\text{ren}}^{U}$
and $\left\langle T_{yy}\right\rangle _{\text{ren}}^{U}$
at two $r$ values in the $a/M=0.8$ case, and performed a (successful)
comparison with the results obtained via the procedure described
and employed in this paper {(}see Appendix~\ref{App:The-analytic-extension}
for more details{)}.  

Finally, we note that in the presentation of our numerical results in the next two subsections, we first begin with the fluxes and then proceed to the field square. We do it in this order mainly because we consider the former to be more physically interesting than the latter. 
(However, in the previous analytical section Sec.~\ref{sec:The-t-splitting-procedure}, we instead followed the reverse order since the
analytical constructions for the fluxes were built on the results for the field square).

\subsection{Plots for the fluxes}

\subsubsection{Between the horizons\label{subsec:fluxes-between-the-horizons}}

\begin{figure}[h!]
\centering \includegraphics[scale=0.3]{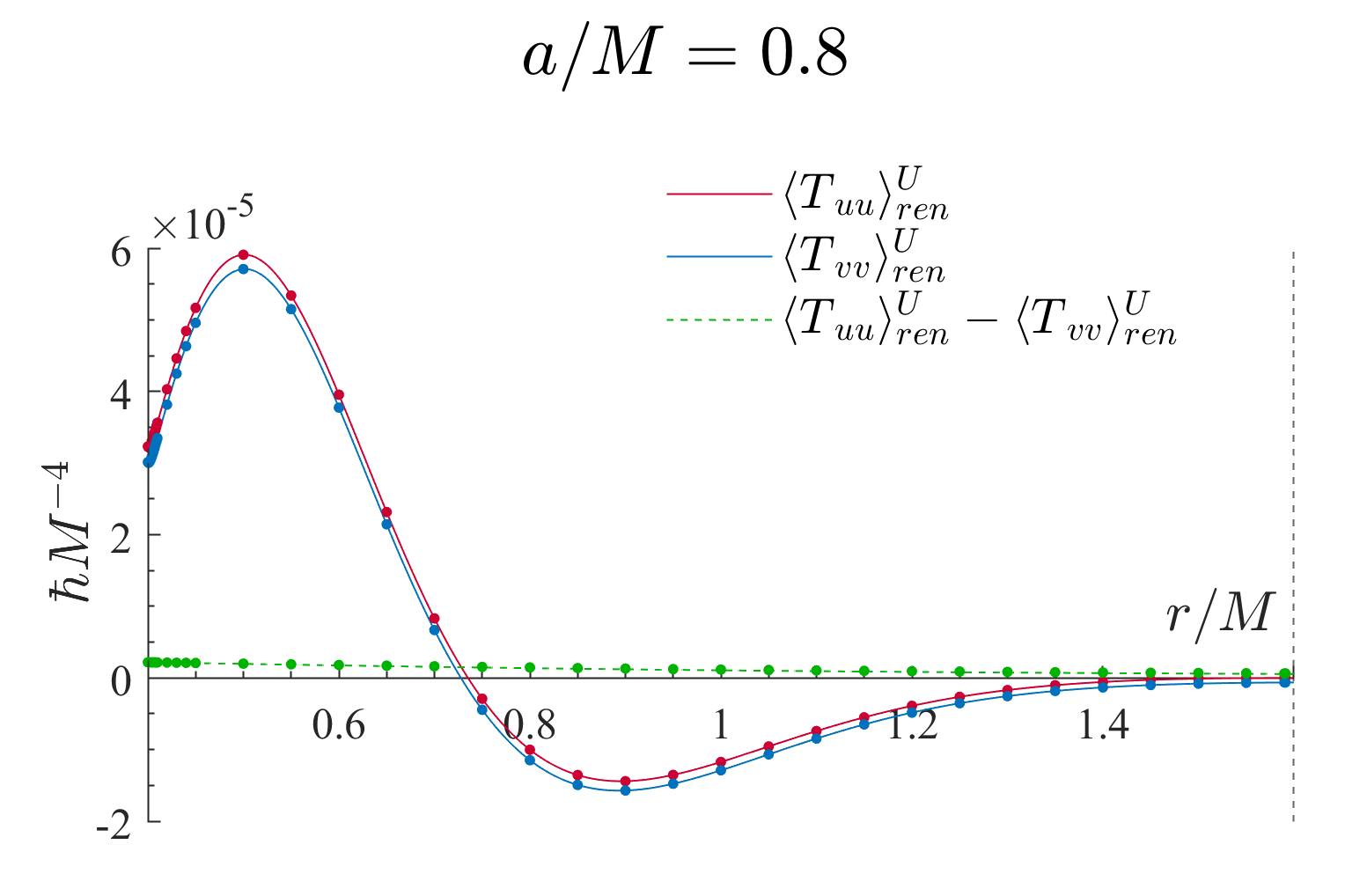}\caption{The Unruh-state fluxes $\left\langle T_{uu}\right\rangle _{\text{ren}}^{U}$
(in red), $\left\langle T_{vv}\right\rangle _{\text{ren}}^{U}$
(in blue) and their difference (in green) at the pole of an $a/M=0.8$
Kerr BH as a function of $r/M$, between the horizons ($r_{-}<r<r_{+}$).
The vertical dashed line marks $r=r_{+}$, and the vertical axis coincides
with $r=r_{-}$. The dots are numerically computed. The red and blue
lines are interpolations of $\left\langle T_{uu}\right\rangle _{\text{ren}}^{U}$
and $\left\langle T_{vv}\right\rangle _{\text{ren}}^{U}$,
respectively. The green dashed line follows the analytic relation
$\left\langle T_{uu}\right\rangle _{\text{ren}}^{U}-\left\langle T_{vv}\right\rangle _{\text{ren}}^{U}=\mathcal{F}_{0}/\left(r^{2}+a^{2}\right)$,
where $\mathcal{F}_{0}$ is the $r$-independent quantity (the Hawking outflux per unit solid angle in the polar direction) given in
Eq.~\eqref{eq:F(0)}. Here, this constant
is $\mathcal{F}_{0}\approx-1.7454781\times10^{-6}\hbar M^{-2}$.}
\label{Fig:a0p8_general}
\end{figure}

\begin{figure}[h!]
\centering \includegraphics[scale=0.3]{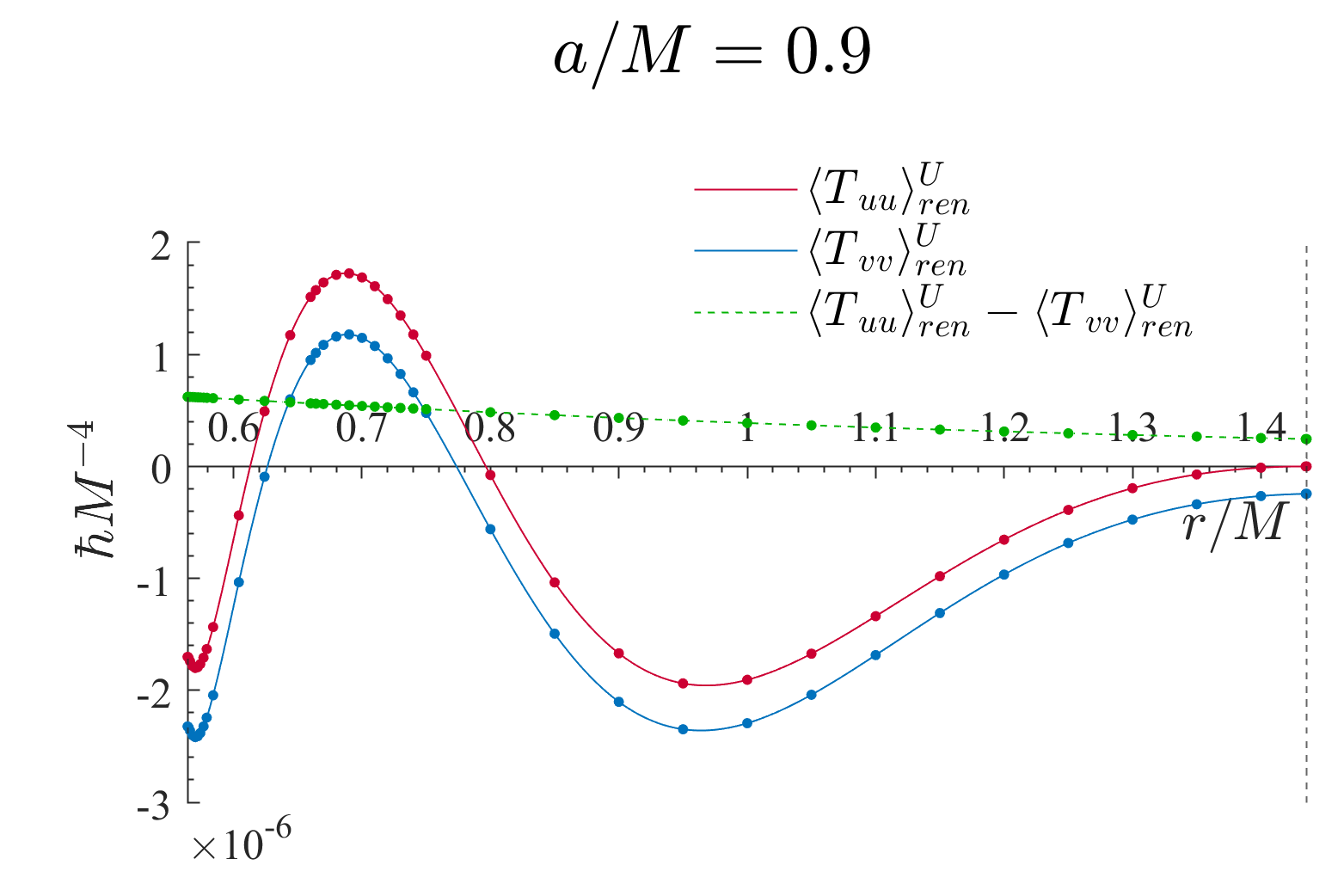}\caption{The same as Fig.~\ref{Fig:a0p8_general}, but for $a/M=0.9$ and
the corresponding conserved quantity is $\mathcal{F}_{0}\approx-7.0150098\times10^{-7}\hbar M^{-2}$.}
\label{Fig:a0p9_general}
\end{figure}

Figures \ref{Fig:a0p8_general} and \ref{Fig:a0p9_general} present
the fluxes $\left\langle T_{uu}\right\rangle _{\text{ren}}^{U}$,
$\left\langle T_{vv}\right\rangle _{\text{ren}}^{U}$ and
their difference between the horizons in the $a/M=0.8$ and $a/M=0.9$
cases, respectively. 

In both cases we see that the fluxes, which start off at the EH
as $\left\langle T_{uu}\right\rangle _{\text{ren}}^{U}=0$
and $\left\langle T_{vv}\right\rangle _{\text{ren}}^{U}<0$
(as should be), initially become increasingly negative as $r$ decreases,
and then follow a non-trivial behavior, with a few trend and sign
changes, until they reach their (either positive or negative) IH values.
(We leave the near-horizon behaviors to be explored in the next subsections.) 

Notably, although the IH values in the $a/M=0.9$ case are non-positive
at both horizon vicinities, there is a region inside the BH for which
they are positive (that is, each flux component vanishes at two points
inside the BH -- unlike the $a/M=0.8$ case, where there is only
one such point separating the negative and positive domains). 

In the $r$-range computed, we find three extrema in the $a/M=0.8$ case
(two are clearly seen in Fig.~\ref{Fig:a0p8_general} and the closest
one to the IH is better seen in Fig.~\ref{Fig:a0p8_IH-1}) and four extrema
in the $a/M=0.9$ case (three are clearly visible in Fig.~\ref{Fig:a0p9_general}
and the one closest to the IH is obtained at around $r=r_{-}+2\times10^{-4}M$
and is too subtle to be clearly seen in the figures in their present
scale).

Generally, the fluxes in the $a/M=0.8$ case are typically an
order of magnitude larger than those in the $a/M=0.9$ case. 

The difference between the fluxes $\left\langle T_{uu}\right\rangle _{\text{ren}}^{U}-\left\langle T_{vv}\right\rangle _{\text{ren}}^{U}$  at the pole equals $\mathcal{F}_{0}/\left(r^{2}+a^{2}\right)$,
where $\mathcal{F}_{0}$ is given in Eq.~\eqref{eq:F(0)} and corresponds to the Hawking
radiation outflux per unit solid angle in the polar direction. For $a/M=0.8$
this conserved quantity is $\mathcal{F}_{0}\approx-1.7454781\times10^{-6}\hbar M^{-2}$,
and for $a/M=0.9$ it is $\mathcal{F}_{0}\approx-7.0150098\times10^{-7}\hbar M^{-2}$
(as computed at $r=r_{-}$ from our state subtraction results of Ref.~\cite{KerrIH:2022}).

\subsubsection{EH vicinity\label{subsec:fluxes_EH-vicinity}}

For a discussion of the EH vicinity, it is useful to recall the parameter $\delta r_+$, which represendts the radial coordinate distance to the EH and is defined as $\delta r_{+}\equiv(r_{+}-r)/M$.
The expected regularity of the Unruh state at the EH implies the vanishing
of $\left\langle T_{uu}\right\rangle _{\text{ren}}^{U}$ there, at
least as $\delta r_{+}^{2}$ {[}otherwise, in the regular Kruskal
$U$ coordinate, $\left\langle T_{UU}\right\rangle _{\text{ren}}^{U}$ would diverge{]}. 

\begin{figure}[h!]
\centering \includegraphics[scale=0.3]{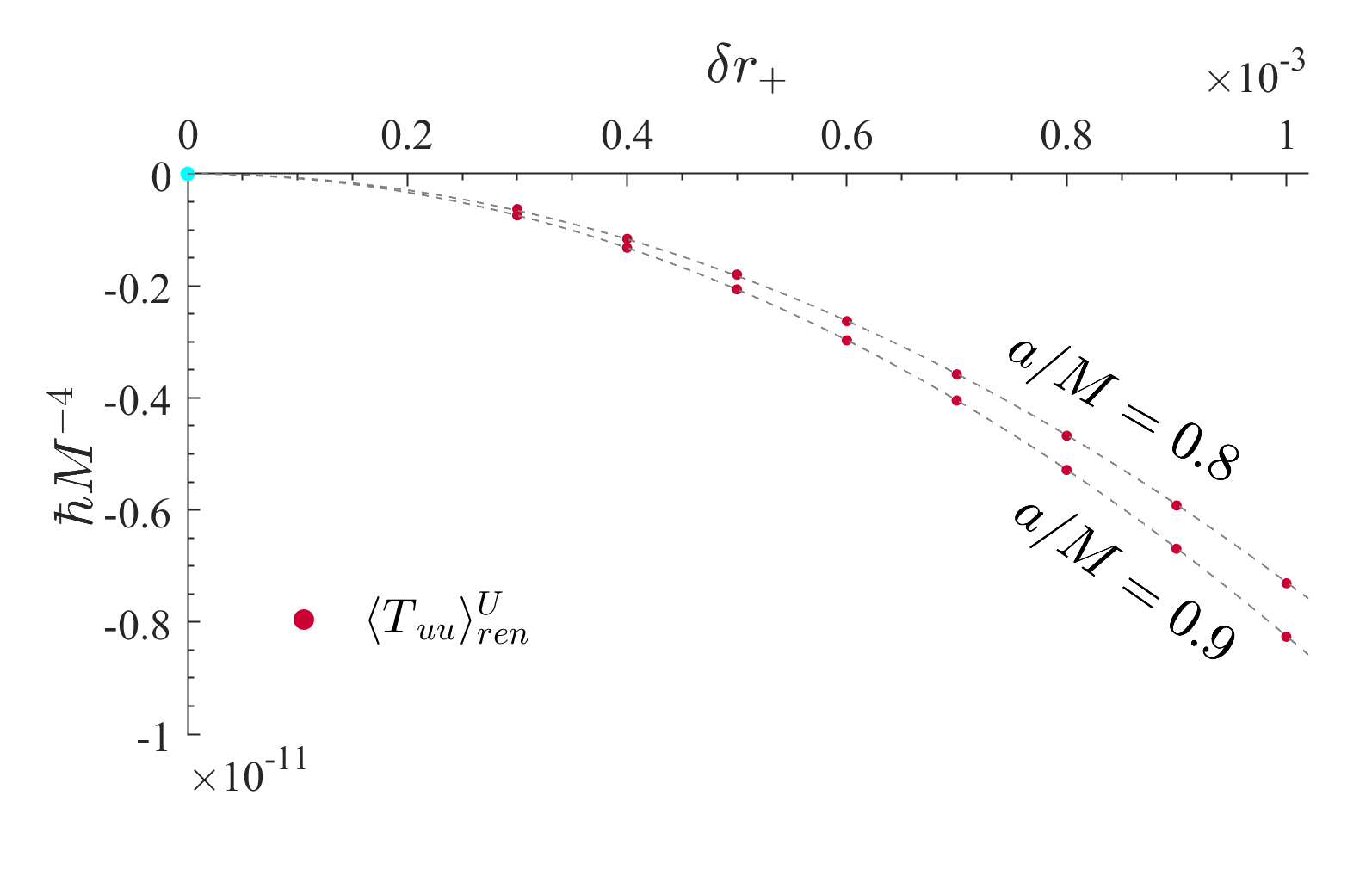}\caption{The Unruh-state $\left\langle T_{uu}\right\rangle _{\text{ren}}^{U}$
at the EH vicinity of $a/M=0.8$ and $a/M=0.9$ BHs (at the pole),
as a function of $\delta r_{+}=\left(r_{+}-r\right)/M$. \textcolor{black}{Unlike
the usual axis orientation, }the EH here corresponds to the leftmost
point ($\delta r_{+}=0$) and moving rightwards amounts to receding
from the EH inwards. That is, this plot corresponds to the reversed
righthand side of Figs.~\ref{Fig:a0p8_general} and \ref{Fig:a0p9_general}.
The red dots are numerically computed, and the continuous dashed lines
correspond to the anticipated behavior $c\cdot\delta r_{+}^{2}$ where
$c\approx-7.29\times10^{-6}\hbar M^{-4}$ for $a/M=0.8$ and $c\approx-8.25\times10^{-6}\hbar M^{-4}$
for $a/M=0.9$. The cyan dot at the origin denotes the anticipated
vanishing value of $\left\langle T_{uu}\right\rangle _{\text{ren}}^{U}$
at the EH.}
\label{Fig:near_EH_both}
\end{figure}

In Fig.~\ref{Fig:near_EH_both} we verify and explore this $\delta r_{+}^{2}$
behavior of $\left\langle T_{uu}\right\rangle _{\text{ren}}^{U}$
at the EH vicinity in both $a/M$ values at the pole. The prefactor
$c$ of $\delta r_{+}^{2}$ for each case is extracted numerically,
and the plot portrays the $c\cdot\delta r_{+}^{2}$ behavior as dashed
lines that run through the numerically computed dots. The two $c$
values are both negative and are of similar magnitude, being $c\approx-7.29\times10^{-6}\hbar M^{-4}$
for $a/M=0.8$ and $c\approx-8.25\times10^{-6}\hbar M^{-4}$ for $a/M=0.9$.

The other flux component, $\left\langle T_{vv}\right\rangle _{\text{ren}}^{U}$,
is not portrayed on this graph, as it does not have any specific anticipated
behavior at the EH (since in particular, unlike $u$, the Eddington
$v$ coordinate does not diverge at the EH). The near EH behavior
of $\left\langle T_{vv}\right\rangle _{\text{ren}}^{U}$
is depicted at the righthand side of the general $r$ plots in the
previous subsection {[}Figs.~\ref{Fig:a0p8_general} and \ref{Fig:a0p9_general}{]},
obtaining {[}from Eq.~\eqref{eq:Tvv_from_Tuu} in Appendix~\ref{App:The-intermediate-divergence}{]}
 finite values at the pole of the EH:
$\left\langle T_{vv}\right\rangle _{\text{ren}}^{U}=-5.454619\times10^{-7}\hbar M^{-4}$
in the $a/M=0.8$ case, and $\left\langle T_{vv}\right\rangle _{\text{ren}}^{U}=-2.442739\times10^{-7}\hbar M^{-4}$
in the $a/M=0.9$ case.

\subsubsection{IH vicinity\label{subsec:fluxes_IH-vicinity}}

To discuss the IH vicinity, we recall the dimensionless parameter $\delta r_{-}\equiv(r-r_-)/M$.

Unlike the EH, which is regular and hence invites a regular behavior
of the fluxes (in particular 
$\left\langle T_{uu}\right\rangle _{\text{ren}}^{U}=O(\delta r_{+}^{2})$,
as seen in the previous section), the IH vicinity offers a non-regular,
more intricate behavior. We shall briefly explore this behavior, up
to the current limitation to our computed points (which is $\delta r_{-}=10^{-5}$
for $a/M=0.8$ and $\delta r_{-}=10^{-4}$ for $a/M=0.9$).

\begin{figure}[h!]
\centering \includegraphics[scale=0.3]{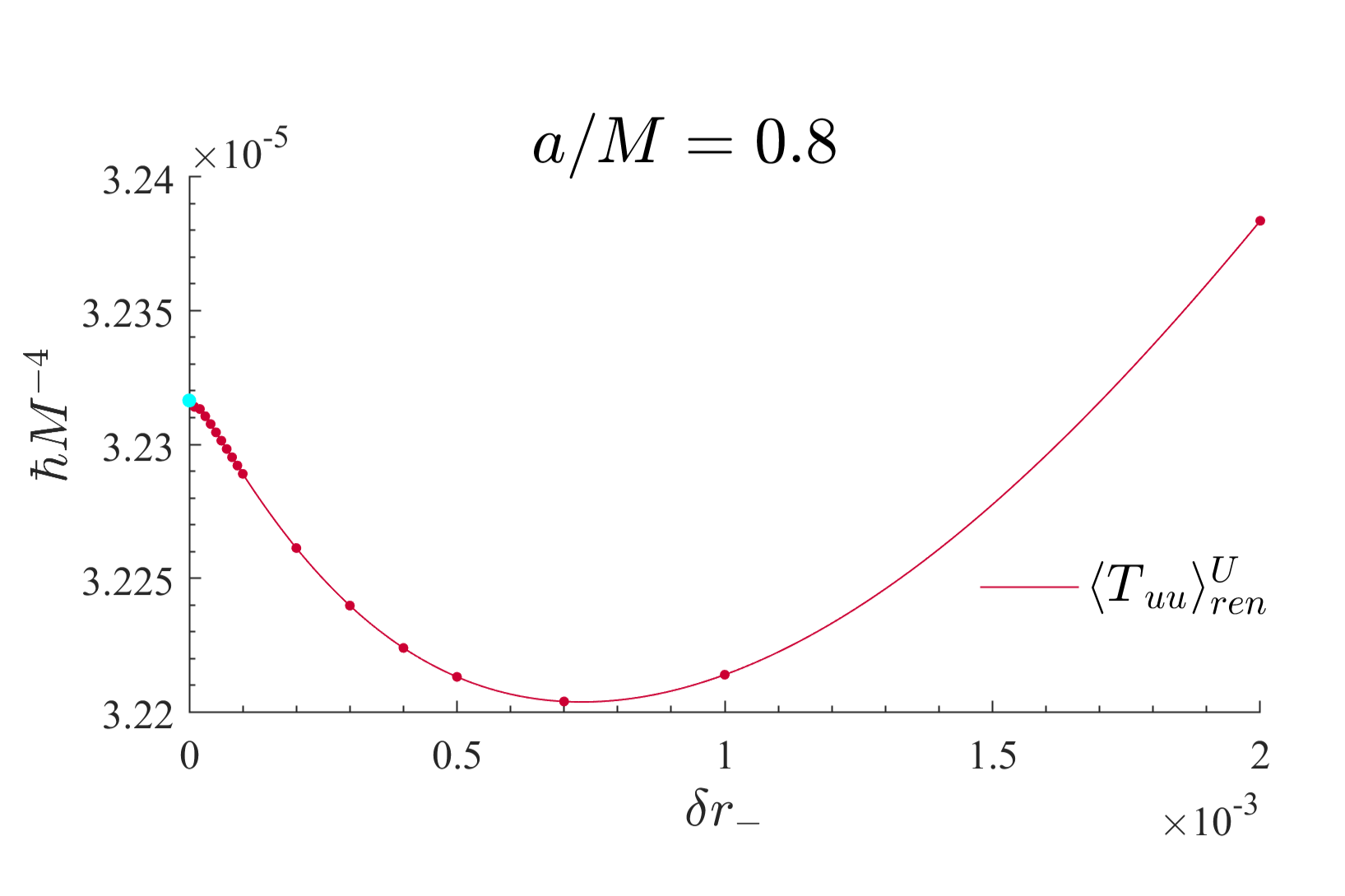}\caption{The Unruh-state $\left\langle T_{uu}\right\rangle _{\text{ren}}^{U}$
at the IH vicinity of an $a/M=0.8$ BH (at the pole), as a function
of $\delta r_{-}=\left(r-r_{-}\right)/M$. The red dots are computed
via $t$-splitting, and the continuous line is interpolated. The cyan
dot denotes the IH value of $\left\langle T_{uu}\right\rangle _{\text{ren}}^{U}$,
computed via state subtraction.}
\label{Fig:a0p8_IH-1}
\end{figure}

\begin{figure}[h!]
\centering \includegraphics[scale=0.3]{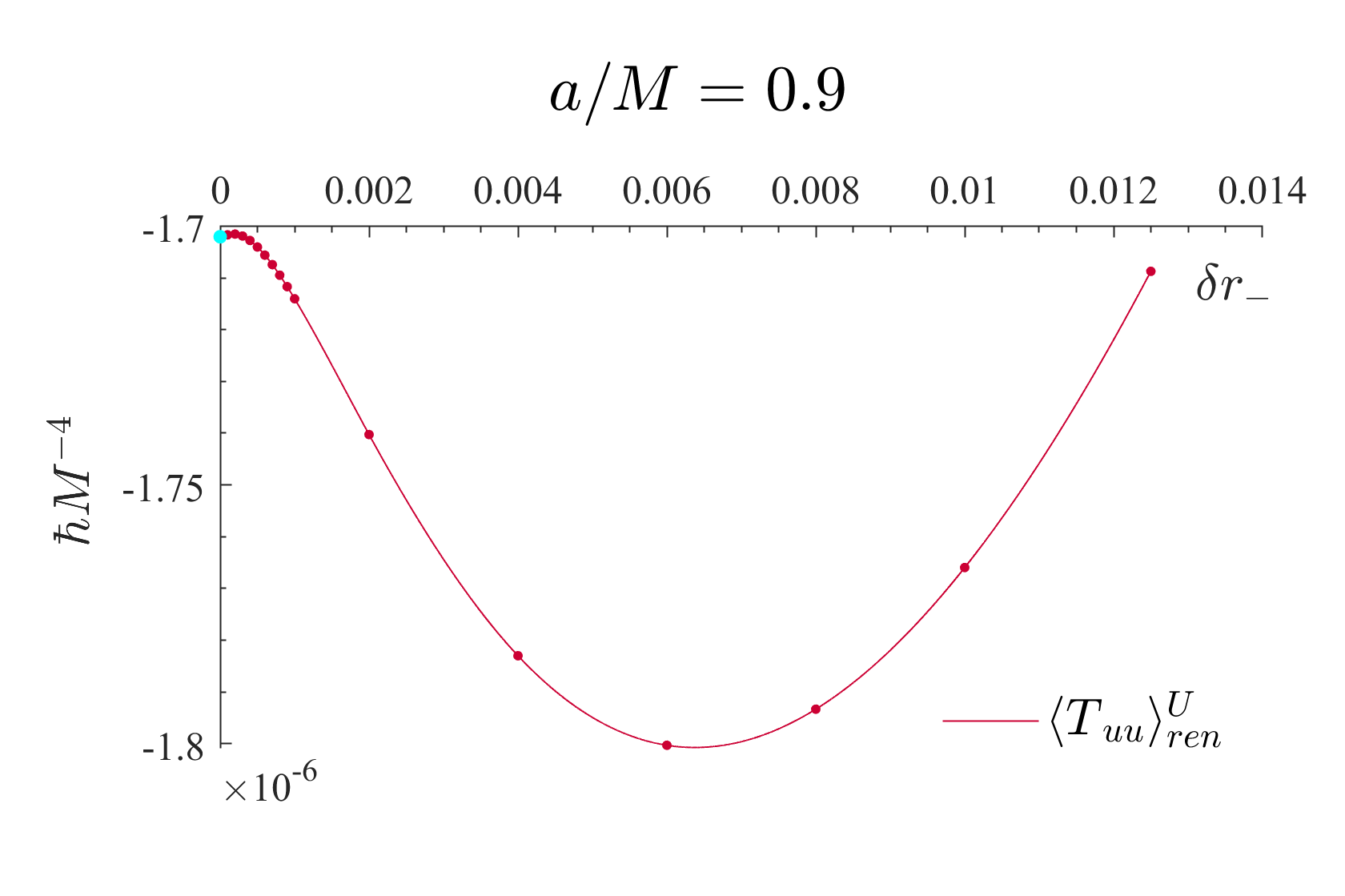}\caption{The same as Fig.~\ref{Fig:a0p8_IH-1}, but for $a/M=0.9$
(note the different horizontal scale).}
\label{Fig:a0p9_IH-1}
\end{figure}

In Figs.~\ref{Fig:a0p8_IH-1} and \ref{Fig:a0p9_IH-1}, we zoom into
the IH vicinity, corresponding to the leftmost side of Figs.~\ref{Fig:a0p8_general}
and \ref{Fig:a0p9_general}, respectively. (We focus on the $\left\langle T_{uu}\right\rangle _{\text{ren}}^{U}$
component for convenience, but $\left\langle T_{vv}\right\rangle _{\text{ren}}^{U}$
shows a similar picture.) This reveals yet another valley in the $a/M=0.8$
case, and another peak at the leftmost side of the $a/M=0.9$ plot
-- both are very close to the IH. These join a sequence of a few
peaks and valleys on approaching the IH in both cases. The presence
of such minima or maxima points so close to the IH exposes a non-regular
behavior (i.e., not a simple power series in $\delta r_{-}$, as in
the analogous EH case).

\begin{figure}[h!]
\centering \includegraphics[scale=0.3]{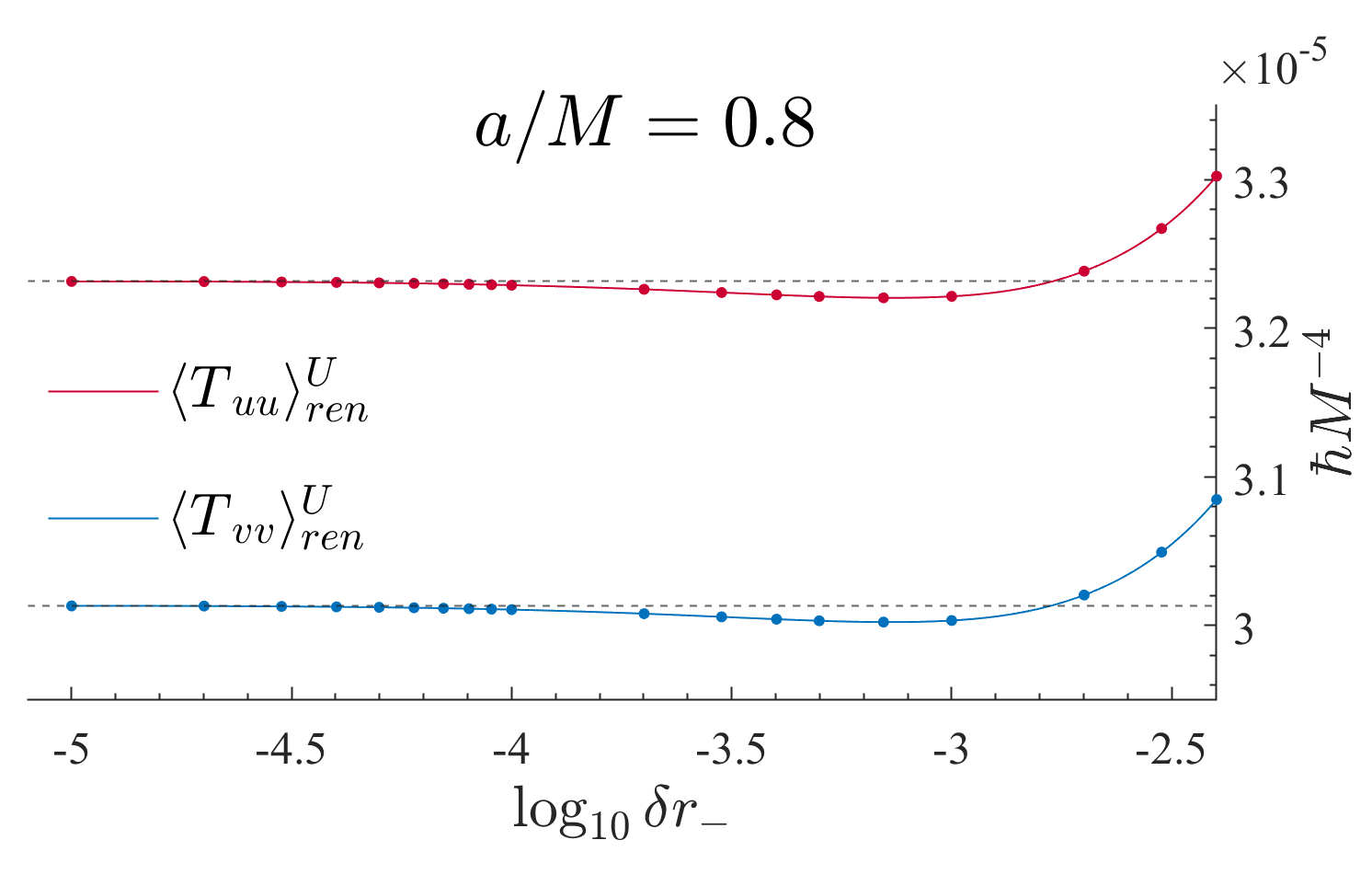}\caption{The Unruh fluxes $\left\langle T_{uu}\right\rangle _{\text{ren}}^{U}$
(in red) and $\left\langle T_{vv}\right\rangle _{\text{ren}}^{U}$
(in blue) at the pole of an $a/M=0.8$ Kerr BH, as a function of $\log_{10}\delta r_{-},$where
$\delta r_{-}\equiv\left(r-r_{-}\right)/M$ denotes the coordinate distance from
the IH. The dots are numerically computed via $t$-splitting, and
the continuous lines are interpolated. The horizontal dashed lines
correspond to the values computed straight at the IH through state-subtraction.
As the IH is approached (proceeding to the left), the point-splitting
values approach the IH values.}
\label{Fig:a0p8_IH}
\end{figure}

\begin{figure}[h!]
\centering \includegraphics[scale=0.3]{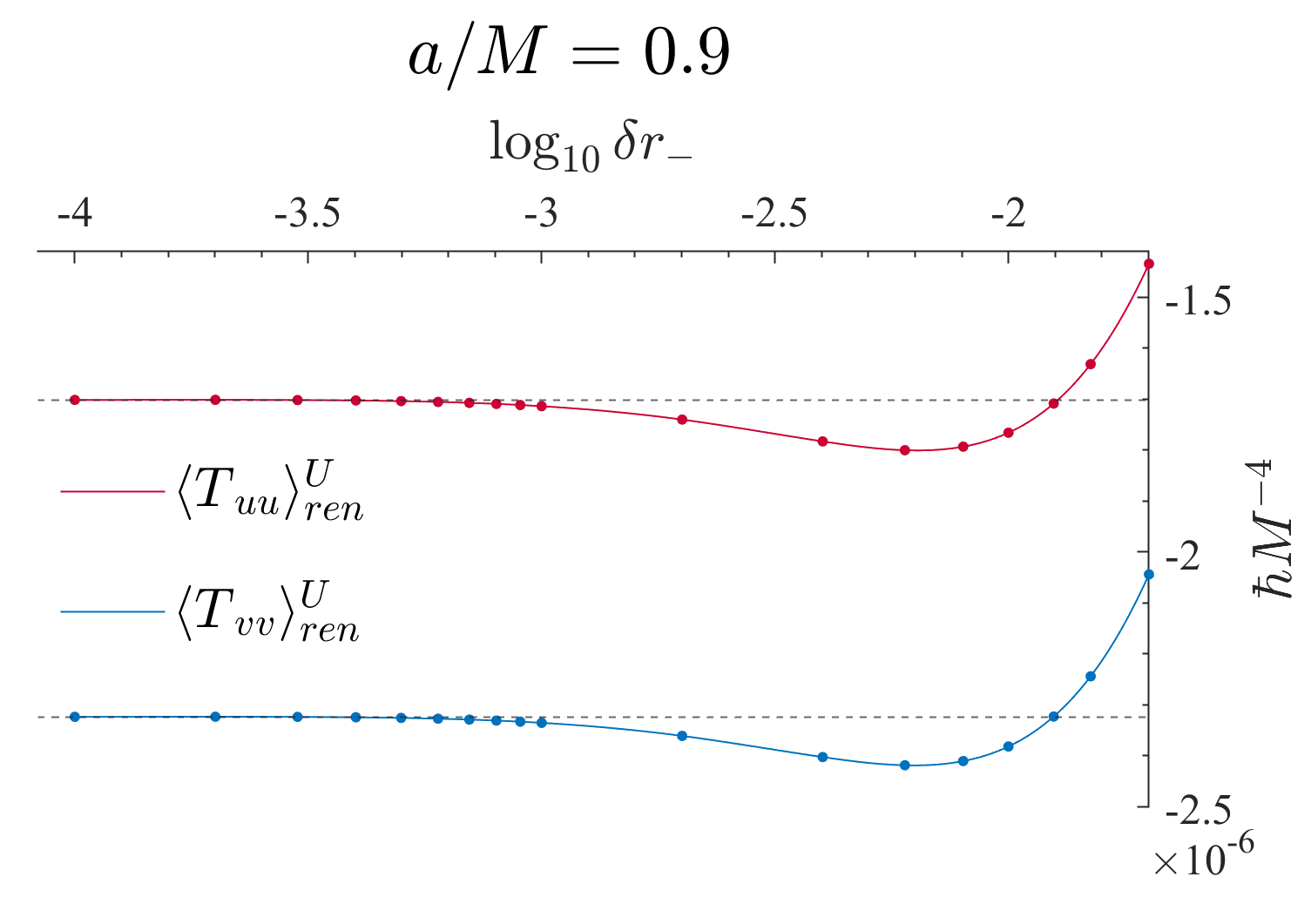}\caption{The same as Fig.~\ref{Fig:a0p8_IH}, but for an $a/M=0.9$ BH.}
\label{Fig:a0p9_IH}
\end{figure}

In a previous work \cite{KerrIH:2022} we computed the
flux components \textit{exactly} at the IH using the -- independent -- method of state subtraction for a variety
of $a/M$ and $\theta$ values. In particular, the values we found
at the pole for $a/M=0.8$ are $\left\langle T_{uu}\right\rangle _{\text{ren}}^{U}=3.23163918\times10^{-5}\hbar M^{-4}$
and $\left\langle T_{vv}\right\rangle _{\text{ren}}^{U}=3.01345442\times10^{-5}\hbar M^{-4}$,
and for $a/M=0.9$ they are $\left\langle T_{uu}\right\rangle _{\text{ren}}^{U}=-1.702041202\times10^{-6}\hbar M^{-4}$
and $\left\langle T_{vv}\right\rangle _{\text{ren}}^{U}=-2.323817852\times10^{-6}\hbar M^{-4}$.
In Sec.~3 in the Supplemental Material of that Letter, we compared
the $r\to r_{-}$ limit of the $t$-splitting results with the state-subtraction
IH result in these two $a/M$ values, reaching an agreement of four
digits of precision. Figs.~\ref{Fig:a0p8_IH} and \ref{Fig:a0p9_IH}
visually portray this agreement.

\subsection{Plots for $\left\langle \Phi^{2}\right\rangle $\label{subsec:Plots-for-Phi^2}}

\begin{figure}[h!]
\centering \includegraphics[scale=0.3]{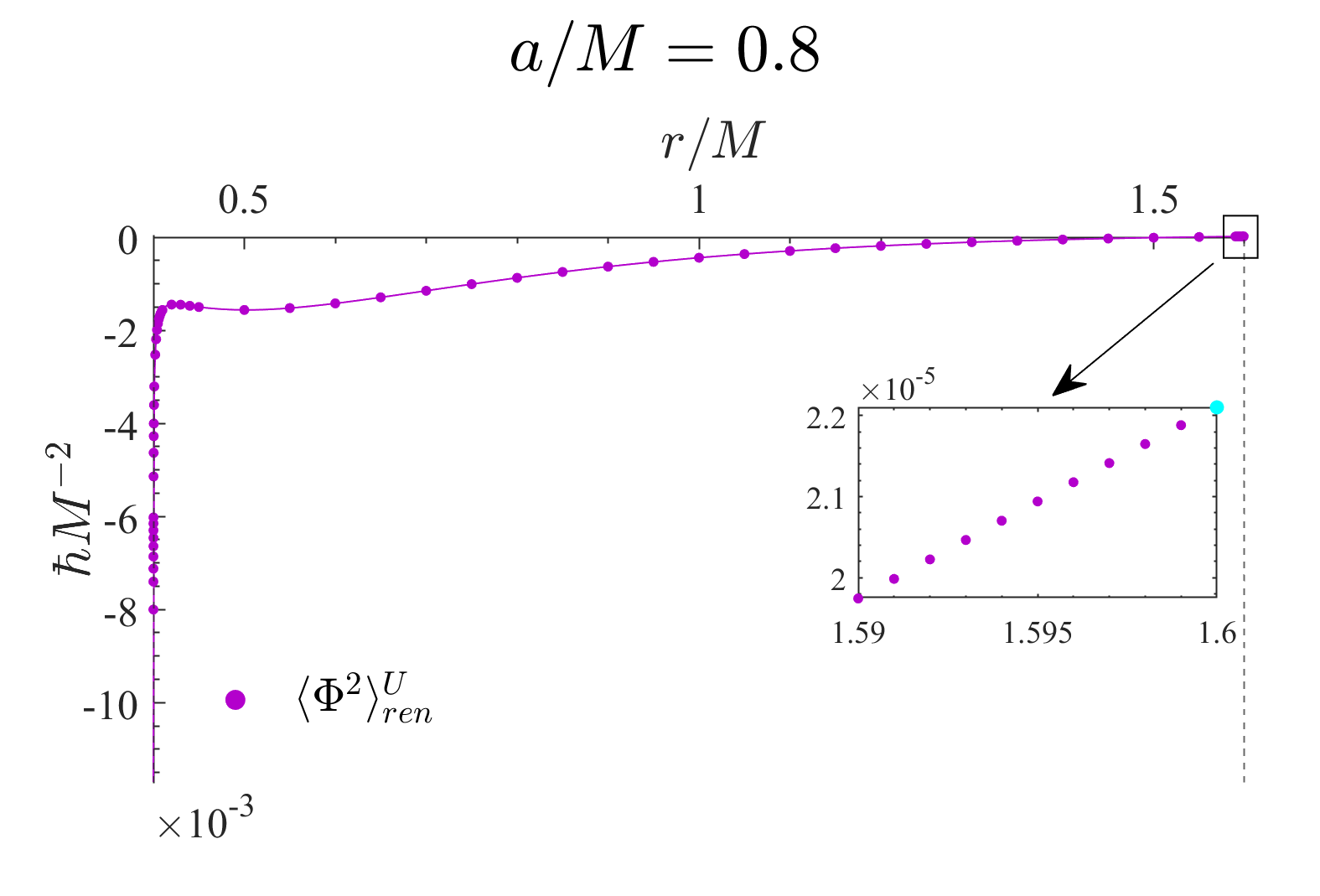}\caption{$\left\langle \Phi^{2}\right\rangle _{\text{ren}}^{U}$ at the pole
of an $a/M=0.8$ Kerr BH as a function of $r$, between the horizons
($r_{-}<r<r_{+}$). The vertical dashed line marks $r=r_{+}$, and
the vertical axis corresponds to $r=r_{-}$. The dots are numerically
computed. The continuous line is interpolated. The near-EH domain
is zoomed-in at the right of the figure, demonstrating a regular behavior
and a change of sign occurring towards $r_{+}$.}
\label{Fig:a0p8_general-phi}
\end{figure}

\begin{figure}[h!]
\centering \includegraphics[scale=0.3]{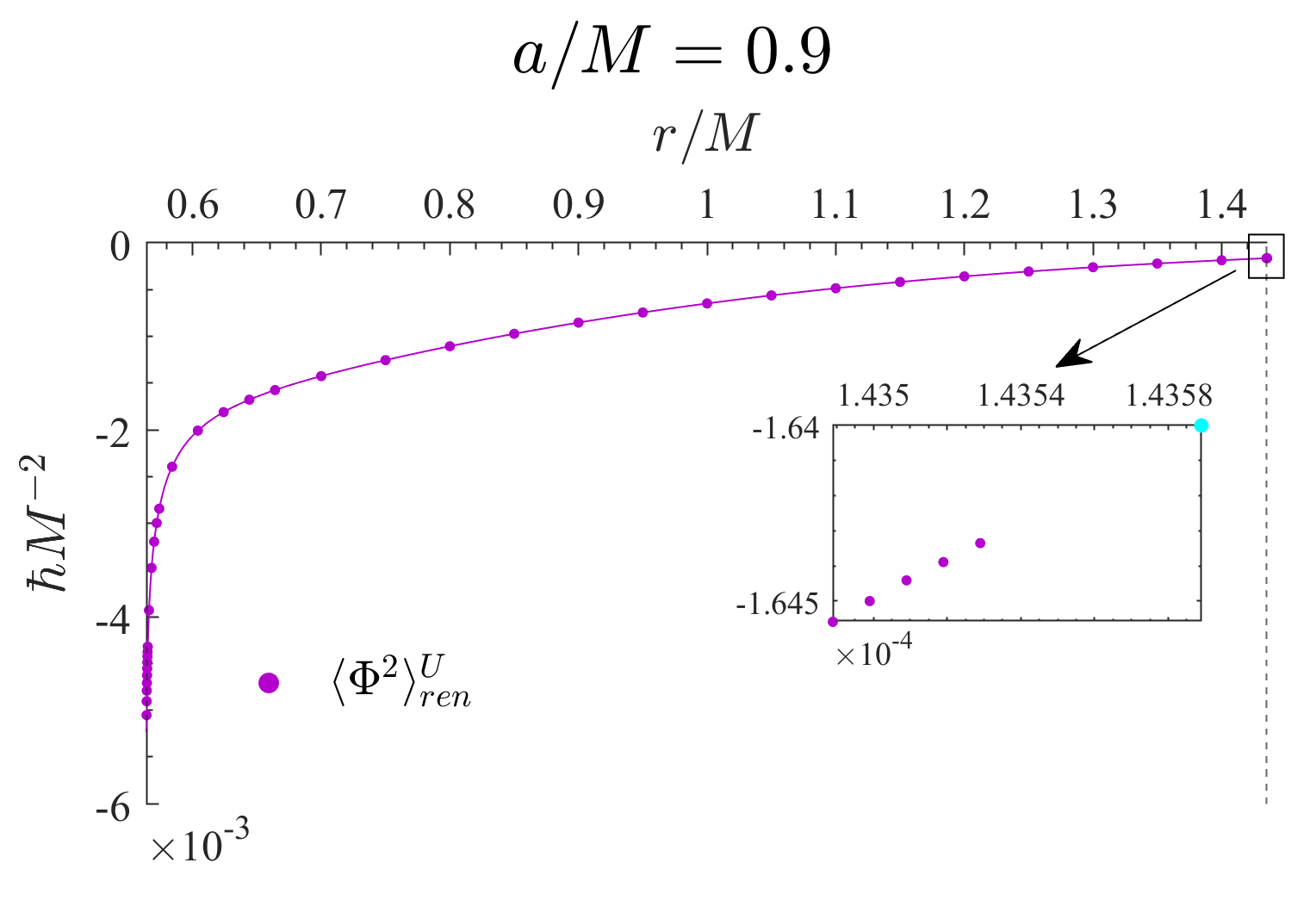}\caption{The same as Fig.~\ref{Fig:a0p8_general-phi}, but for $a/M=0.9$.
It is difficult to approach $r_{+}$ closer than given here, but the
existing data indeed shows regularity on approaching the EH.}
\label{Fig:a0p9_general-phi}
\end{figure}

\begin{figure}[h!]
\centering \includegraphics[scale=0.3]{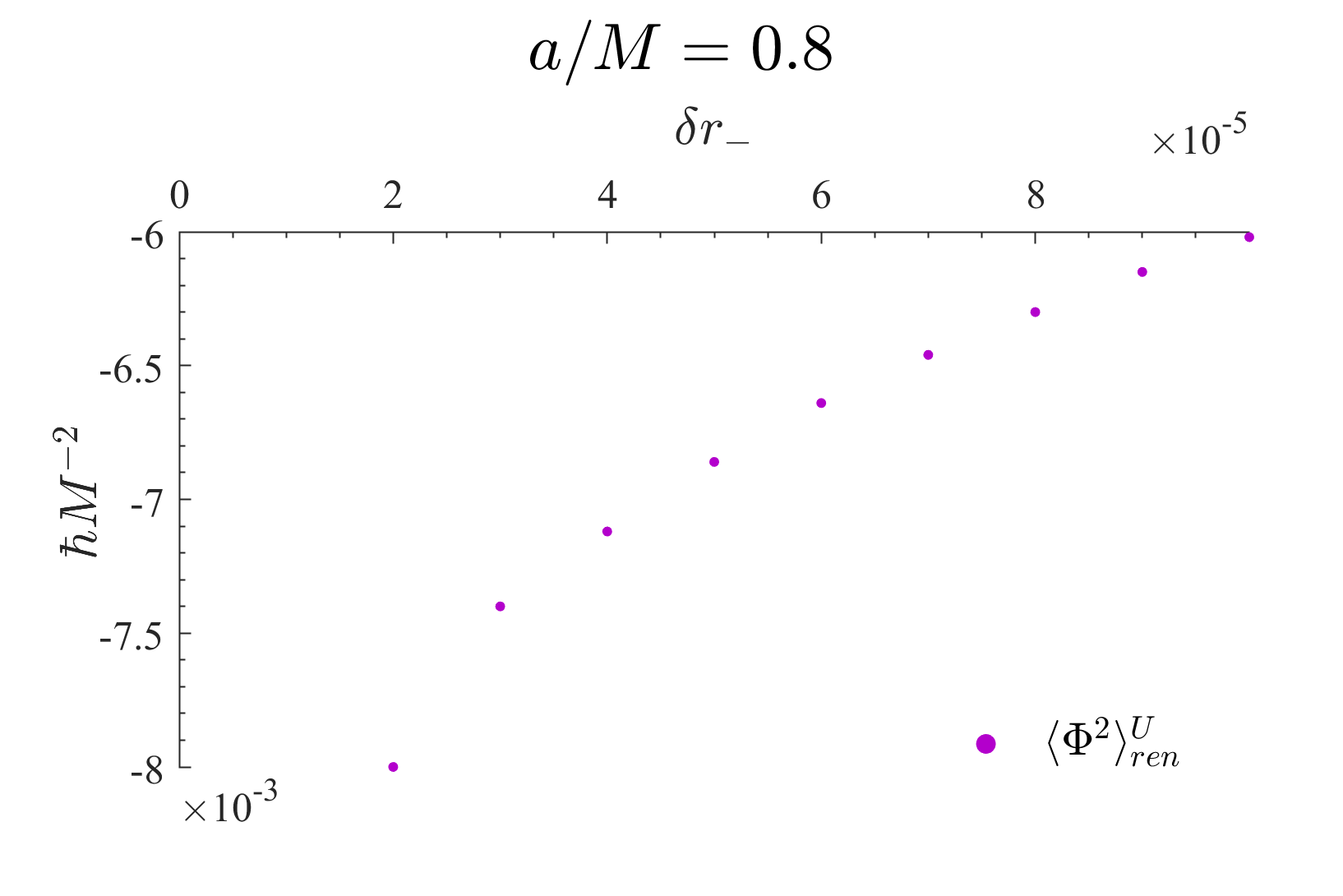}\caption{$\left\langle \Phi^{2}\right\rangle _{\text{ren}}^{U}$ at the pole
of an $a/M=0.8$ Kerr BH near the IH, as a function of $\delta r_{-}\equiv\left(r-r_{-}\right)/M$.}
\label{Fig:a0p8_near_IH-phi}
\end{figure}

\begin{figure}[h!]
\centering \includegraphics[scale=0.3]{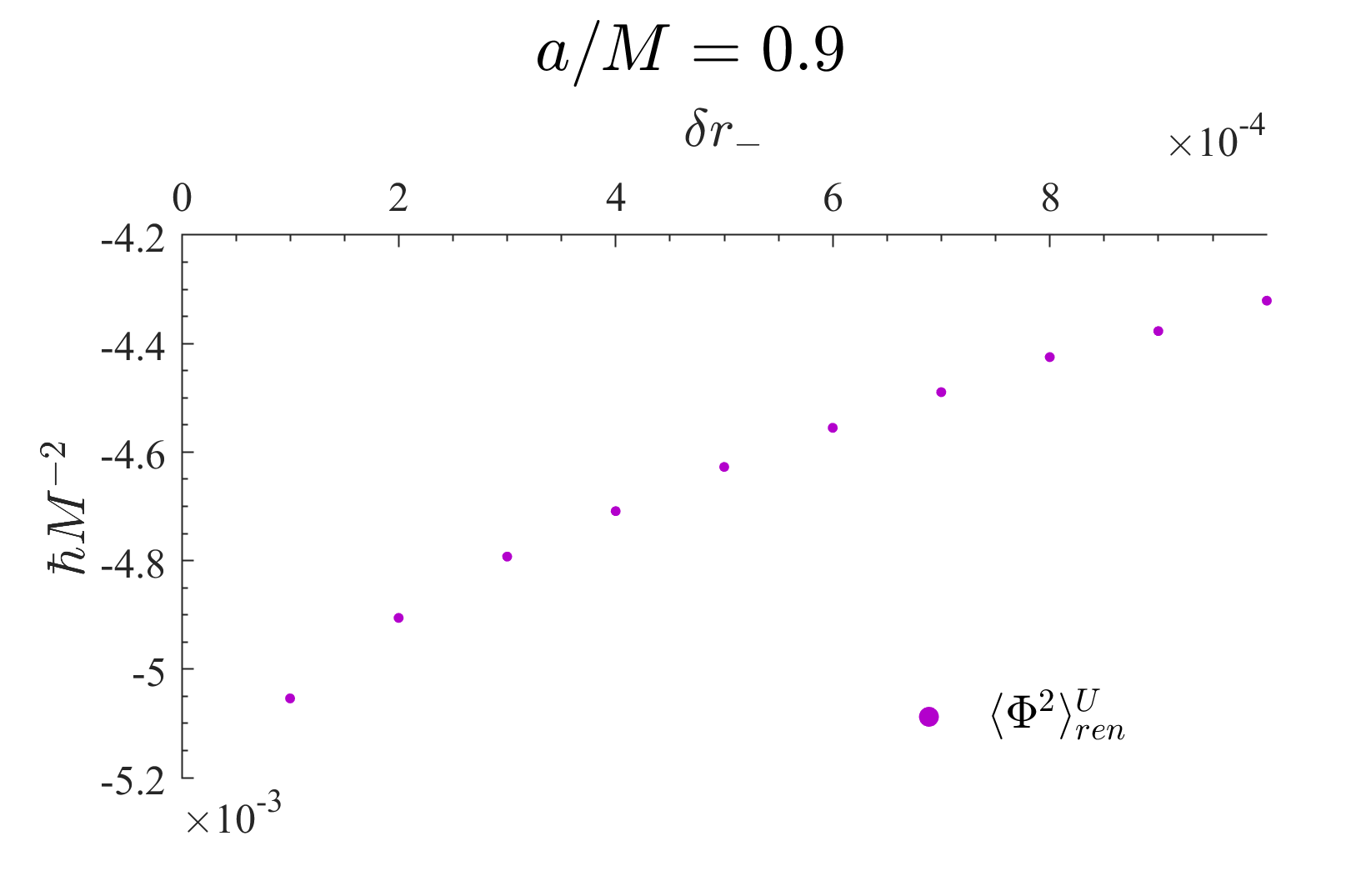}\caption{The same as Fig.~\ref{Fig:a0p8_near_IH-phi}, but for $a/M=0.9${{}
(note the different horizontal scale).}}
\label{Fig:a0p9_near_IH-phi}
\end{figure}

Figures \ref{Fig:a0p8_general-phi} and \ref{Fig:a0p9_general-phi}
portray $\left\langle \Phi^{2}\right\rangle _{\text{ren}}^{U}$ as
a function of $r/M$ between the horizons at the pole of $a/M=0.8$
and $a/M=0.9$ Kerr BHs, respectively, with the EH vicinity zoomed-in.
In Figs.~\ref{Fig:a0p8_near_IH-phi} and \ref{Fig:a0p9_near_IH-phi}
we zoom into the IH vicinity, corresponding to the left side of Figs.
\ref{Fig:a0p8_general-phi} and \ref{Fig:a0p9_general-phi}. 

At the EH the behavior of $\left\langle \Phi^{2}\right\rangle _{\text{ren}}^{U}$
seems perfectly regular, approaching a (positive) extrapolated value
of $2.21\times10^{-5}\hbar M^{-2}$ at the EH in the $a/M=0.8$ case and a
(negative) extrapolated value of $-1.64\times10^{-4}\hbar M^{-2}$ at the EH
in the $a/M=0.9$ case (see also Sec.~\ref{subsec:Frolov-state}
which mentions this different sign issue for the field square at the EH on the axis of rotation, although in a different quantum
state). 
However, at the IH vicinity $\left\langle \Phi^{2}\right\rangle _{\text{ren}}^{U}$
seems to follow a less trivial behavior, in particular a rapid decrease
towards the IH which may raise the suspicion of a divergence there.

We note, however, that the computed results do not in fact reveal the asymptotic form of  $\left\langle \Phi^{2}\right\rangle _{\text{ren}}^{U}$ at the IH vicinity. By employing state subtraction (as in Ref.~\cite{KerrIH:2022}) and considering the regularity of the reference state at the IH, we managed to determine that $\left\langle \Phi^{2}\right\rangle _{\text{ren}}^{U}$ in fact reaches a \emph{finite} asymptotic value in the limit $r\to r_-$ along the axis of rotation, which it approaches with an $r_*^{-3}$ tail. (The actual limiting value of $\left\langle \Phi^{2}\right\rangle _{\text{ren}}^{U}$ at the IH remains unknown, since we only computed the state \emph{difference}). 
In addition, while $\left\langle \Phi^{2}\right\rangle _{\text{ren}}^{U}$ asymptotes to a finite value at the IH on the pole, we have numerical indications that \emph{off} the pole ($\theta\neq0,\pi$) $\left\langle \Phi^{2}\right\rangle _{\text{ren}}^{U}$ actually diverges like $r_*$ on approaching the IH. However, this requires further investigation, and this entire  very-near IH exploration of $\left\langle \Phi^{2}\right\rangle _{\text{ren}}^{U}$ remains beyond the scope of this paper (in particular, since this exploration was not done using $t$-splitting, which is the topic of the current paper).   

It is
worth comparing these plots to the corresponding plots in the analogous
RN case with charge parameter
$Q/M=0.8$, presented and explored in Ref.~\cite{GroupPhiRN:2019}.
In the RN case examined there, $\left\langle \Phi^{2}\right\rangle _{\text{ren}}^{U}$
 sharply increases towards the IH, as may be seen in Fig. 1 therein.
(Note that here we have a sharp decrease rather than increase on approaching
the IH, but this difference does not seem to be significant for the
present discussion). This trend, continuing up to about $\delta r_{-}\sim10^{-6}$,
raises the suspicion of a divergence at the IH. However, beyond this
point the trend changes, and a peak is obtained at about $\delta r_{-}\sim10^{-9}$
(see Fig.~2 therein). Then, a surprising intricate behavior follows,
eventually leading to a \emph{finite} value at the IH. This finite
asymptotic value is obtained after a sequence of a few minima and
maxima, followed by a final inverse-power decay which is only exposed
as deep as $\delta r_{-}\sim10^{-175}$. {[}Reaching this very small
$\delta r_{-}$ domain in the RN case was not done through a straightforward
numerical computation of the radial function, but using a certain
approximation (the so-called \emph{semi-asymptotic approximation})
that was tailored to the $\theta$-splitting method used there. Here
we use $t$-splitting instead (generally using $\theta$-splitting
is not pragmatic in Kerr due to the lack of spherical symmetry), hence
the method that allowed us to approach extremely close to the IH in
the RN case is not available in the Kerr case.

\subsubsection*{Testing the EH value of $\left\langle \Phi^{2}\right\rangle _{\text{ren}}$:
the polar Hartle-Hawking state \label{subsec:Frolov-state}}

 We wish to test the $t$-splitting computation of $\left\langle \Phi^{2}\right\rangle _{\text{ren}}^{U}$, 
but we do not have any value known a priori in the Unruh state (unlike the anchors we had for the fluxes, as described above). Hence, we may search for simple tests available in other states. The Hartle-Hawking state does not exist in Kerr, due to the existence of superradiant modes~\cite{Kay:Wald,OttewillWinstanley:2000}. However, at the pole, there are no superradiant modes and so we may define a formal ``polar
Hartle-Hawking'' analog, which we denote with a superscript $H$; see Refs.~\cite{Frolov:Thorne,OttewillWinstanley:2000} for mode-sum expressions for the HTPF in this state, which contain a pole singularity at a real frequency value for $\theta\neq 0,\pi$  but are well-defined for $\theta= 0,\pi$. 
Using Wick rotation, Frolov~\cite{Frolov:1982}  found a closed form expression for
$\left\langle \Phi^{2}\right\rangle _{\text{ren}}^{H}$ at the pole ($\theta= 0,\pi$) of the EH
of an electrically-charged and rotating, Kerr-Newman BH {[}see Eq.~(11) therein{]}, reading
\begin{equation}
\left\langle \Phi^{2}\left(r_{+}\right)\right\rangle _{\text{ren}}^{H}=\frac{1}{24\pi^{2}\left(r_{+}^{2}+a^{2}\right)}\left(r_{+}\kappa_{+}-\frac{a^{2}}{r_{+}^{2}+a^{2}}\right)\,.\label{eq:Frolov}
\end{equation}
Note that in the RN case, with $a=0$, this quantity is always positive
-- whereas in the Kerr case (BH charge $Q=0$) this quantity is positive for
$a/M<\sqrt{3}/2$ and nonpositive otherwise. We see something similar
happening in the Unruh state (whose difference from the polar-Hartle-Hawking
state is regular): $\left\langle \Phi^{2}\right\rangle _{\text{ren}}^{U}$
at the pole of the EH is positive for $a/M=0.8$ and negative for $a/M=0.9$ {[}see
Sec.~\ref{subsec:Plots-for-Phi^2}{]}. However, we  have only considered these two spin values, so we do not know whether there is a spin value in which $\left\langle \Phi^{2}\right\rangle _{\text{ren}}^{U}$ transits from positive to negative, and if so, what it is. 

Using the bare mode-sum expression corresponding to the polar-Hartle-Hawking
state at the pole (outside the scope of this paper), and regularizing
via the standard $t$-splitting procedure, we obtained $\left\langle \Phi^{2}\right\rangle _{\text{ren}}^{H}$
at the pole for several $r$ values approaching the EH, for both spin values. Extrapolating
these $t-$splitting values to the EH in both cases, we reached an
agreement of at least 4 figures with the analytical result of Eq.~\eqref{eq:Frolov} (being $\left\langle \Phi^{2}\left(r_{+}\right)\right\rangle _{\text{ren}}^{H}\approx1.319\times10^{-4}\hbar M^{-2}$
for $a/M=0.8$ and $\left\langle \Phi^{2}\left(r_{+}\right)\right\rangle _{\text{ren}}^{H}\approx-9.425\times10^{-5}\hbar M^{-2}$
for $a/M=0.9$).

\section{Discussion}\label{sec:discussion}

In this paper, we employ the method of point splitting -- specifically, $t$-splitting -- to calculate the Unruh-state fluxes $\left\langle T_{uu}\right\rangle _{\text{ren}}^{U}$, $\left\langle T_{vv}\right\rangle _{\text{ren}}^{U}$, as well as the field square $\left\langle \Phi^2\right\rangle _{\text{ren}}^{U}$, for a minimally-coupled massless scalar field, in the interior of a spinning BH on the axis of rotation. These fluxes are particularly crucial for understanding backreaction near the IH, as discussed in Ref.~\cite{KerrIH:2022}.
We calculated $\left\langle T_{uu}\right\rangle^U_{\text{ren}}$, $\left\langle T_{vv}\right\rangle^U_{\text{ren}}$ and $\left\langle \Phi^{2}\right\rangle^U_{\text{ren}}$ for two different BH spin values, at the pole $\theta=0$, spanning from near the EH to near the IH. Our results, displayed in the various figures of Sec.~\ref{sec:Numerical-results}, provide a quantitative picture of the behavior of $\left\langle T_{uu}\right\rangle^U_{\text{ren}}$, $\left\langle T_{vv}\right\rangle^U_{\text{ren}}$ and $\left\langle \Phi^{2}\right\rangle^U_{\text{ren}}$ in the BH interior. In particular, our results include a focus on the IH vicinity, validating our state-subtraction results computed directly at the IH (see Ref.~\cite{KerrIH:2022}). In addition, our computations at the EH vicinity provide numerical support for the anticipated regularity of the Unruh state at the EH.

The method presented and employed in this paper is the interior counterpart of the $t$-splitting method formulated in the BH exterior, generally introduced in Ref.~\cite{AAt:2015} and employed outside a Kerr BH in Ref.~\cite{LeviEilonOriMeentKerr:2017} ($\varphi$-splitting was also use for computations outside a Kerr BH in that work, but this variant is technically more difficult and in particular is not applicable at the pole). Note, however, that employing $t$-splitting inside a Kerr BH is significantly harder and involves unique challenges not present in the BH exterior. This is rooted in the form of the centrifugal potential (in particular at large l), see Fig.~\ref{Fig:potential}. Both inside and outside the BH, the large-$l$ effective potential generally scales as $l^2$. In the exterior, it acts as a potential barrier. As a consequence, the $l$-series converges exponentially fast, and the summation over $l$ is easily performed. Inside the BH, however, the potential at large $l$ acts as a potential well. As a consequence, the $l$-series does not converge, strictly speaking. This constitutes the so-called \emph{intermediate divergence} (ID) problem. To overcome this problem, we introduce a ``small split" in $\theta$, that is taken to vanish before the coincidence limit in the $t$ direction is taken. With the aid of this additional limiting process, we can identify a certain $l$-sequence (to which we sometimes refer as the ID) which properly captures the large-$l$ divergent piece of the original sequence – and which should be subtracted from that original $l$-sequence before summation, in order to obtain the correct renormalized result.
As part of this treatment, we show (partly analytically and partly numerically) that the ID attains a specific form, given in Eq.~\eqref{eq:Ewl_div}. 
  This ID subtraction procedure is illustrated through various figures (in particular, Figs.~\ref{Fig:phi_vs_l_4tiles} and ~\ref{Fig:phi_integrand_2tiles} for the field square and Figs.~\ref{Fig:Tuu_vs_l_4tiles}, ~\ref{Fig:Tuu_vs_w_2tiles} and ~\ref{Fig:Tuu_integrand_2tiles} for $T_{uu}$) and justified in an appendix.
After this ID is subtracted, the remaining regular $l$-series does converge (and thereby yields the basic $\omega$-integrand). However, even post-subtraction, the convergence of the regularized $l$-series is rather slow, proceeding as $1/l$ (that is, $1/l^2$ for the $l$-sequence
). This presents a significant numerical challenge, which we overcome by computing the $l$-sequence up to $l=300$, and then fitting the sequence as a series of inverse powers $1/l^k$,\footnote{For a more precise account of the numerical implementation, see Appendix~\ref{App:Numerical-implementation}.} typically reaching $k\sim100$. For this high-order fit to succeed, we had to compute the individual $l$-contributions with a high precision of more than $\sim250$ decimal figures. This, in turn, required the computation of the reflection coefficient $\rho_{\omega l}^\text{up}$ and the radial function $\psi_{\omega l}^\text{int}$  [as well as its derivative $\psi_{\omega l,r}^\text{int}$ and also the spheroidal eigenfunctions $S_{\omega l}(0)$] at that level of precision. 
Upon successful computation of the $l$-sum, we obtained the $\omega$-integrand, paving the way to integration (after subtracting the known PMR counterterms). 
This eventually allowed the renormalized quantity (either the field square $\left\langle \Phi^{2}\right\rangle _{\text{ren}}^{U}$ or the fluxes $\left\langle T_{uu}\right\rangle _{\text{ren}}^{U}$ and $\left\langle T_{vv}\right\rangle _{\text{ren}}^{U}$) to be computed at a wide range of $r$ values, spanning from (just off) the EH to (just off) the IH.

Close to the horizons (located at $r_{+}$ and $r_{-}$), the $t$-splitting computation at the pole becomes more challenging (mainly due to the divergence of counterterms). In particular, $t$-splitting on the axis of rotation is inapplicable directly at the horizons. Nevertheless, we managed to compute the fluxes $\left\langle T_{vv}\right\rangle _{\text{ren}}^{U}$ and $\left\langle T_{uu}\right\rangle _{\text{ren}}^{U}$ sufficiently close to the horizons, so as to obtain the horizon limits of these two quantities by extrapolation.
In particular, our results extrapolated to the IH (as demonstrated in Figs.~\ref{Fig:a0p8_IH} and \ref{Fig:a0p9_IH}) show remarkable agreement with those obtained using the state-subtraction method directly at the IH ~\cite{KerrIH:2022}, hence validating the state-subtraction method used therein.
In addition, on approaching the EH, $\left\langle T_{uu}\right\rangle _{\text{ren}}^{U}$ decays to zero as expected (like $\delta r_{+}^2$, see Fig.~\ref{Fig:near_EH_both}).

For $\left\langle\Phi^{2}\right\rangle_{\text{ren}}^{U}$, while we were able to successfully obtain the EH limit,
 this was not possible at the IH. Obtaining the IH limit of $\left\langle \Phi^{2}\right\rangle _{\text{ren}}^{U}$ poses difficulties, likely attributed to a non-trivial, non-monotonic behavior near the IH, as observed in the analogous RN case of Ref.~\cite{GroupPhiRN:2019}.\footnote{Remarkably, using state subtraction, and taking advantage of the presumed regularity of the reference quantum state of Ref.~\cite{KerrIH:2022}, we indeed found that $\left\langle \Phi^{2}\right\rangle _{\text{ren}}^{U}$ reaches a \emph{finite} value (which is yet unknown) at the IH, approaching it as $r_{*}^{-3}$ (as in the analogous RN case; this analysis also reproduced the non-trivial and non-monotonic behavior of $\left\langle \Phi^{2}\right\rangle _{\text{ren}}^{U}$ on approaching the IH, similarly to that observed in the aforementioned RN case.)}

Still, as our method involves the unique step of ID subtraction, it would be worth testing it against another independent method at general $r$ values inside the BH. 
In Appendix~\ref{App:The-analytic-extension}, we present an alternative variant of the $t$-splitting PMR method, dubbed the \emph{analytic extension} method. This method leverages an analytic extension of the HTPF mode contributions from the exterior to the interior of the BH, building on the analyticity of the background geometry at the EH. While this approach circumvents the ID problem, it introduces challenges of dealing with growing oscillatory behavior in $l$ and $\omega$, which can be managed by our ``oscillation cancellation" procedure. Our results at two specific $r$ values reveal a remarkable agreement between the two variants, which, in particular, provides a crucial test for the $t$-splitting PMR method described in this manuscript.

As previously discussed, our calculations necessitated computing contributions for $l$ up to approximately $l_{\text{max}} = 300$. However, since our analysis was confined to the pole, we only needed to compute these contributions for $m = 0$ modes. To extend our study off the pole, one must compute all $m$ modes across the entire domain $-l\leq m\leq l$. Consequently, this will significantly increase the number of modes that need to be computed –- and consequently, the required computational resources –- typically by a factor of $l_{\text{max}} = 300$. Such an increase seems impractical, at least with our current methods.\footnote{Note, however, that for a computation directly on the IH we can use the state-subtraction method, which is much more efficient numerically. It requires a smaller $l$ range (say, up to a few dozen), and hence is practical to apply off the pole as well. Indeed, we have performed this computation for an $a/M=0.8$ BH in the entire range $0\leq\theta\leq\pi/2$, see Ref.~\cite{KerrIH:2022}.}

In Table~\ref{Tbl:numValues} we summarize the values of the various quantities at the two horizons at the pole of a Kerr BH which we have given in the main text.

\begin{table}
\begin{tabular}{|c|c|c|}
\hline 
 & At the IH ($r=r_{-}$), for $a/M=\{0.8,0.9\}$ & At the EH ($r=r_{+}$), for $a/M=\{0.8,0.9\}$ \tabularnewline
\hline 
\hline 
$\frac{M^{4}}{\hbar}\left\langle T_{uu}\right\rangle _{\text{ren}}^{U}$ & $\{3.23163918\times10^{-5},\,-1.702041202\times10^{-6}\}$ & $\{-7.29\times10^{-6}\,\delta r_{+}^{2},\,-8.25\times10^{-6}\,\delta r_{+}^{2}\}$\tabularnewline
\hline 
$\frac{M^{4}}{\hbar}\left\langle T_{vv}\right\rangle _{\text{ren}}^{U}$ & $\{3.01345442\times10^{-5},\,-2.323817852\times10^{-6}\}$ & $\{-5.454619\times10^{-7},\,-2.442739\times10^{-7}\}$\tabularnewline
\hline 
$\frac{M^{2}}{\hbar}\mathcal{F}_{0}$  &  \multicolumn{2}{c|}{ $\{-1.7454781\times10^{-6},\,-7.0150098\times10^{-7}\}$}\tabularnewline
\hline 
$\frac{M^{2}}{\hbar}\left\langle \Phi^{2}\right\rangle _{\text{ren}}^{U}$ & Still unknown & $\{2.21\times10^{-5},\,-1.64\times10^{-4}\}$\tabularnewline
\hline 
\end{tabular}\label{Tbl:numValues}

\caption{Summary of the numerical values of various
quantities at the pole of a Kerr BH at the two horizons with $a/M={0.8,0.9}$.
The behavior $c\cdot\delta r_{+}^{2}$
of $M^4\hbar^{-1}\left\langle T_{uu}\right\rangle _{\text{ren}}^{U}$
in the 1st row of the 2nd column is, of course, the leading \textit{asymptotic}
behavior as $r=r_{+}$ is approached. The values in the first column
of $\left\langle T_{vv}\right\rangle _{\text{ren}}^{U}$ and $\left\langle T_{uu}\right\rangle _{\text{ren}}^{U}$ at (exactly) $r=r_{-}$ were obtained in
\cite{KerrIH:2022} via the state-subtraction method, agreeing with what we find by extrapolation of $t$-splitting results. We give the values
of $M^2\hbar^{-1}\mathcal{F}_{0}$ [where $\mathcal{F}_{0}$ is the Hawking outflux per unit solid angle in the polar direction, given in Eq.~\eqref{eq:F(0)}]
on both columns, since they are actually constant for all values of the radius. The values
of $\left\langle T_{vv}\right\rangle _{\text{ren}}^{U}$
and $\left\langle \Phi^{2}\right\rangle _{\text{ren}}^{U}$ at $r=r_{+}$
are extrapolations from the results for $r<r_{+}$ (the value of $\left\langle \Phi^{2}\right\rangle _{\text{ren}}^{U}$
cannot be reliably extrapolated to $r=r_{-}$, as discussed).}
\end{table}

There exist several compelling directions for further research. An immediate extension would be to study a wider range of values for the spin parameter $a/M$ beyond the presently treated values of $a/M=0.8$ and $a/M=0.9$. It would also be valuable to extend the method for off-pole computations, although, as mentioned, currently this seems to be numerically challenging (without the adoption of more efficient methods such as state subtraction, which is presently only applicable for computations at the IH). Furthermore, a fuller picture of semiclassical effects inside a Kerr BH will be gained by extending our analysis to other components of the RSET.

In addition, it is highly valuable to extend our analysis to other quantum fields. An exploration of other scalar fields may include allowing a non-vanishing coupling parameter $\xi$. Of higher physical relevance is the study of the quantum electromagnetic field, which is the actual field observed in nature. It is also worth noting the importance of the linearized-gravitational semiclassical contribution, which should be no less significant than its electromagnetic counterpart (but likely to be technically and conceptually more intricate).

\begin{acknowledgments} 

A.O. and N.Z. were supported by the Israel Science Foundation under
Grant No. 600/18. N.Z. also acknowledges support by the Israeli Planning
and Budgeting Committee.

\end{acknowledgments}


\appendix

\section{Large-$l$ asymptotics\label{App:Large-l-asymptotics}}

In this Appendix we briefly describe the large-$l$ analysis of the
interior radial function $\psi_{\omega l}^{\text{int}}$ [see Eq.~\eqref{eq:psi_int_BC}] and the reflection coefficient
$\rho^{\text{up}}_{\omega l}$ [see Eq.~\eqref{eq:psiUP_asym}]  in a Kerr BH, focusing on the $m=0$ case, given that only $m=0$ modes are relevant at the pole. The analysis we describe yields the leading
order expressions which are subsequently used in Sec.~\ref{subsec:IBS_phi2}
for the ID computation of $\left\langle \Phi^{2}\right\rangle $.

Generally speaking (for a general $m$ value), the problem of modes
traveling on a Kerr spacetime is very different from the analogous
RN problem. In particular, the fact that the effective potential in
Kerr approaches three different asymptotic values: $\omega_{-}^{2}$
at the IH, $\omega_{+}^{2}$ at the EH, and $\omega^{2}$ at infinity
{[}see Eq.~\eqref{eq:asymPot}{]} -- whereas in RN all three coincide
-- is the root to a cascade of differences between the two cases.
However, the particular instance of $m=0$ modes in Kerr, in which
all three asymptotic frequencies coincide ($\omega_{+}=\omega_{-}=\omega$),
turns out to be  analogous to the RN case (for most relevant problems).
There are still some tiny differences between the polar Kerr and RN
cases, e.g. the form of the surface gravity parameters \eqref{eq:kappa_pm}
or the effective potential \eqref{eq:KerrPot}, involving $\omega$
in a non-trivial way through the angular eigenvalue.

Our analysis for the polar Kerr case thus follows closely the one
presented in Ref.~\cite{Sela:2018} (mainly in sections III and V
therein) for the RN case, with slight modifications to be pointed
out.

\subsection{The interior radial function in the WKB approximation}

Our goal here is to apply the WKB approximation to the $m=0$ interior
radial function $\psi_{\omega l}^{\text{int}}$ at large $l$. To this end, we follow
closely the analysis in Sec.~V in Ref.~\cite{Sela:2018}. The validity
condition of the WKB approximation should be
\[
\sqrt{V_{\omega l}}M\gg1\,.
\]
To pinpoint the difference between the RN and polar Kerr analyses,
we start with the large-$l$ form of the $m=0$ effective potential
$V_{\omega l}$ given in Eq.~\eqref{eq:pot_large_l}, and compare
it to its RN counterpart given in Eq.~(5.9) in Ref.~\cite{Sela:2018}
(note that, due to the way they are defined in the corresponding radial
equation, the RN potential $V_{l}$ is analogous to \emph{minus} the
Kerr potential $V_{\omega lm}$). Clearly, the large-$l$ $V_{\omega l}$
is analogous to its RN counterpart, in which the denominator $\left(r^{2}+a^{2}\right)^{2}$
is replaced by $r^{4}$ and $\Delta$ is written out as in Eq.~\eqref{eq:Delta}.
Furthermore, it suffices in this leading-order analysis to replace $l\left(l+1\right)$
by $\tilde{l}^{2}$, where $\tilde{l}=l+1/2$.

Then, the large-$l$ WKB approximation is valid as long as
\begin{equation}\label{eq:WKB lim}
\frac{\tilde{l}M}{r^{2}+a^{2}}\sqrt{\left(r_{+}-r\right)\left(r-r_{-}\right)}\gg1\,.
\end{equation}

The general WKB form of the radial function is
\[
\psi_{\omega l}^{\text{WKB}}=\frac{1}{V_{\omega l}^{1/4}\left(r\right)}\left[a_{+}\exp\left(i\int^{r_{*}}\sqrt{V_{\omega l}\left(r\left(\tilde{r}_{*}\right)\right)}d\tilde{r}_{*}\right)+a_{-}\exp\left(-i\int^{r_{*}}\sqrt{V_{\omega l}\left(r\left(\tilde{r}_{*}\right)\right)}d\tilde{r}_{*}\right)\right]
\]
(no lower limits of integration were applied on the phase functions
$\pm\int^{r_{*}}\sqrt{V_{\omega l}\left(r\left(\tilde{r}_{*}\right)\right)}d\tilde{r}_{*}$,
since any constants of integration can be absorbed into the prefactors
$a_{\pm}$). In computing the phase functions, note that since $dr_{*}/dr$
cancels the $r^{2}+a^{2}$ factor arising from $V_{\omega l}$, we
obtain precisely as in RN: 

\[
\int^{r_{*}}\sqrt{V_{\omega l}\left(r\left(\tilde{r}_{*}\right)\right)}d\tilde{r}_{*}\simeq-\tilde{l}\arctan\left(\frac{r-M}{\sqrt{\left(r_{+}-r\right)\left(r-r_{-}\right)}}\right)+\text{const}\,.
\]
Hence
\begin{equation}
\psi_{\omega l}^{\text{WKB}}\simeq\frac{\sqrt{r^{2}+a^{2}}}{\sqrt{\tilde{l}}\left[\left(r_{+}-r\right)\left(r-r_{-}\right)\right]^{1/4}}\left[a_{+}e^{-i\tilde{l}g\left(r\right)}+a_{-}e^{i\tilde{l}g\left(r\right)}\right]\label{eq:psi_pm}
\end{equation}
where we denote 
\begin{equation}
g(r)\equiv\arctan\left(\frac{r-M}{\sqrt{\left(r_{+}-r\right)\left(r-r_{-}\right)}}\right)\label{eq:g(r)}
\end{equation}
and the constants of integration are absorbed into the prefactors
$a_{\pm}$, to be determined. The prefactors $a_{\pm}$ are then found
by matching with the near-EH solution, which is precisely as its RN
counterpart given in Eq.~(5.3) in Ref.~\cite{Sela:2018} (as turns
out, when expressed in terms of $\kappa_{+}$). Finally, we obtain
the 
large-$l$ WKB approximated form of $\psi_{\omega l}^{\text{int }}$:
\begin{equation}
\psi_{\omega l}^{\text{WKB}}\simeq\frac{\sqrt{r^{2}+a^{2}}}{\sqrt{\tilde{l}}\left[\left(r_{+}-r\right)\left(r-r_{-}\right)\right]^{1/4}}\sqrt{\frac{\kappa_{+}}{2\pi}}e^{i\frac{\pi}{4}}\tilde{l}{}^{\frac{i\omega}{\kappa_{+}}}e^{-i\omega r_{+}}\Gamma\left(1-\frac{i\omega}{\kappa_{+}}\right)\left(i^{\tilde{l}-1}e^{-\frac{\pi\omega}{2\kappa_{+}}}e^{-i\tilde{l}g(r)}+\left(-i\right)^{\tilde{l}}e^{\frac{\pi\omega}{2\kappa_{+}}}e^{i\tilde{l}g(r)}\right)\,,\label{eq:psiWKB}
\end{equation}
which is identical to the RN counterpart given in Eqs.~(5.13)-(5.15) in Ref.~\cite{Sela:2018},
up to a prefactor being $\sqrt{r^{2}+a^{2}}$ instead of $r$.

\subsection{$\rho_{\omega l}^{\text{up}}$ at large $l$}

Following the analysis presented in Sec.~III in Ref.~\cite{Sela:2018},
keeping track of all potential factors that may cause a difference
between RN and polar Kerr (e.g. certain powers of $r^{2}+a^{2}$
replacing $r^{2}$), the reflection coefficient
$\rho_{\omega l}^{\text{up}}$ at large $l$ \footnote{The {\it leading} order in the large-$l$ limit obviously coincides with that in the large-$\tilde{l}$ limit (recall $\tilde{l}\equiv l+1/2$). More specifically, the leading order in the expansion of $\rho_{\omega l}^{\text{up}}$  for $l\to \infty$ (which is our concern here) precisely matches the corresponding leading order in its expansion for $\tilde{l}\to \infty$. In the expressions below – and in fact throughout the paper – we freely use either $l$ or $\tilde{l}$, as convenient in each place.}
is found to have precisely
the same form as its RN counterpart {[}given in Eq.~(3.17) therein{]}:
\begin{equation}
\rho_{\omega l}^{\text{up}}\simeq\tilde{l}^{-2i\omega/\kappa_{+}}e^{2i\omega r_{+}}\frac{\Gamma\left(i\omega/\kappa_{+}\right)}{\Gamma\left(-i\omega/\kappa_{+}\right)}\,.\label{eq:rho_largeL}
\end{equation}
We note that in Eq.~(3.5.13) in Ref.~\cite{th:CasalsPhD}\footnote{\label{ftn:large-l}We note typographical errors in Eq.~(3.5.12)~\cite{th:CasalsPhD}, which should read:
\begin{equation*} 
I_{\tilde{\omega}}\equiv e^{-i\tilde{\omega}r_+}\left[\left(4Mr_+-2Q^2\right)\kappa_+\right]^{-\frac{i\tilde{\omega}}{2\kappa_+}}\left[-\left(4Mr_--2Q^2\right)\kappa_-\right]^{-\frac{i\tilde{\omega}}{2\kappa_-}}.
\end{equation*}
This expression follows the notation in Ref.~\cite{th:CasalsPhD} and so, in particular the $\kappa_-$ in this expression is equal to {\it minus} the $\kappa_-$ in Eq.~\eqref{eq:kappa_pm} of the current paper.}, an expression for the large-$l$ asymptotics of the radial reflection coefficient is given in the more general case of a mode for generic azimuthal number $m$ of a field with arbitrary spin on  Kerr-Newman space-time. 
We have checked that our Eq.~\eqref{eq:rho_largeL} indeed agrees with Eq.~(3.5.12) in Ref.~\cite{th:CasalsPhD} (when taking into account footnote \ref{ftn:large-l} as well as the fact that some quantities -- such as the tortoise coordinate -- are defined differently here and in Ref.~\cite{th:CasalsPhD}), for the particular case of $m=0$, zero field spin (i.e., scalar field) and zero black hole charge (i.e., Kerr spacetime).
See also Appendix A in Ref.~\cite{Candelas:1980}, where a similar large-$l$ asymptotic analysis was done in the Schwarzschild case. We have checked that our asymptotic results reduce to their Schwarzschild counterparts given in Ref.~\cite{Candelas:1980}.

\section{The intermediate divergence problem\label{App:The-intermediate-divergence}}

As described in Sec.~\ref{sec:The-t-splitting-procedure}, a crucial
step of our method is the subtraction of a divergent piece ${E}_{\omega l}^{\text{div}}$ (the ID)
of the form given in Eq.~\eqref{eq:Ewl_div}, prior to summation
and integration. For $\left\langle \Phi^{2}\right\rangle ^{U}$ we
demonstrated that $c_{1}=c_{2}=0$ {[}leaving only an $\omega$- and $l$-independent
ID, for which we found an analytic expression in Eq.~\eqref{eq:Ewl_div_phi2}{]}.
For the fluxes $\left\langle T_{yy}\right\rangle ^{U}$ (for which the ID was specifically denoted by  $T_{yy\left(\omega l\right)}^{\text{div}}$)
we have all three prefactors $c_{0},c_{1}$ and $c_{2}$ generally
non-vanishing {[}with $c_{2}$  analytically computed in Eq.~\eqref{eq:c2(r)}, and also with an empirical expression for $c_1$given in Eq.~\eqref{eq:c1}{]}.

In this Appendix, we provide a justification for the ID
subtraction. We shall use the same notation as in the beginning of Sec.~\ref{sec:The-t-splitting-procedure}, treating both the vacuum polarization and the fluxes as the generic quantity $P$ (and similarly for all related quantities, such as $E_{\omega l}$ to denote the bare mode contribution, $E_{\omega l}^\text{div}$ to denote the ID, etc.).

To overcome the ID problem, we shall introduce an additional split in the $\theta$
direction, denoted by $\delta\equiv\theta'-\theta$. This
 split is ``small'' in the sense that we shall always
take the limit $\delta\to0$ before taking the limit $\varepsilon\to0$
(see below). 

The analytical treatment of the ID becomes simpler if we express the
large-$l$ singular behavior in terms of the angular eigenvalue $\lambda_{\omega l}\equiv\lambda_{\omega l\left(m=0\right)}$
{[}instead of $l(l+1)${]}.\footnote{Anywhere else in the manuscript, we use the notation $\lambda_{l}\left(a\omega\right)$
{[}or $\lambda_{lm}\left(a\omega\right)$ for a general $m$, as in
Eq.~\eqref{eq:KerrPotFun}{]}, since the spheroidal function's dependence
on $\omega$ is always through the combination $a\omega$. Here, however, we
prefer to separate the dependence on $\omega$,$l$ from the dependence
on the BH parameters $a$,$M$. Hence we use a different notation,
$\lambda_{\omega l}$, highlighting only the $\omega,l$ indices,
keeping the dependence on the parameters $a$,$M$ implicit (such
as for the other quantities appearing in this Appendix).} In order to relate the two large-$l$ expressions, we recall that $\lambda_{\omega l}$  takes the following large $l$ asymptotic form \cite{RokhlinXiao:2007} (stating again Eq.~\eqref{eq:lambda-1} for convenience)
\begin{equation}
\lambda_{\omega l}=l\left(l+1\right)+\frac{1}{2}a^{2}\omega^{2}+\hat{\lambda}_{\omega l},\label{eq:lambda}
\end{equation}
with the remainder term $\hat{\lambda}_{\omega l}$ vanishing like $1/l(l+1)$  as $l\to\infty$.
Thus, we may equally well rewrite the large-$l$ asymptotic behavior
of $E_{\omega l}$ as 
\begin{equation}
c_{0}+\tilde{c}_{1}\omega^{2}+c_{2}\lambda_{\omega l}\equiv\tilde{E}_{\omega l}^{\text{div}}\label{eq:Ediv_tilde}
\end{equation}
(plus a term whose infinite sum over $l$ converges), where $\tilde{c}_{1}\equiv c_{1}-a^{2}c_{2}/2$.

As already mentioned, the analytical treatment of the intermediate
divergence becomes simpler if expressed in terms of $\tilde{E}_{\omega l}^{\text{div}}$.
However, the actual numerical mode-sum procedure becomes more convenient
when done in terms of $E_{\omega l}^{\text{div}}$ {[}namely, using
$l(l+1)$ instead of $\lambda_{\omega l}$, as in Eq.~\eqref{eq:Ewl_div}{]}.
For this reason, in Sec.~\ref{sec:Basic-IBS-analysis} (presenting
the basic analytical ID treatment) we use the large-$l$ asymptotic
form \eqref{eq:Ediv_tilde}. Then, in Sec.~\ref{sec:Replacing-lambda-by-l(l+1)}
we describe the translation of the ID analysis from $\tilde{E}_{\omega l}^{\text{div}}$
to $E_{\omega l}^{\text{div}}$.

\subsection{Basic ID analysis\label{sec:Basic-IBS-analysis}}

The mode contribution $E_{\omega l}$ may be written (at coincidence)
with the radial and angular parts factored out:
\begin{equation}
E_{\omega l}=\left[S_{\omega l}\left(0\right)\right]^{2}H_{\omega l}\left(r\right)\,,\label{eq:Ewl}
\end{equation}
 for some radial function $H_{\omega l}\left(r\right)$.

As mentioned, inside the BH we introduce a small split $\delta$ in the polar angle
$\theta$ (in addition to the ``primary'' split $\varepsilon$ in
$t$). Since we focus on the pole, we have $\theta=0$ and $\theta'=\delta$.
 The renormalized quantity $P_{\text{ren}}$ is then given by:
\begin{equation}
P_{\text{ren}}=\lim_{\varepsilon\to0}\lim_{\delta\to0}\left[\int_{0}^{\infty}\text{d}\omega\,\cos\left(\omega\varepsilon\right)\left(\sum_{l=0}^{\infty}H_{\omega l}S_{\omega l}\left(0\right)S_{\omega l}\left(\delta\right)\right)-C\left(\varepsilon\right)\right]\label{eq:P(H)}
\end{equation}
where, recall, $C\left(\varepsilon\right)$ is the counterterm (with the dependence on $x$ henceforth suppressed). Using
$E_{\omega l}$, this may be written
as 
\begin{equation}
P_{\text{ren}}=\lim_{\varepsilon\to0}\lim_{\delta\to0}\left[\int_{0}^{\infty}\text{d}\omega\,\cos\left(\omega\varepsilon\right)\left(\sum_{l=0}^{\infty}E_{\omega l}\frac{S_{\omega l}\left(\delta\right)}{S_{\omega l}\left(0\right)}\right)-C\left(\varepsilon\right)\right]\,.\label{eq:P(E)}
\end{equation}

The large-$l$ divergent piece of $E_{\omega l}$ is $\tilde{E}_{\omega l}^{\text{div}}$
given in Eq.~\eqref{eq:Ediv_tilde}. The residual, denoted $\hat{E}_{\omega l}\equiv E_{\omega l}-\tilde{E}_{\omega l}^{\text{div}}$,
has a convergent sum over $l$. We next substitute 
\begin{equation}
E_{\omega l}=\hat{E}_{\omega l}+\tilde{E}_{\omega l}^{\text{div}}\label{eq:E_split}
\end{equation}
in Eq.~\eqref{eq:P(E)}. Accordingly, we define

\begin{equation}
P_{\text{ren}}=\hat{P}+P_{\text{div}}\label{eq:  P_split}
\end{equation}
where
\begin{equation}
\hat{P}=\lim_{\varepsilon\to0}\lim_{\delta\to0}\left[\int_{0}^{\infty}\cos\left(\omega\varepsilon\right)\left(\sum_{l=0}^{\infty}\hat{E}_{\omega l}\frac{S_{\omega l}\left(\delta\right)}{S_{\omega l}\left(0\right)}\right)\text{d}\omega-C\left(\varepsilon\right)\right]\label{eq:Phat}
\end{equation}
and 
\begin{equation}
P_{\text{div}}=\lim_{\varepsilon\to0}\lim_{\delta\to0}\int_{0}^{\infty}\cos\left(\omega\varepsilon\right)\left(\sum_{l=0}^{\infty}\tilde{E}_{\omega l}^{\text{div}}\frac{S_{\omega l}\left(\delta\right)}{S_{\omega l}\left(0\right)}\right)\text{d}\omega\,.\label{eq:Pdiv}
\end{equation}

Our goal in this section is to show $P_{\text{div}}=0$, hence establishing $P_{\text{ren}}=\hat{P}$.

For $\hat{P}$, since  the series $\sum_l\hat{E}_{\omega l}$ is regular,
we assume that we may readily take the $\delta\to0$ limit (i.e.,
"closing" the small split in $\theta$) in $\hat{E}_{\omega l}\frac{S_{\omega l}\left(\delta\right)}{S_{\omega l}\left(0\right)}$
already prior to summation. Then,
\begin{equation}
\hat{P}=\lim_{\varepsilon\to0}\left[\int_{0}^{\infty}\cos\left(\omega\varepsilon\right)\left(\sum_{l=0}^{\infty}\hat{E}_{\omega l}\right)\text{d}\omega-C\left(\varepsilon\right)\right]\,\label{eq:Phat1}
\end{equation}
{[}which resembles Eq.~\eqref{eq:Pren_out} for the BH exterior{]}. 

Plugging the form \eqref{eq:Ediv_tilde} of $\tilde{E}_{\omega l}^{\text{div}}$
into $P_{\text{div}}$, we have:
\begin{equation}
P_{\text{div}}=c_{0}P_{\text{div}}^{\left(0\right)}+\tilde{c}_{1}P_{\text{div}}^{\left(1\right)}+c_{2}P_{\text{div}}^{\left(2\right)}\label{eq:Pdiv1}
\end{equation}
where
\begin{equation}
P_{\text{div}}^{\left(0\right)}=\lim_{\varepsilon\to0}\lim_{\delta\to0}\int_{0}^{\infty}\cos\left(\omega\varepsilon\right)\sum_{l=0}^{\infty}\frac{S_{\omega l}\left(\delta\right)}{S_{\omega l}\left(0\right)}\text{d}\omega\,,\label{eq:P0}
\end{equation}
\begin{equation}
P_{\text{div}}^{\left(1\right)}=\lim_{\varepsilon\to0}\lim_{\delta\to0}\int_{0}^{\infty}\omega^{2}\cos\left(\omega\varepsilon\right)\sum_{l=0}^{\infty}\frac{S_{\omega l}\left(\delta\right)}{S_{\omega l}\left(0\right)}\text{d}\omega\,\label{eq:P1}
\end{equation}
and
\begin{equation}
P_{\text{div}}^{\left(2\right)}=\lim_{\varepsilon\to0}\lim_{\delta\to0}\int_{0}^{\infty}\cos\left(\omega\varepsilon\right)\sum_{l=0}^{\infty}\left(\lambda_{\omega l}\frac{S_{\omega l}\left(\delta\right)}{S_{\omega l}\left(0\right)}\right)\text{d}\omega\,.\label{eq:P2}
\end{equation}
We shall now analyze each of these three terms $P_{\text{div}}^{\left(k\right)}$, $k=0,1,2$, separately.

We begin with $P_{\text{div}}^{\left(0\right)}$. Denoting 
\begin{equation}
\Sigma_{0}\left(\delta,\omega\right)\equiv\sum_{l=0}^{\infty}\frac{S_{\omega l}\left(\delta\right)}{S_{\omega l}\left(0\right)}\,,\label{eq:S0def}
\end{equation}
we have
\begin{equation}
P_{\text{div}}^{\left(0\right)}=\lim_{\varepsilon\to0}\lim_{\delta\to0}\int_{0}^{\infty}\cos\left(\omega\varepsilon\right)\Sigma_{0}\left(\delta,\omega\right)\text{d}\omega\,.\label{eq:P0(S0)}
\end{equation}
We numerically obtained
the following simple equality: 
\begin{equation}
\Sigma_{0}\left(\delta,\omega\right)=\frac{1}{2\sin\left(\delta/2\right)}J_{0}\left(\omega\sin\delta\right)\,.\label{eq:S0res}
\end{equation}
(We numerically verified this relation for various $\delta,\omega$ pairs of
values, with more than 20 decimals, using the software {\rm Mathematica} \cite{Mathematica13.2}.) In particular, the dependence
on $\omega$ is only through the combination $\omega\sin\delta$
in the argument of the Bessel function. We now show that the right
hand side of Eq.~\eqref{eq:P0(S0)} vanishes.

The following equality is known to hold at $\sin\delta\neq\varepsilon$
\footnote{Note that in Eq.~\eqref{eq:int(J0)}, and similarly in Eqs.~\eqref{eq:int(w^2*J0)} and \eqref{eq:int(w*J1)}
below, the resulting integral may contain an additional distribution with sharp support
at $\sin\delta=\varepsilon$. However, this is irrelevant to our analysis,
being outside of the $\left(\delta,\varepsilon\right)$ domain that concerns
us.} {[}see Eq.~(7) in Sec.~13.42 in Ref.~\cite{WatsonBessel}{]}\footnote{We shall assume that both $\varepsilon$ and $\delta$ (and $\sin\delta$
likewise) are positive. This assumption is unnecessary, but if relaxed,
we should replace the step function $\Theta\left(\sin\delta-\varepsilon\right)$
by $\Theta\left(\sin^{2}\delta-\varepsilon^{2}\right)$.} :
\begin{equation}
\int_{0}^{\infty}\cos\left(\omega\varepsilon\right)J_{0}\left(\omega\sin\delta\right)\text{d}\omega=\frac{1}{\sqrt{\sin^{2}\delta-\varepsilon^{2}}}\Theta\left(\sin\delta-\varepsilon\right)\,.\label{eq:int(J0)}
\end{equation}
Then, in the relevant domain of $\delta$ and $\varepsilon$, being
$\sin\delta<\varepsilon$ (because of the order of the limits  that
we take), we find
\begin{equation}
\int_{0}^{\infty}\cos\left(\omega\varepsilon\right)\Sigma_{0}\left(\delta,\omega\right)\text{d}\omega=0\,.\label{eq:int(S0)}
\end{equation}

This (partly numerically, partly analytically) establishes that
\begin{equation}
P_{\text{div}}^{\left(0\right)}=0\,,\label{eq:P0}
\end{equation}
hence its subtraction is justified.  This settles the $\left\langle \Phi^{2}\right\rangle $
case (for which $\tilde{c}_{1}=c_{2}=0$). However, for the fluxes 
we still need to address $P_{\text{div}}^{\left(1\right)}$ and $P_{\text{div}}^{\left(2\right)}$.

We next consider $P_{\text{div}}^{\left(1\right)}$, which we write 
by plugging Eq.~\eqref{eq:S0def} into Eq.~\eqref{eq:P1} and using
Eq.~\eqref{eq:S0res} as
\begin{align}
P_{\text{div}}^{\left(1\right)} & =\lim_{\varepsilon\to0}\lim_{\delta\to0}\frac{1}{2\sin\left(\delta/2\right)}\left[\int_{0}^{\infty}\omega^{2}\cos\left(\omega\varepsilon\right)J_{0}\left(\omega\sin\delta\right)\text{d}\omega\right]\,.\label{eq:P1-1}
\end{align}

Now, 
using a formal manipulation (interchanging the derivative with the
integration), we may write:
\begin{equation}
\int_{0}^{\infty}\omega^{2}\cos\left(\omega\varepsilon\right)J_{0}\left(\omega\sin\delta\right)\text{d}\omega=-\frac{\partial^{2}}{\partial\varepsilon^{2}}\int_{0}^{\infty}\cos\left(\omega\varepsilon\right)J_{0}\left(\omega\sin\delta\right)\text{d}\omega.\label{eq:deriv_relation}
\end{equation}
Thus,
by taking minus the second derivative with respect to $\varepsilon$ of 
Eq.~\eqref{eq:int(J0)}, we obtain the following distributional identity valid for  $\sin \delta\neq \varepsilon$ \footnote{\label{fn: integrals}In this footnote we present an alternative derivation of Eq.~\eqref{eq:int(w^2*J0)}. Strictly speaking, this integral is not well
defined because the $\omega^{2}$ factor {[}or the $\omega$ factor
in the analogous situation of the integral appearing in Eq.~\eqref{eq:int(w*J1)}{]}
spoils convergence at large $\omega$. Nevertheless, the integral
(like some other integrals involved in our $t$-splitting method)
becomes well-defined with the procedure of ``oscillation cancellation"
(see Ref. \cite{AAt:2015}), and it is this ``generalized integral'' that we
want to compute here. An alternative (and equivalent) formulation
of this generalized integral is via multiplying the integrand by a
regulating function $\exp(-c\, \omega)$ (with $c>0$) and then taking
the limit $c\to0$ after integration. This alternative definition,
involving the $\exp(-c\, \omega)$ factor, is more commonly used in
the mathematical physics literature; however, it is much harder to
numerically implement it (compared to oscillation cancellation). Nevertheless,
for the \emph{analytical} derivation of this (generalized) integral
-- which is our concern here -- this second method {[}i.e. using
the regulating function $\exp(-c\, \omega)${]} is very convenient.
Thus, the relevant integrand now becomes $\omega^{2}\cos\left(\omega\varepsilon\right)J_{0}\left(\omega\sin\delta\right)\exp(-c\, \omega)$.
Unfortunately, {\rm Mathematica} is unable to directly perform this integral.
However,  when reduced to the evaluation of $\Re\lim_{c\to0}\int_{0}^{\infty}\omega^{2}\exp\left(i\omega\varepsilon\right)J_{0}\left(\omega\sin\delta\right)\exp(-c\, \omega)$,
{\rm Mathematica} successfully yields the desired result given in Eq.~\eqref{eq:int(w^2*J0)}.}:

\begin{equation}
\int_{0}^{\infty}\omega^{2}\cos\left(\omega\varepsilon\right)J_{0}\left(\omega\sin\delta\right)\text{d}\omega=-\frac{2\varepsilon^{2}+\sin^{2}\delta}{\left(\sin^{2}\delta-\varepsilon^{2}\right)^{5/2}}\Theta\left(\sin\delta-\varepsilon\right)\,.\label{eq:int(w^2*J0)}
\end{equation}

Then,  for $\sin\delta<\varepsilon$ (being again the relevant domain
of $\delta$ and $\varepsilon$), we have
\begin{equation}
\int_{0}^{\infty}\omega^{2}\cos\left(\omega\varepsilon\right)\Sigma_{0}\left(\delta,\omega\right)\text{d}\omega=0\,,\label{eq:int(w^2*S0)}
\end{equation}
hence also
\begin{equation}
P_{\text{div}}^{\left(1\right)}=0\,.\label{eq:P1}
\end{equation}

Finally, we deal with $P_{\text{div}}^{\left(2\right)}$. We denote
\begin{equation}
\Sigma_{2}\left(\delta,\omega\right)\equiv\sum_{l=0}^{\infty}\left(\lambda_{\omega l}\frac{S_{\omega l}\left(\delta\right)}{S_{\omega l}\left(0\right)}\right)\,,\label{eq:S2def}
\end{equation}
so that
\begin{equation}
P_{\text{div}}^{\left(2\right)}=\lim_{\varepsilon\to0}\lim_{\delta\to0}\left[\int_{0}^{\infty}\cos\left(\omega\varepsilon\right)\Sigma_{2}\left(\delta,\omega\right)\text{d}\omega\right]\,.\label{eq:P2-1}
\end{equation}
In order to evaluate $\Sigma_{2}\left(\delta,\omega\right)$, we recall
the angular equation satisfied by $S_{\omega l}$ {[}Eq.~\eqref{eq:angKerr}
for $m=0${]}:
\begin{equation}
\frac{1}{\sin\theta}\frac{d}{d\theta}\left(\sin\theta\frac{d}{d\theta}S_{\omega l}(\theta)\right)+\left(-a^{2}\omega^{2}\sin^{2}\theta+\lambda_{\omega l}\right)S_{\omega l}(\theta)=0\,.\label{eq:ang_eq}
\end{equation}
We may therefore write (upon replacing the independent variable $\theta$
by $\delta$)
\[
\lambda_{\omega l}S_{\omega l}\left(\delta\right)=D_{\omega}\left[S_{\omega l}\left(\delta\right)\right]\,,
\]
where $D_{\omega}$ is the linear differential operator defined by
\begin{equation}
D_{\omega}\equiv-\frac{1}{\sin\delta}\frac{d}{d\delta}\left(\sin\delta\frac{d}{d\delta}\right)+a^{2}\omega^{2}\sin^{2}\delta\,.\label{eq:  D_operator}
\end{equation}
We may then express $\Sigma_{2}\left(\delta,\omega\right)$ as:
\[
\Sigma_{2}\left(\delta,\omega\right)=\sum_{l=0}^{\infty}\left(\frac{D_{\omega}\left[S_{\omega l}\left(\delta\right)\right]}{S_{\omega l}\left(0\right)}\right)=\sum_{l=0}^{\infty}\left(D_{\omega}\left[\frac{S_{\omega l}\left(\delta\right)}{S_{\omega l}\left(0\right)}\right]\right)\,.
\]
Noting that the linear operator $D_{\omega}$ is independent of $l$,
 we may interchange it with the sum over $l$, which yields
\begin{equation}
\Sigma_{2}\left(\delta,\omega\right)=D_{\omega}\left[\sum_{l=0}^{\infty}\left(\frac{S_{\omega l}\left(\delta\right)}{S_{\omega l}\left(0\right)}\right)\right]=D_{\omega}\left[\Sigma_{0}\left(\delta,\omega\right)\right]=D_{\omega}\left[\frac{1}{2\sin\left(\delta/2\right)}J_{0}\left(\omega\sin\delta\right)\right]\,,\label{eq:  Sigma_2}
\end{equation}
where we have used Eqs.~\eqref{eq:S0def} and \eqref{eq:S0res}. Explicitly,
we obtain 
\begin{equation}
\Sigma_{2}\left(\delta,\omega\right)=\alpha_{0}\left(\delta\right)J_{0}\left(\omega\sin\delta\right)+\alpha_{1}\left(\delta\right)\omega J_{1}\left(\omega\sin\delta\right)+\alpha_{2}\left(\delta\right)\omega^{2}J_{0}\left(\omega\sin\delta\right)\label{eq:S2res}
\end{equation}
with the ($\omega$-independent) coefficients
\begin{align}
\alpha_{0}\left(\delta\right) & =\frac{1}{16}\left(\sin\left(\delta/2\right)\right)^{-3}\left[-3+\cos\delta\right]\,,\nonumber \\
\alpha_{1}\left(\delta\right) & =-\frac{\sin\delta}{4}\left(\sin\left(\delta/2\right)\right)^{-3}\,,\nonumber \\
\alpha_{2}\left(\delta\right) & =\frac{1}{4}\left(\sin\left(\delta/2\right)\right)^{-3}\left[1-\cos\delta\right]\,.\label{eq:coefS2}
\end{align}

We note that the above derivation of $\Sigma_{2}\left(\delta,\omega\right)$
involved a formal manipulation (interchange of the $D_{\omega}$ differential
operator with the sum over $l$), which we are unable to justify rigorously;
nevertheless, we also verified Eq.~\eqref{eq:S2res} numerically (with
more than 40 digits of precision), for several $\delta,\omega$ pairs.
\footnote{The numerical evaluation of $\Sigma_{2}\left(\delta,\omega\right)$
given as an infinite sum in Eq.~\eqref{eq:S2def} involved the procedure
of oscillation cancellation (on the partial sums over $l$). This
procedure is further explained (and implemented as part of the analytic extension method) in Appendix~\ref{subsec:regularization_l_ae}. Here -- and in the analytic extension -- we encounter the need for oscillation cancellation
of a \emph{discrete} sequence, differing from the continuous case discussed
in  Ref. \cite{AAt:2015}.}

Taking the expression for $\Sigma_{2}\left(\delta,\omega\right)$
given in Eq.~\eqref{eq:S2res} and plugging it into the $\omega$
integral in Eq.~\eqref{eq:P2-1}, we note that both the $\propto J_{0}\left(\omega\sin\delta\right)$
and $\propto\omega^{2}J_{0}\left(\omega\sin\delta\right)$ terms have
already been shown to not contribute to the integral in the relevant
domain $\sin\delta<\varepsilon$ {[}see Eqs.~\eqref{eq:int(J0)} and
\eqref{eq:int(w^2*J0)} respectively{]}. We are left with the $\propto\omega J_{1}\left(\omega\sin\delta\right)$
term in $\Sigma_{2}$. 
First, we note that (see, e.g., Eq.~(10.6.6) in Ref.~\cite{NIST:DLMF})
\begin{equation}
\omega J_{1}(\omega \sin\delta)=-\frac{\partial}{\partial(\sin\delta)}J_{0}(\omega \sin\delta).
\end{equation}
Thus, similarly to Eq.~\eqref{eq:deriv_relation}, we may formally write:
\begin{equation}
\int_{0}^{\infty}
\omega \cos\left(\omega\varepsilon\right)J_{1}\left(\omega\sin\delta\right)\text{d}\omega=-\frac{\partial}{\partial\left(\sin\delta\right)}
\int_{0}^{\infty}\cos\left(\omega\varepsilon\right)J_{0}\left(\omega\sin\delta\right)\text{d}\omega\,.
\end{equation}
Finally, by differentiating Eq.~\eqref{eq:int(J0)} by minus $\sin\delta$, we obtain at $\sin\delta\neq\varepsilon$: \footnote{As an alternative derivation, we also obtained Eq.~\eqref{eq:int(w*J1)} with the aid of {\rm Mathematica} and the methods 
described in footnote \ref{fn: integrals}.}
\begin{equation}
\int_{0}^{\infty}\omega\cos\left(\omega\varepsilon\right)J_{1}\left(\omega\sin\delta\right)\text{d}\omega=\frac{\sin\delta}{\left(\sin^{2}\delta-\varepsilon^{2}\right)^{3/2}}\Theta\left(\sin\delta-\varepsilon\right)\,.\,\label{eq:int(w*J1)}
\end{equation}
Again, at $\sin\delta<\varepsilon$ we find that this integral vanishes.

All in all, we find that, for $\sin\delta<\varepsilon$,
\begin{equation}
\int_{0}^{\infty}\cos\left(\omega\varepsilon\right)\Sigma_{2}\left(\delta,\omega\right)\text{d}\omega=0\,,\label{eq:int(S2)}
\end{equation}
implying the vanishing of the integral on the RHS of Eq.~\eqref{eq:P2-1},
and hence: 
\begin{equation}
P_{\text{div}}^{\left(2\right)}=0\,.\label{eq:P2}
\end{equation}

We here mention another, formal
derivation of Eq.~\eqref{eq:int(S2)}: Recalling
the relation $\Sigma_{2}\left(\delta,\omega\right)=D_{\omega}\left[\Sigma_{0}\left(\delta,\omega\right)\right]$
{[}see Eq.~\eqref{eq:  Sigma_2}{]}, we may simply apply the differential
operator $D_{\omega}$ to $\int_{0}^{\infty}\cos\left(\omega\varepsilon\right)\Sigma_{0}\left(\delta,\omega\right)\text{d}\omega$.
This last integral vanishes for $\sin\delta<\varepsilon$ [see Eq.~\eqref{eq:int(S0)}], hence
so does the  integral in Eq.~\eqref{eq:int(S2)}.

To conclude, putting together Eqs.~\eqref{eq:Pdiv1}, \eqref{eq:P0},
\eqref{eq:P1} and \eqref{eq:P2}, we find that 
\begin{equation}
P_{\text{div}}=0\,.\label{eq:Pdiv}
\end{equation}
 Recalling Eq.~\eqref{eq:  P_split}, this in turn implies that 
\begin{equation}
P_{\text{ren}}=\hat{P}\,.\label{eq:  final}
\end{equation}
Stated in other words, we have shown -- partly numerically, partly analytically -- that the subtraction of the ID
--- namely the singular piece $\tilde{E}_{\omega l}^{\text{div}}$
as given in Eq.~\eqref{eq:Ediv_tilde} --- is justified.

\subsection{Replacing $\lambda_{\omega l}$ by $l\left(l+1\right)$\label{sec:Replacing-lambda-by-l(l+1)}}

We would like to convert the above results (justifying the ID subtraction)
from $\tilde{E}_{\omega l}^{\text{div}}$ to $E_{\omega l}^{\text{div}}$,
where, recall, 
\begin{equation}
\tilde{E}_{\omega l}^{\text{div}}=c_{0}+\tilde{c}_{1}\omega^{2}+c_{2}\lambda_{\omega l}\,,\;\;E_{\omega l}^{\text{div}}=c_{0}+c_{1}\omega^{2}+c_{2}l(l+1)\label{eq:  IBS}
\end{equation}
(and $\tilde{c}_{1}=c_{1}-a^{2}c_{2}/2$).

The analysis of the previous section established that
\begin{equation}
P_{\text{ren}}=\hat{P}=\lim_{\varepsilon\to0}\left[\int_{0}^{\infty}\cos\left(\omega\varepsilon\right)\left(\sum_{l=0}^{\infty}\hat{E}_{\omega l}\right)\,\text{d}\omega-C\left(\varepsilon\right)\right]\,.\label{eq:Ptilde11}
\end{equation}
We rewrite this result as
\begin{equation}
P_{\text{ren}}=\lim_{\varepsilon\to0}\left[\int_{0}^{\infty}\cos\left(\omega\varepsilon\right)\hat{\Sigma}\,\text{d}\omega\,-C\left(\varepsilon\right)\right]\,\label{eq:P-1}
\end{equation}
where
\[
\hat{\Sigma}\equiv\sum_{l=0}^{\infty}\hat{E}_{\omega l}=\sum_{l=0}^{\infty}\left(E_{\omega l}-\tilde{E}_{\omega l}^{\text{div}}\right)\,,
\]
or more explicitly,
\begin{equation}
\hat{\Sigma}=\sum_{l=0}^{\infty}\left[E_{\omega l}-c_{0}-\tilde{c}_{1}\omega^{2}-c_{2}\lambda_{\omega l}\right]\,.\label{eq:  Sigma_tilde}
\end{equation}

 Using the large-$l$ asymptotic behavior of $\lambda_{\omega l}$
given in Eq.~\eqref{eq:lambda}, we obtain
\[
\hat{\Sigma}=\sum_{l=0}^{\infty}\left[E_{\omega l}-c_{0}-\left(\tilde{c}_{1}+\frac{a^{2}}{2}c_{2}\right)\omega^{2}-c_{2}l\left(l+1\right)-c_{2}\hat{\lambda}_{\omega l}\right]=\sum_{l=0}^{\infty}\left(E_{\omega l}-E_{\omega l}^{\text{div}}-c_{2}\hat{\lambda}_{\omega l}\right).
\]
We split this sum into two (regular) pieces as 
\begin{equation}
\hat{\Sigma}=\sum_{l=0}^{\infty}\left(E_{\omega l}-E_{\omega l}^{\text{div}}\right)-c_{2}\sum_{l=0}^{\infty}\hat{\lambda}_{\omega l}\,.\label{eq:  Sigma_tilde1}
\end{equation}
Note that, unlike the first sum on the RHS, the second sum $\sum_{l=0}^{\infty}\hat{\lambda}_{\omega l}$
 does not depend at all on the dynamics of the problem (namely on
$\psi_{\omega l}$, $\rho^\text{up}_{\omega l}$, etc). We shall now focus on this last sum, writing it explicitly as
\begin{equation}
\sum_{l=0}^{\infty}\hat{\lambda}_{\omega l}=\sum_{l=0}^{\infty}\left[\lambda_{\omega l}-l\left(l+1\right)-\frac{1}{2}a^{2}\omega^{2}\right]\,.\label{eq:sum(lambda_tilde)}
\end{equation}
Numerically exploring this quantity we find, quite surprisingly, that
this sum actually \emph{vanishes} (we are unaware of a source that
demonstrates this analytically, but we have verified it, for a variety of
$a\omega$ values, up to 40 decimals). This implies that 
\begin{equation}
\hat{\Sigma}=\sum_{l=0}^{\infty}\left(E_{\omega l}-E_{\omega l}^{\text{div}}\right)\,.\label{eq:  Sigma_tilde2}
\end{equation}
Substituting this result back into Eq.~\eqref{eq:P-1}, we find that
the desired renormalized quantity $P_{\text{ren}}$ is given by
\begin{equation}
P_{\text{ren}}=\lim_{\varepsilon\to0}\left[\int_{0}^{\infty}\cos\left(\omega\varepsilon\right)\sum_{l=0}^{\infty}\left(E_{\omega l}-E_{\omega l}^{\text{div}}\right)\text{d}\omega-C\left(\varepsilon\right)\right]\,,\label{eq:P_final2}
\end{equation}
as stated in Eq.~\eqref{eq:P_ren} along with Eq.~\eqref{eq:E_basic_def}.

Thus, when we come to sum the series $E_{\omega l}$ over $l$, we
are allowed to subtract its large-$l$ singular piece in either of
the forms $E_{\omega l}^{\text{div}}\equiv c_{0}+c_{1}\omega^{2}+c_{2}l(l+1)$
or $\tilde{E}_{\omega l}^{\text{div}}\equiv c_{0}+\tilde{c}_{1}\omega^{2}+c_{2}\lambda_{\omega l}$
-- both subtraction schemes yield the (same) correct result.

Returning to the scheme considered in this paper (described in Sec.~\ref{sec:The-t-splitting-procedure}), we conclude that we are indeed
allowed to simply drop the ID, i.e. the term $E_{\omega l}^{\text{div}}$
in Eq.~\eqref{eq:Ewl_div}, prior to summation and integration.

\section{Numerical methods: the computational side\label{App:Numerical-implementation}}

This section deals with the numerical implementation of the $t$-splitting
regularization procedure in practice (which we perform using the software {\rm Mathematica} \cite{Mathematica13.2}) --- from the computation of
the basic ingredients $\rho_{\omega l}^{\text{up}}$, $\psi_{\omega l}^{\text{int}}$
and $\psi_{\omega l,r}^{\text{int}}$ (as well as $S_{\omega l}(\theta=0)$)  
to the performance of the various
steps (including various challenges that arise), leading to the renormalized
quantities $\left\langle \Phi^{2}\right\rangle _{\text{ren}}^{U}$
and $\left\langle T_{yy}\right\rangle _{\text{ren}}^{U}$
in the cases under consideration (detailed in Sec.~\ref{sec:Numerical-results}).

It is worth recalling the following fact: while outside the BH the
mode contribution per $l$ and $\omega$ decays exponentially in $l$  for fixed $\omega$ (owing
to the potential barrier, as already mentioned in Sec.~\ref{sec:The-t-splitting-procedure}),
in the BH interior the sum over $l$ does not converge. (Indeed, for
$T_{yy}$ the sequence in $l$ diverges like $l\left(l+1\right)$,
while for $\Phi^{2}$ it approaches a non-vanishing constant -- implying
a divergent $l$-sum in both cases.) We named this large-$l$ divergent
piece the ID {[}i.e. $E_{\omega l}^{\text{div}}$ of the form given
in Eq.~\eqref{eq:Ewl_div}{]}. After subtracting the ID, the resulting
sequence decays at large $l$ like $1/l\left(l+1\right)$, yielding
a convergent $l$-sum.
Still, this slow decay rate poses a challenge to the feasibility of numerical implementation. In order to achieve sufficient accuracy, which is necessary given that many orders of magnitude of precision are still to be lost in the subsequent steps of the regularization procedure, a vast $l$-range would be required. For instance, it would need to be on the order of $\sim10^{4}$, whereas outside the BH, an order of $\sim10^{1}$ or at most $\sim10^{2}$ would suffice.
We overcome this difficulty by performing a fit on the sequence of
partial sums (to be further described in what follows) with $\sim100$
(or even more) orders in $1/l$ (or in $1/\left(l+1\right)$).  This,
in turn, requires taking an $l$ range reaching $l=300$. In addition,
performing such a high-order fit also requires computing the individual
mode contributions $E_{\omega l}$, and in turn their basic
constituents $\rho^\text{up}_{\omega l}$, $\psi_{\omega l}^{\text{int}}$,
$\psi_{\omega l,r}^{\text{int}}$ and $S_{\omega l}(\theta=0)$, to a typical precision of hundreds of figures (at least $\sim250$).

In what follows, we provide a detailed description of our computation of these basic components with the required high precision. We then discuss the implementation of the summation over $l$ and the subsequent numerical integration over $\omega$.

\subsection{Computation of $\rho^\text{up}_{\omega l}$, $\psi_{\omega l}^{\text{int}}$, $\psi_{\omega l,r}^{\text{int}}$ and $S_{\omega l}(\theta=0)$\label{subsec:Computation-of-raw-data}}

In order to calculate the outside reflection coefficient $\rho^\text{up}_{\omega l}$ (defined via \eqref{eq:m=0 quantities} and \eqref{eq:psiUP_asym}) and the interior radial solution $\psi_{\omega l}^{\text{int}}$ (defined via \eqref{eq:m=0 quantities} and \eqref{eq:psi_int_BC}), as well as its derivative $\psi_{\omega l,r}^{\text{int}}$, we used various methods, which served as a check of our results. 
We note that the radial ODE \eqref{eq:radKerr} satisfied by  $\psi_{\omega l}^{\text{int}}$ has two regular singular points (at $r=r_{\pm}$) and one irregular singular point (at $r=\infty$), and so it is a confluent Heun equation.
This greatly facilitates the computation of the solution between the IH and EH, corresponding to the interval between the two regular singular points.
The radial ODE requires the calculation of the angular eigenvalue $\lambda_{l}(a\omega)$, which we did via the in-built ``{\rm SpheroidalEigenvalue}" function in the software {\rm Mathematica}. We next briefly describe the various methods we employed to calculate $\rho^\text{up}_{\omega l}$, $\psi_{\omega l}^{\text{int}}$ and  $\psi_{\omega l,r}^{\text{int}}$.

In the first and main method, we expressed the solution 
$\psi_{\omega lm}^{\text{int}}$
in terms of the 
in-built ``{\rm HeunC}" function in {\rm Mathematica}:
\begin{align}\label{eq:psi-HeunC}
\psi_{\omega lm}^{\text{int}}(r)&=\sqrt{\frac{r^2 + a^2}{r_+^2 + a^2}}e^{-  i  (1 + \kappa) (\varepsilon - m  q)/2}e^{i  \varepsilon  x  \kappa}   (1 - 
    x)^{i  (\varepsilon - \tau)/2}  x ^{- i  (\varepsilon + \tau)/2} \text{HeunC}(q_H, \alpha_H, \gamma_H, \delta_H, \varepsilon_H,
  x),
\end{align}
where $\varepsilon=2M\omega$, $\tau = (\varepsilon  - m  q)/\kappa$, $x=(r_+ - r)/(2 M  \kappa)$, $\kappa=\sqrt{1-q^2}$, $q=a/M$ and 
\begin{align*}
q_H&=\lambda_{lm} +\tau ^2-\varepsilon ^2+i (\tau +\kappa  \varepsilon ),\\
\alpha_H&=2 \kappa  \varepsilon  (\tau -\varepsilon +i),\\
\gamma_H&=-i \tau -i \varepsilon +1,\\
\delta_H&=-i \tau +i \varepsilon +1,\\
\varepsilon_H&=2 i \kappa  \varepsilon.
\end{align*}
For obtaining $\psi_{\omega l}^{\text{int}}$, merely set $m=0$ in Eq.~\eqref{eq:psi-HeunC}.

The second method that we used is the so-called MST method (after the original authors, Mano, Suzuki and Takasugi; see~\cite{Sasaki-Tagoshi:2003} for a review) which was derived for the exterior of the BH and which we extended to the interior. It essentially consists of writing the solutions to the radial ODE \eqref{eq:radKerr} as infinite series of hypergeometric functions and finding the coefficients in the series as solutions to three-term recurrence relations. By matching a series representation which converges everywhere outside the EH except at $r=\infty$ with another one which converges everywhere except at $r=r_+$, a series representation for the outside scattering coefficients, which includes $\rho^\text{up}_{\omega l}$, is obtained. We used such MST series representation to calculate $\rho^\text{up}_{\omega l}$.
We then used our extension inside the EH of the MST series representation for $\psi_{\omega l}^{\text{int}}$ in order to obtain this solution and its $r$-derivative inside the EH.

As a third and last method, we also used simple power series expansions for $\psi_{\omega l}^{\text{int}}$ about $r=r_+$ and about $r=r_-$. In the former case, the boundary condition~\eqref{eq:psi_int_BC} for $\psi_{\omega l}^{\text{int}}$ at $r=r_+$ is readily incorporated. In the latter case, in order to incorporate the boundary for $\psi_{\omega l}^{\text{int}}$ at $r=r_-$, we calculated the scattering coefficients of $\psi_{\omega l}^{\text{int}}$ at $r=r_-$ (namely, $A_{\omega lm}$ and $B_{\omega lm}$ in Eq.~(3.23) of Ref.~\cite{HTPF:2022}) by using the extension inside the EH mentioned above of the MST method.
Clearly, the expansion about $r=r_+$ thus requires much less computational work than the expansion about $r=r_-$ but, on the other hand, it converges a lot more slowly near $r=r_-$. We note that the coefficients in both expansions about $r=r_{\pm}$ satisfy four-term recurrence relations.

Finally, the spheroidal eigenfunctions $S_{\omega l}(\theta=0)$ [defined via Eqs.~\eqref{eq:m=0 quantities},~\eqref{eq:norm S} and~\eqref{eq:angKerr}] were computed using {\rm Mathematica}'s built-in function ``{\rm SpheroidalPS}". In particular, taking normalization into account, our $S_{\omega l}(\theta=0)$ corresponds to $\sqrt{(2l+1)/2}\;\text{SpheroidalPS} [l,0,i\,a\, \omega,1]$. The numerical evaluation of the spheroidal eigenfunctions was done to 300 significant digits.

\subsection{Numerical methods\label{subsec:numerical_methods}}

As described above, we
have computed $\rho_{\omega l}^{\text{up}}$, $\psi_{\omega l}^{\text{int}}\left(r\right)$, $\psi_{\omega l,r}^{\text{int}}\left(r\right)$  and
$S_{\omega l}\left(\theta=0\right)$ for the chosen $a/M$ (and $r/M$) values,
typically reaching up to $l=300$ and $\omega=10/M$ with an increment of
$d\omega=1/200M$.\footnote{Closer to the horizons (i.e., in the regions corresponding to about 
$\delta r_{\pm}\sim10^{-3}-10^{-5}$), the usable range in $\omega$ (with
the same, fixed $l$ range) drops to $\omega\gtrsim2/M$.} 

The raw material {[}$\rho_{\omega l}^{\text{up}}$, $\psi_{\omega l}^{\text{int}}\left(r\right)$,
$\psi_{\omega l,r}^{\text{int}}\left(r\right)$ and $S_{\omega l}\left(\theta=0\right)${]}
is then used to construct the bare mode contribution {[}given in Sec.
\ref{sec:The-t-splitting-procedure} or Eq.~\eqref{eq:Tyy_wl_expanded}){]}
per $\omega,l$ in the mentioned ranges. This bare mode contribution
is then treated as described in Sec.~\ref{sec:The-t-splitting-procedure},
performing the regularization steps while summing and integrating,
to finally reach the desired renormalized quantity. We now provide
a more detailed account of the numerical implementation of this regularization
procedure.

First, in principle, the ID needs to be subtracted from the bare mode
contribution, to be followed by a summation over
$l$ (per $\omega$) to produce the basic integrand function {[}defined
in Eq.~\eqref{eq:E_basic_def}{]}. For $\Phi^{2}$ this ID subtraction
stage (demonstrated in Fig.~\ref{Fig:phi_vs_l_4tiles}) merely includes the removal of the analytically-known $O\left(l^{0}\right)$
leading order $c_{0}$ {[}see Eq.~\eqref{eq:Ewl_div_phi2}{]}, leaving
a remainder that decays like $1/l\left(l+1\right)$, which is to be
summed over. As already mentioned above, we overcome this slow decay
by fitting the sequence of partial sums (namely $\sum_{l'=0}^{l}\left(E_{\omega l'}-E_{\omega l'}^{\text{div}}\right)$)
with $\sim100$ (or more) orders in $1/l$ (starting with order $l^{0}$). \footnote{\label{fn:fit} We perform this fit by inputting the last $N$ elements of the sequence and the $N$ coefficients of a power series in $1/l$ (starting from the zeroth order). This results in an algebraic system of $N$ equations and $N$ coefficients, which is solved using {\rm Mathematica}'s ``{\rm Solve}” function. The value of $N$ considered depends on $r$, but it is typically around $~100$. In a sample of $r$ and $\omega$ values, we also extracted these coefficients using a least-square fit (using {\rm Mathematica}’s built-in ``{\rm LinearModelFit}" function), and the results aligned exceptionally well with the explicit computation.
}
 The fitted coefficient of this $l^{0}$ term then constitutes the
desired $l$-sum. This allows the basic integrand $E^{\text{basic}}\left(\omega\right)$
to be extracted with a precision of typically $\gtrsim50$ digits, depending
on $r$  (this level of precision, reflecting the smallness of the $l$-sum error, was in practice evaluated by comparing fits with different numbers of orders in $1/l$, i.e. varying the number $N$ mentioned in footnote~\ref{fn:fit}).

Let us now turn to the fluxes. First we note that we only directly
calculate $T_{uu}$, whereas $T_{vv}$ is calculated
from it via the conserved quantity given in Eq.~\eqref{eq:conserved_quantity}
with $\theta=0$, noting Eq.~\eqref{eq:Tuu-Tvv(bar=00003Dstan)},
and whose value we obtain at $r=r_{-}$ using our state subtraction
results of Ref.~\cite{KerrIH:2022}. Specifically, we calculate $T_{vv}$
as

\begin{equation}
\left\langle T_{vv}\left(r,0\right)\right\rangle _{\text{ren}}=\left\langle T_{uu}\left(r,0\right)\right\rangle _{\text{ren}}+\frac{\left(r_{-}^{2}+a^{2}\right)}{\left(r^{2}+a^{2}\right)}\left(\left\langle T_{vv}\left(r_{-},0\right)\right\rangle _{\text{ren}}-\left\langle T_{uu}\left(r_{-},0\right)\right\rangle _{\text{ren}}\right)\,.\label{eq:Tvv_from_Tuu}
\end{equation}

Hence, we now focus on the direct computation of $\left\langle T_{uu}\right\rangle _{\text{ren}}^{U}$.
For $T_{uu}$, the large-$l$ divergent piece $T_{uu\left(\omega l\right)}^{\text{div}}$
has an analytically-known leading-order piece  {[}$c_{2}l\left(l+1\right)$,
with $c_{2}$ given in Eq.~\eqref{eq:c2(r)}{]} and an additional
  $O\left(l^{0}\right)$ piece {[}namely $c_{01}\left(\omega\right)$
defined in Eq.~\eqref{eq:c01_flux_def} {]}. This quantity $c_{01}\left(\omega\right)$
is, in principle, not known analytically, but can be extracted numerically. Thus,
the straightforward procedure would be to compute the $l$-sum in
two stages: first, numerically extracting $c_{01}\left(\omega\right)$,
and using it to obtain the regularized sequence $T_{uu\left(\omega l\right)}-T_{uu\left(\omega l\right)}^{\text{div}}$;
and then  summing this regularized sequence over $l$ to obtain $T_{uu}^{\text{basic}}\left(\omega\right)$. This procedure is demonstrated in Fig.~\ref{Fig:Tuu_vs_l_4tiles}. However, we
find it simpler and more efficient (and significantly more accurate)
to perform the desired (regularized) mode sum in a single step, in
a manner that does not require the knowledge of $c_{01}\left(\omega\right)$.
To this end, we construct the sequence of partial sums  $\sum_{l'=0}^{l}\left[T_{uu\left(\omega l'\right)}-c_{2}l'\left(l'+1\right)\right]$
and  fit it with $\sim100$ (or more) orders in $1/\left(l+1\right)$,
starting with order $\left(l+1\right)^{1}$ (this fit is performed in a similar fashion to that applied in the $\Phi^2$ case described above). The coefficient of $\left(l+1\right)^{0}$
 constitutes the desired regularized mode sum, namely the quantity
$T_{uu}^{\text{basic}}\left(\omega\right)$, which is extracted
to a precision of $\gtrsim50$ digits (depending on $r$). 

As a side product, this  fit also yields the unknown parameter $c_{01}\left(\omega\right)$
(as the coefficient of the leading order term $\left(l+1\right)^{1}$),
to a typical precision of $\sim10^{2}$ figures (depending on $r$).
This allows us to numerically explore its dependence on $\omega$,
and hence to verify its highly-accurate parabolic form (which is crucial
for the justification of the ID subtraction, see Sec.
\ref{sec:The-t-splitting-procedure}), as demonstrated in Fig.~\ref{Fig:Tuu_vs_w_2tiles}.
In fact, we routinely repeat this parabolicity test in all cases computed.
(An alternative test is taking the third-order numerical derivative
of $c_{01}\left(\omega\right)$ with respect to $\omega$, observing
its extremely small deviation from zero.)

In the next step, portrayed in Fig.~\ref{Fig:phi_integrand_2tiles} for $\Phi^{2}$ and in Fig.~\ref{Fig:Tuu_integrand_2tiles} for $T_{uu}$, we subtract the $\omega$-dependent singular piece
(the PMR counterterm $E^{\text{sing}}\left(\omega\right)$ for $\Phi^{2}$
or $T_{yy}^{\text{sing}}\left(\omega\right)$ for $T_{yy}$,
as given in Sec.~\ref{subsec:Counterterms}) from the basic integrand.
Integrating the resulting integrand also proved somewhat challenging due to its slow convergence, necessitating fitting to a series of inverse powers $1/\omega^k$.
We carry out the integration over $\omega$, and finally subtract
the finite counterterm as described in Eq.~\eqref{eq:P_ren_PMR}.
The integration range is as described above ($\omega\in\left[0,10/M\right]$
for a typical $r$ value, decreasing towards the horizon vicinities
up to around $\omega\in\left[0,2/M\right]$).

The results of this computation for the various cases are presented
in Sec.~\ref{sec:Numerical-results} and in the figures within. The
absolute error differs from figure to figure, depending on the presented
quantity and range, but overall it is generally less than $1\%$ from the vertical scale range spanned in each
figure.  (We do not mention a relative error in the quantities themselves,
since some cross the horizontal axis in the presented ranges). In
particular, we should emphasize that the error in all points shown
in these figures is too small to be visually discernible on the displayed
figures.

\section{The analytic extension variant \label{App:The-analytic-extension}}

In this appendix, we develop a variant of the PMR $t$-splitting method in the
BH interior which differs from the one presented in the rest of the
manuscript. We refer to this variant as the\emph{ analytic extension
}method. This approach is subsequently utilized for comparison and
cross-verification of the results provided in the main text. This
variant of $t$-splitting is based on the fundamental idea that the
physical quantities we aim to compute -- which are regular functions
of $r$, such as $\left\langle \Phi^{2}\right\rangle _{\text{ren}}^{U}$
and $\left\langle T_{yy}\right\rangle _{\text{ren}}^{U}$ 
-- must exhibit analytic behavior at the EH (see Sec.~\ref{subsec:Field-quantization-Unruh}) \footnote{As far as we know, this analyticity has not been rigorously proven for the Unruh state in Kerr. However, this assumption  (commonly accepted in the study of BHs) forms the basis of our approach, and it is crucial for justifying the method adopted in this appendix. In Sec.~\ref{subsec:Field-quantization-Unruh} we give some evidence towards this assumption.}. 

This principle suggests that the expressions for the HTPF mode contributions
can be extended past the EH into the BH interior region (a similar
idea was considered in related works, see Refs.~\cite{DamourRuffini:1976,BussCasals}).
We proceed by deriving explicit expressions for the mode contributions
inside the BH. This involves analytically extending the expressions
for the external HTPF mode contributions into the BH interior. The
resultant mode contributions display oscillatory behavior at large
values of angular momentum ($l$) and frequency ($\omega$). However,
this issue, can be addressed through an  ``oscillation
cancellation" procedure, to be described. Moreover,
large-$\omega$ oscillations are now accompanied by exponential growth,
but this too can be mitigated using our ``oscillation
cancellation" technique. Nevertheless, this behavior
introduces significant numerical challenges, especially when attempting
to decrease $r$, not to mention approaching the IH. Therefore, we
have implemented this method only at two relatively manageable $r$
values, namely $r=1.4M$ and $r=1.5M$ (at the pole) inside a Kerr
BH of spin $a/M=0.8$. 

Comparing this \emph{analytic extension} variant of $t$-splitting
with the \emph{standard} variant described in the main text reveals
a significant fact: the difficulty encountered in one method is absent
in the other. Specifically, the intermediate divergence in $l$ present
in the standard variant does not exist in the analytic extension variant.
Moreover, the oscillatory behavior (accompanied by exponential growth)
in $\omega$ observed in the analytic extension variant is absent
in the standard variant. In other words, the ID subtraction step is
not required in the application of the analytic extension variant,
whereas the oscillation cancellation step is not needed in the application
of the standard variant. Consequently, comparing the results obtained
using both methods serves as a robust tool for verifying the validity
of each, with their unique and non-trivial intermediate steps.

We shall now turn to develop the method, ending with some numerical
results. The method was initially devised and tested in the RN case
before being adapted to Kerr (for general $\theta$). Then, while
this appendix focuses on the pole of a Kerr BH, in accordance with
the overall manuscript, it is worth noting that the method can be
readily adapted to these other cases and was successfully applied for computations in the RN case as well. 

\subsection{Analytic extension of the HTPF}

We start with the HTPF at the exterior of a Kerr BH, given in Eq.~(3.22) in Ref.~\cite{FrolovThorne:1989} (and in our notation, in
Eq.~(5.4) in Ref.~\cite{HTPF:2022}) in terms of the exterior Eddington
modes $f_{\omega lm}^{\text{in}}$ and $f_{\omega lm}^{\text{up}}$
of Eqs.~\eqref{eq:EddModesIn} and \eqref{eq:EddModesUp}:

\begin{equation}
G^{(1)}_{U}\left(x,x'\right)=\hbar\int_{0}^{\infty}\text{d}\omega\sum_{l=0}^{\infty}\sum_{m=-l}^{l}\left\{ f_{\omega lm}^{\text{in}}\left(x\right),f_{\omega lm}^{\text{in}*}\left(x'\right)\right\} +\hbar\int_{0}^{\infty}\text{d}\omega_{+}\sum_{l=0}^{\infty}\sum_{m=-l}^{l}\coth\left(\frac{\pi\omega_{+}}{\kappa_{+}}\right)\left\{ f_{\omega lm}^{\text{up}}\left(x\right),f_{\omega lm}^{\text{up}*}\left(x'\right)\right\} \,,\label{eq:HTPF_out}
\end{equation}
where, recall, curly brackets denote symmetrization with respect to the spacetime point, that is, for two functions $\zeta$ and $\xi$, we define $\{\zeta(x),\xi(x')\}\equiv\zeta(x)\xi(x')+\xi(x)\zeta(x')$. Concentrating on the polar axis ($\theta=0$, hence only $m=0$ survives),
this reduces to
\begin{equation}
G^{(1)}_{U}\left(x,x'\right)=\hbar\int_{0}^{\infty}\text{d}\omega\sum_{l=0}^{\infty}\left[\left\{ f_{\omega l}^{\text{in}}\left(x\right),f_{\omega l}^{\text{in}*}\left(x'\right)\right\} +\coth\left(\frac{\pi\omega}{\kappa_{+}}\right)\left\{ f_{\omega l}^{\text{up}}\left(x\right),f_{\omega l}^{\text{up}*}\left(x'\right)\right\} \right]\,,\,\,\,\,\,\,\,(r>r_{+},\theta=0)\label{eq:HTPF_out_pole}
\end{equation}
where, as stated earlier, eliminating the $m$ index in our notation
is equivalent to taking $m=0$ (for quantities defined with $\omega lm$
indices).

To extend the ``in'' and ``up'' mode contributions analytically beyond
the EH and into the BH, we will examine the contribution of each mode
near the EH (where $u$ diverges) at some constant $v$. Starting
with $f_{\omega l}^{\text{in}}$ and considering the asymptotic form
of $\psi_{\omega l}^{\text{in}}$ as $r_{*}\to-\infty$ {[}by setting
$m=0$ in Eq.~\eqref{eq:psiIN_asym}{]}, we find that it assumes the
following simple asymptotic form at the EH vicinity:
\begin{equation}
f_{\omega l}^{\text{in}}\simeq\frac{S_{\omega l}\left(0\right)}{\sqrt{8\pi^{2}\left|\omega\right|\left(r_{+}^{2}+a^{2}\right)}}\tau_{\omega l}^{\text{in}}e^{-i\omega v}\,,\,\,\,\,\,\,\,\,\text{(EH vicinity)}\,. \label{eq:fin_rp}
\end{equation}

This form corresponds to a purely ingoing wave which may be extended
smoothly beyond the EH. However, taking the $r_{*}\to-\infty$ asymptotic
behavior of $\psi_{\omega l}^{\text{up}}$ {[}by setting $m=0$ in
Eq.~\eqref{eq:psiUP_asym}{]} in the general form of $f_{\omega l}^{\text{up}}$,
we find that $f_{\omega l}^{\text{up}}$ has two distinct contributions at
the EH: a smooth ingoing $\propto e^{-i\omega v}$ term, and an outgoing
$\propto e^{-i\omega u_{\text{ext}}}$ term which is infinitely oscillatory
at the EH, as $u_{\text{ext}}\to\infty$ there. (This $\propto e^{-i\omega u_{\text{ext}}}$
term exists in the entire vicinity of $r=r_{+}$, which includes the
vicinity of the EH.) The asymptotic form of the ``up'' modes at the EH vicinity 
is then \footnote{
One may notice that the form~\eqref{eq:fup_rp} presented here for the asymptotic behavior of $f_{\omega l}^{\text{up}}$ in the EH limit differs from Eq.~(3.15) in Ref.~\cite{HTPF:2022} by an additional term proportional to $e^{-i\omega u_{\text{ext}}}$. As elaborated in Ref.~\cite{HTPF:2022}, Eq.~(3.15) therein aligns more closely with the intuitive interpretation of a typical wavepacket originating from $H_\text{past}$, then being reflected to the EH and transmitted to future null infinity (as illustrated in Fig.~2 therein).
However, note that while Eq.~(3.15) in Ref.~\cite{HTPF:2022} conforms with the more intuitive picture commonly accepted in the literature, the form~\eqref{eq:fup_rp} presented here is the {\it exact} form (as obtained by simply plugging Eq.~\eqref{eq:psiUP_asym} into Eq.~\eqref{eq:EddModesUp}, taking $m=0$). It is this form that facilitates analytic continuation, as described in the current appendix.
\label{fn:fup_asymp}}
\begin{equation}
f_{\omega l}^{\text{up}}\simeq\frac{S_{\omega l}\left(0\right)}{\sqrt{8\pi^{2}\left|\omega\right|\left(r_{+}^{2}+a^{2}\right)}}\left(\rho_{\omega l}^{\text{up}}e^{-i\omega v}+e^{-i\omega u_{\text{ext}}}\right)\,,\,\,\,\,\,\,\,\,\text{(EH vicinity)}\,. \label{eq:fup_rp}
\end{equation}

On the other side of the EH, within the BH interior, there are the
``right'' and ``left'' Eddington modes of Eq.~\eqref{eq:EddModesRL}
(in which we set $m=0$). Considering the asymptotic behavior of $\psi_{\omega l}^{\text{int}}$
given in Eq.~\eqref{eq:psi_int_BC}, one obtains the asymptotic behavior
of $f_{\omega l}^{R}$ and $f_{\omega l}^{L}$ at the EH vicinity: \footnote{The same remark made in footnote~\ref{fn:fup_asymp} regarding the asymptotic behavior of $f^\text{up}_{\omega l}$ at the EH is valid here for $f^L_{\omega l}$, comparing the above Eq.~\eqref{eq:fL_rp} with Eq.~(3.19) of Ref.~\cite{HTPF:2022}. As manifested in the latter (see also Fig.~2 therein), the $f^L_{\omega l}$ Eddington modes are commonly thought of as arising from $H_L$ and having zero initial data at the EH. However, the form given here constitutes the exact asymptotic behavior at $r\to r_+$, 
and hence applies to $H_L$ as well as to the EH. While indeed the $f^L_{\omega l}$ modes are more naturally tied to $H_L$ and may be intuitively thought of as arising from $H_L$ (where $u_\text{int}$ varies), in the current context of analytical extension we restrict our attention to the asymptotic behavior at the EH vicinity.}
\begin{equation}
f_{\omega l}^{R}\simeq\frac{S_{\omega l}\left(0\right)}{\sqrt{8\pi^{2}\left|\omega\right|\left(r_{+}^{2}+a^{2}\right)}}e^{-i\omega v}\,,\,\,\,\,\,\,\,\,\text{(EH vicinity)}\,\label{eq:fR_rp}
\end{equation}
\begin{equation}
f_{\omega l}^{L}\simeq\frac{S_{\omega l}\left(0\right)}{\sqrt{8\pi^{2}\left|\omega\right|\left(r_{+}^{2}+a^{2}\right)}}e^{i\omega u_{\text{int}}}\,,\,\,\,\,\,\,\,\,\text{(EH vicinity)}\,\label{eq:fL_rp}
\end{equation}
{[}where $u_\text{int}$ 
is now the internal coordinate given in Eq.~\eqref{eq:intEddCoor}{]}.

We now wish to match the exterior Eddington modes with the interior
ones, analytically extending beyond the EH. We denote the extension
of exterior quantities to interior quantities by $\mapsto$. 

Comparing Eq.~\eqref{eq:fin_rp} with Eqs.~\eqref{eq:fR_rp} and \eqref{eq:fL_rp},
it is clear that the ``in'' modes extend through the EH in a regular
manner, with $f_{\omega l}^{\text{in}}$ matched to $\tau_{\omega l}^{\text{in}}f_{\omega l}^{R}$:
\begin{equation}
f_{\omega l}^{\text{in}}\mapsto\tau_{\omega l}^{\text{in}}f_{\omega l}^{R}\,.\label{eq:fin_ext}
\end{equation}
The extension of the ``up'' modes is trickier. The $\propto\rho_{\omega l}^{\text{up}}e^{-i\omega v}$
term in Eq.~\eqref{eq:fup_rp} passes through the EH regularly (as
$v$ remains regular there), and matches to $\rho_{\omega l}^{\text{up}}f_{\omega l}^{R}$
in the BH interior. However, as $u_{\text{ext}}\to\infty$ at the EH, the term $\propto e^{-i\omega u_{\text{ext}}}$ does not pass regularly
to the BH interior. To proceed, we write $u_{\text{ext}}=v-2r_{*}$
{[}see Eq.~\eqref{eq:extEddCoor}{]} with $r_{*}$ as given in Eq.~\eqref{eq:rstarKerr}.
Approaching $r_{+}$ from the BH exterior, $r_{*}$ has the form
\begin{equation}
r_{*}\simeq r_{+}+\frac{1}{2\kappa_{+}}\log\left(\frac{r-r_{+}}{r_{+}-r_{-}}\right)\,,\,\,\,\,r\to r_+^{(+)}. \label{eq:rstar_EH}
\end{equation}
where $r\to r_+^{(+)}$ denotes the limit of $r$ approaching $r_+$ from above.
Hence, at the EH vicinity, we may write
\[
e^{-i\omega u_{\text{ext}}}\simeq A\cdot z^{i\alpha}\,,
\]
where
\begin{equation}
A\equiv e^{-i\omega\left(v-2r_{+}\right)}\,,\,\,z\equiv\frac{r-r_{+}}{r_{+}-r_{-}}\,,\,\,\alpha\equiv\frac{\omega}{\kappa_{+}}\,.\label{eq:A,z,alpha_def}
\end{equation}
The $z^{i\alpha}$ term is singular as $z$ vanishes at $r=r_{+}$.
We shall now analytically extend it through the EH.

At $z>0$ we have the original function $g\left(z\right)\equiv\exp\left[i\alpha\,\ln\left(z\right)\right]$,
and we want to analytically extend it to the negative real axis of
$z$. We wish to express the resultant analytically-extended function
in the form $q\,\exp\left[i\alpha\,\ln\left(-z\right)\right]$, where
$q$ is a pre-factor to be determined. To this end, we express $z$
as $z=|z|\,e^{i\phi}$ where $\phi\equiv\text{arg}(z)$.  Then $\ln\left(z\right)=\ln|z|+i\phi$,
and hence the original function becomes
\[
g\left(z\right)=e^{i\alpha\,\left(\ln|z|+i\phi\right)}=e^{i\alpha\,\ln|z|}e^{-\alpha\phi}\,.
\]
Evaluating this function at $z=-|z|$, which corresponds to taking
$\phi=\pi$~\footnote{Here we analytically extend along a curve in the complex plane bypassing the $r=r_+$ singularity from above. A second option would be to bypass the $r=r_+$ singularity from below. (Note that since the quantity of interest, the HTPF, is  real and analytic across the EH, one may analytically extend along any curve of choice in the complex plane, and the result should be independent of the curve chosen.) This second option would result in replacing $e^{-\alpha \pi}$ in what follows by $e^{\alpha\pi}$. However,  the real part of the final mode-sum expression given in Eq.~\eqref{eq:Gana}  is in fact invariant under this choice of curve. (The imaginary part changes its sign under this change of curve, but this does not concern us, as discussed in footnote~\ref{fn:ima_Gana}).\label{fn:curve}}, we obtain 
\[
g\left(z\right)=e^{i\alpha\,\ln(-z)}e^{-\alpha\pi},
\quad z<0,
\]
and therefore the sought-after ``de-amplification factor'' $q$ is
\[
q=e^{-\alpha\pi}
\]
and the analytic extension of $e^{-i\omega u_{\text{ext}}}$ is
\[
e^{-i\omega u_{\text{ext}}}\mapsto Ae^{-\alpha\pi}\left(-z\right)^{i\alpha}\,.
\]

Approaching the EH from the BH interior, we have
\begin{equation}
r_{*}\simeq r_{+}+\frac{1}{2\kappa_{+}}\log\left(\frac{r_{+}-r}{r_{+}-r_{-}}\right)\,,\,\,\,\,r\to r_{+}^{(-)}\label{eq:rstar_EH-}
\end{equation}
where $r\to r_+^{(-)}$ denotes the limit of $r$ approaching $r_+$ from below.
Hence we have here
\[
e^{i\omega u_{\text{int}}}=A\left(-z\right)^{i\alpha}
\]
with $A,$ $z$ and $\alpha$ given in Eq.~\eqref{eq:A,z,alpha_def},
and the matching is
\[
e^{-i\omega u_{\text{ext}}}\mapsto e^{-\alpha\pi}e^{i\omega u_{\text{int}}}\,.
\]
The ``up'' mode is then analytically extended to the BH interior
as {[}see Eq.~\eqref{eq:fL_rp}{]}
\begin{align}
f_{\omega l}^{\text{up}} & \mapsto\rho_{\omega l}^{\text{up}}f_{\omega l}^{R}+e^{-\frac{\pi\omega}{\kappa_{+}}}f_{\omega l}^{L}\,.\label{eq:fup_ext}
\end{align}

Similarly, one obtains
\[
e^{i\omega u_{\text{ext}}}\mapsto A^{*}e^{\alpha\pi}\left(-z\right)^{-i\alpha}=e^{\alpha\pi}e^{-i\omega u_{\text{int}}}
\]
and
\begin{align}
f_{\omega l}^{\text{up}*} & \mapsto\rho_{\omega l}^{\text{up}*}f_{\omega l}^{R*}+e^{\frac{\pi\omega}{\kappa_{+}}}f_{\omega l}^{L*}.\label{eq:fup*_ext}
\end{align}

Now, equipped with the extension of the modes through the EH, we may
return to the exterior expression of the HTPF at the pole {[}given
in Eq.~\eqref{eq:HTPF_out_pole}{]} and carry it to the BH interior.
Using Eqs.~\eqref{eq:fin_ext}, \eqref{eq:fup_ext} and \eqref{eq:fup*_ext}
and simplifying by hypergeometric identities, one obtains the interior
HTPF in the analytic extension variant (at the pole), denoted $G_{\text{ae}}^{U}\left(x,x'\right)$ (hereafter, a subscript/superscript ``ae" denotes the analytic extension):
\begin{align}
G_{\text{ae}}^{U}\left(x,x'\right) & =\hbar\int_{0}^{\infty}\text{d}\omega\sum_{l=0}^{\infty}\left[\coth\left(\frac{\pi\omega}{\kappa_{+}}\right)\left(\left\{ f_{\omega l}^{L}\left(x\right),f_{\omega l}^{L*}\left(x'\right)\right\} +\left|\rho_{\omega l}^{\text{up}}\right|^{2}\left\{ f_{\omega l}^{R}\left(x\right),f_{\omega l}^{R*}\left(x'\right)\right\} \right)\right.\label{eq:Gana} \\
 & +2\left[\text{cosech}\left(\frac{\pi\omega}{\kappa_{+}}\right)+\sinh\left(\frac{\pi\omega}{\kappa_{+}}\right)\right]\Re\left(\rho_{\omega l}^{\text{up}}\left\{ f_{\omega l}^{R}\left(x\right),f_{\omega l}^{L*}\left(x'\right)\right\} \right)+\left|\tau_{\omega l}^{\text{in}}\right|^{2}\left\{ f_{\omega l}^{R}\left(x\right),f_{\omega l}^{R*}\left(x'\right)\right\} \nonumber \\
 & \left.+i\cosh\left(\frac{\pi\omega}{\kappa_{+}}\right)\Im\left(\rho_{\omega l}^{\text{up}}\left\{ f_{\omega l}^{R}\left(x\right),f_{\omega l}^{L*}\left(x'\right)\right\} \right)\right]\,\nonumber. 
\end{align}
As this expression should yield a manifestly real result, the imaginary
part appearing in the third line of Eq.~\eqref{eq:Gana} should vanish
(after integration and summation). \footnote{Note that the positive sign of the imaginary part appearing in the third line of Eq.~\eqref{eq:Gana}  was obtained following our  choice of analytically extending along a curve bypassing the $r=r_+$ singularity \emph{from above}. If one were to bypass the $r=r_+$ singularity from below, the sign of this imaginary term would be negative.  This sign ambiguity does not concern us, given
the expectation for the vanishing of this imaginary piece after summation and integration.\label{fn:ima_Gana}}

We may write $G_{\text{ae}}^{U}$ as
\begin{align*}
G_{\text{ae}}^{U}\left(x,x'\right) & =G_{\text{stn}}^{U}\left(x,x'\right)+G_{\text{dif}}^{U}\left(x,x'\right),
\end{align*}
where $G_{\text{stn}}^{U}\left(x,x'\right)$ denotes the interior
HTPF in the standard variant {[}at the pole, which is the $\theta=0$, $m=0$ version
of Eqs.~\eqref{eq:Gmodesum} and \eqref{eq:Gwlm}{]}, and $G_{\text{dif}}^{U}\left(x,x'\right)$
is their difference, which reads

\begin{equation}
G_{\text{dif}}^{U}\left(x,x'\right)=2\hbar\int_{0}^{\infty}\text{d}\omega\sum_{l=0}^{\infty}\left[\sinh\left(\frac{\pi\omega}{\kappa_{+}}\right)\Re\left(\rho_{\omega l}^{\text{up}}\left\{ f_{\omega l}^{R}\left(x\right),f_{\omega l}^{L*}\left(x'\right)\right\} \right)+i\cosh\left(\frac{\pi\omega}{\kappa_{+}}\right)\Im\left(\rho_{\omega l}^{\text{up}}\left\{ f_{\omega l}^{R}\left(x\right),f_{\omega l}^{L*}\left(x'\right)\right\} \right)\right]\,.\label{eq:DeltaG}
\end{equation}

The entire $G_{\text{dif}}^{U}\left(x,x'\right)$ quantity is expected to \emph{vanish} in order for the two variants to yield the same results -- see footnote \ref{fn:ae_vanishing}.

Since we know that the HTPF should be real, from this point on we  concentrate only on the real part of $G^U_{\text{ae}}$ (and of $G^U_{\text{dif}}$).

\subsection{Individual mode contributions at coincidence\label{subsec:Expressions-in-coincidence}}

We now provide computationally-amenable expressions for the mode contributions in the
analytic extension variant, at coincidence ($x'\to x$), for both
$\left\langle \Phi^{2}\right\rangle ^{U}$ and $\left\langle T_{yy}\right\rangle ^{U}$. It is in fact more compact to explicitly give the \emph{difference} in the $\omega l$-mode contributions between the analytic extension and standard variants, derived from (the real part of) $G_{\text{dif}}^{U}\left(x,x'\right)$  of Eq.~\eqref{eq:DeltaG}. We denote this difference by $E_{\omega l}^{\text{dif}}$ for the field square [given in Eq.~\eqref{eq:Edif} below] and by $T_{yy\left(\omega l\right)}^{\text{dif}}$ for the fluxes [given in Eq.~\eqref{eq:Tdif} below]. 
The summation and integration (with appropriate regularization) of these ``dif" quantities are expected
to \emph{vanish}, in order for the two variants to yield the same result. Since both quantities, $E_{\omega l}^{\text{dif}}$ and $T_{yy\left(\omega l\right)}^{\text{dif}}$, diverge exponentially with $\omega$  (due to the $\sinh(\pi\omega/\kappa_+)$ function appearing in both expressions), this vanishing a priori seems far from trivial. In the next subsections we explore and numerically show that the difference between the analytic extension variant and standard variant indeed vanishes for both the field square and the fluxes. (In practice, we present the computation of the full analytic extension renormalized quantities and show their agreement with their standard-variant counterparts). 
The fact that the difference indeed vanishes then serves as a robust test of
both variants of $t$-splitting, the standard and the analytic extension
variants, each involving its own unique procedure (as discussed above, as well as summarized in Sec.~\ref{subsec:ae_summary}). \footnote{Here (and in the rest of this appendix) we focus on the quantities of interest, the field square and the fluxes (or their difference), derived from the real part of $G_{\text{dif}}^{U}\left(x,x'\right)$ followed by the coincidence limit  $x'\to x$. It is worth noting, however, that the entire $G_{\text{dif}}^{U}\left(x,x'\right)$ quantity, prior to taking the coincidence limit, is expected to vanish.
The individual mode contribution to $G_{\text{dif}}^{U}\left(x,x'\right)$, given in Eq.~\eqref{eq:DeltaG},
includes two parts, a real part and an imaginary part, both growing
exponentially with $\omega$ as dictated by the hyperbolic functions $\sinh\left(\pi\omega/\kappa_{+}\right)$ and $\cosh\left(\pi\omega/\kappa_{+}\right)$. We would expect that both real and imaginary parts will vanish separately. However, here we focus on verifying this expectation numerically only for the real part in the coincidence limit.
\label{fn:ae_vanishing}}

\subsubsection{Individual mode contribution to $\left\langle \Phi^{2}\right\rangle ^{U}$ }

Using the explicit forms of the interior Eddington mode functions
of Eq.~\eqref{eq:EddModesRL}, one obtains the mode contribution to
$\left\langle \Phi^{2}\right\rangle ^{U}$ in the analytic extension,
denoted $E_{\omega l}^{\text{ae}}$:
\begin{equation}
E_{\omega l}^{\text{ae}}=E_{\omega l}^{\text{stn}}+E_{\omega l}^{\text{dif}}\label{eq:Eana}
\end{equation}
where $E_{\omega l}^{\text{stn}}$ is the standard expression used
in this paper, given in Eq.~\eqref{eq:Ewl_phi2}, and $E_{\omega l}^{\text{dif}}$
is the difference given by
\begin{equation}
E_{\omega l}^{\text{dif}}=\hbar\frac{\left[S_{\omega l}\left(0\right)\right]^{2}}{4\pi^{2}\omega\left(r^{2}+a^{2}\right)}\sinh\left(\frac{\pi\omega}{\kappa_{+}}\right)\Re\left[\rho_{\omega l}^{\text{up}}\left(\psi_{\omega l}^{\text{int}}\right)^{2}\right]\,.\label{eq:Edif}
\end{equation}

\subsubsection{Individual mode contribution to $\left\langle T_{yy}\right\rangle ^{U}$}

Deriving the flux components from the HTPF of the analytical extension
variant given in Eq.~\eqref{eq:Gana} is done in the exact same manner
as in Appendix B in Ref.~\cite{HTPF:2022}. This yields the individual
mode contribution to $\left\langle T_{yy}\right\rangle ^{U}$
in the analytic extension variant, which we write as
\begin{equation}
T_{yy\left(\omega l\right)}^{\text{ae}}=T_{yy\left(\omega l\right)}^{\text{stn}}+T_{yy\left(\omega l\right)}^{\text{dif}},\label{eq:Tana}
\end{equation}
where $T_{yy\left(\omega l\right)}^{\text{stn}}$ is the
standard expression used in the main manuscript {[}given in Eqs.~\eqref{eq:Tyy_wl_expanded}-\eqref{eq:TC}{]}, and $T_{yy\left(\omega l\right)}^{\text{dif}}$
is the difference between the two variants, given by
\begin{equation}
T_{yy\left(\omega l\right)}^{\text{dif}}=\hbar\frac{\left[S_{\omega l}\left(0\right)\right]^{2}}{16\pi^{2}\omega\left(r^{2}+a^{2}\right)}\sinh\left(\frac{\pi\omega}{\kappa_{+}}\right)\Re\left(\rho_{\omega l}^{\text{up}}\left[\omega^{2}\left(\psi_{\omega l}^{\text{int}}\right)^{2}+\left(\psi_{\omega l,r_{*}}^{\text{int}}\right)^{2}-2\frac{\Delta r}{\left(r^{2}+a^{2}\right)^{2}}\psi_{\omega l}^{\text{int}}\psi_{\omega l,r_{*}}^{\text{int}}+\frac{\Delta^{2}r^{2}}{\left(r^{2}+a^{2}\right)^{4}}\left(\psi_{\omega l}^{\text{int}}\right)^{2}\right]\right)\,.\label{eq:Tdif}
\end{equation}

\subsection{The regularization procedure in the analytic extension variant\label{subsec:The-renormalization-procedure}}

The quantities $E_{\omega l}^{\text{ae}}$, given in Eqs.~\eqref{eq:Eana} and \eqref{eq:Edif},
and $T_{yy\left(\omega l\right)}^{\text{ae}}$, given in
Eqs.~\eqref{eq:Tana} and \eqref{eq:Tdif}, constitute the individual mode
contributions to $\left\langle \Phi^{2}\right\rangle ^{U}$ and $\left\langle T_{yy}\right\rangle ^{U}$,
respectively, within the analytic extension variant. However, the mode sums of these quantities are clearly
divergent. These mode sums may be regularized within the analytic extension
variant of $t$-splitting in several steps, outlined briefly in this
section {[}it may be compared with the standard PMR $t$-splitting procedure, outlined in Sec.~\ref{sec:The-t-splitting-procedure} and
summarized in Eqs.~\eqref{eq:E_basic_def} and \eqref{eq:P_ren_PMR}{]}.

\subsubsection{Regularization of the $l$ sum}\label{subsec:regularization_l_ae}

As in the standard variant, the first step involves summing over $l$.
However, whereas the standard variant encounters a diverging sum that
is addressed by the ID subtraction,
in the analytic extension variant the $l$-sum's failure to converge is due
to growing oscillations. 

To obtain these oscillations analytically, we look into the large-$l$ limit of $E_{\omega l}^{\text{ae}}$ and $T_{yy(\omega l)}^{ae}$ (as we did in Sec.~\ref{subsec:Intermediate-blindspots}  for the individual mode contributions in the standard variant). While the leading order large-$l$ behavior of $E_{\omega l}$ and  $T_{yy\left(\omega l\right)}$
in the standard variant is   $\propto l^0$  and $\propto l^2$, respectively, with $r$-dependent coefficients found analytically in Eqs.~\eqref{eq:Ewl_div_phi2} and  ~\eqref{eq:c2(r)}, the analytic extension counterpart has an extra multiplicative factor of $\pm\cosh\left(\pi\omega/\kappa_{+}\right)\left(-1\right)^{l}\cos\left[\left(l+\frac{1}{2}\right)s\left(r\right)\right]$, where $s(r)$ is a function of $r$ to be given and the minus sign goes with the  $E_{\omega l}^{\text{ae}}$ case. Explicitly, one finds to leading order in $l$,

\begin{equation}
E_{\omega l}^{\text{ae}}\simeq\frac{-1}{4\pi^{2}\sqrt{\left(r-r_-\right)\left(r_+-r\right)}}\cosh\left(\frac{\pi\omega}{\kappa_{+}}\right)\left(-1\right)^{l}\cos\left[\left(l+\frac{1}{2}\right)s\left(r\right)\right]\,,\,\,\,l\gg1,\label{eq:E_ae_large_l}
\end{equation}
and

\begin{equation}
T_{yy\left(\omega l\right)}^{\text{ae}}\simeq\frac{l^2\sqrt{\left(r_{+}-r\right)\left(r-r_{-}\right)}}{16\pi^{2}\left(r^{2}+a^{2}\right)^{2}}\cosh\left(\frac{\pi\omega}{\kappa_{+}}\right)\left(-1\right)^{l}\cos\left[\left(l+\frac{1}{2}\right)s\left(r\right)\right]\,,\,\,\,l\gg1,\label{eq:Tyy_ae_large_l}
\end{equation}
where 
\begin{equation}
s\left(r\right)\equiv2\arctan\left[\frac{r-M}{\sqrt{\left(r_{+}-r\right)\left(r-r_{-}\right)}}\right]\,.\label{eq:s(r)}
\end{equation}
Clearly, this introduces oscillatory behavior of the mode contributions, with an amplitude growing as $l^2$ in the $T_{yy\left(\omega l\right)}^{\text{ae}}$ case. This large-$l$ behavior is confirmed numerically for both  $E_{\omega l}^{\text{ae}}$ and $T_{yy\left(\omega l\right)}^{\text{ae}}$.

The wavelength of the oscillation at large $l$, which we hereby denote by $\tilde{\lambda}_{l}\left(r\right)$, can be read from Eqs.~\eqref{eq:E_ae_large_l}-\eqref{eq:s(r)}. Oscillations with this same wavelength clearly appear also in the sequence of partial sums, whose limit at $\infty$ is the (generalized) sum we are interested in.
Remarkably, this oscillation may be damped to reveal the sum by the method of \emph{oscillation cancellation}, which includes operating on the sequence of partial sums with a modified
version (to be adapted to the discrete case) of the \emph{self cancellation} operator introduced in Ref.~\cite{AAt:2015}.
For a function $f(x)$ of a variable $x$ we define the operator
\begin{equation}
O_{\Delta{x}}\left[f\left(x\right)\right]\equiv\frac{f\left(x\right)+f\left(x+\Delta x\right)}{2}\,,\label{eq:osc_canc_Delta_x}
\end{equation}
where $\Delta x$ is some chosen increment. The limit $x\to\infty$ of $O_{\Delta{x}}\left[f\left(x\right)\right]$ coincides with the $x\to\infty$ limit of $f\left(x\right)$, if the latter exists. If it does not exist, $O_{\Delta{x}}\left[f\left(x\right)\right]$ may be used to define a generalized limit of the original function.
For a function exhibiting oscillatory behavior of wavelength $\tilde{\lambda}$  at large $x$ (possibly times a non-exponential function of $x$), an application of this operator acts to damp the oscillations while leaving the non-oscillatory content of the function at $x\to\infty$ unaffected, hence producing a generalized limit at infinity. In particular, if $x$ is a continuous variable, it is clearly most beneficial to take $\Delta x$ to be $\tilde{\lambda}/2$. Then, one application of $O_{\tilde{\lambda}/2}$ suffices to ``kill" the oscillation completely in the case of constant amplitude, or more generally, reduce the amplitude to its $x$ derivative if it is a function of $x$. Applying this operator (repeatedly, in the case of non-constant amplitude) on an accumulation function produces  the \emph{generalized} infinite integral. Similarly, in the discrete case (which is what we have here, as $l$ is a discrete variable), applying this operator on the sequence of partial sums may yield the generalized infinite sum. However, a slight difficulty arises in this case, since half the wavelength is not a whole number and hence can not be taken as the increment for averaging. In that case, we may take $\Delta x$ to be the closest integer to $\tilde{\lambda}/2$ (given that $\tilde{\lambda}$ is well-enough covered by the discrete set of points). This may affect the convergence rate, perhaps slightly decreasing the efficiency of the damping (i.e. increasing the number of required repetitions). In the cases we computed, in which the wavelength in $l$ is sufficiently long to be well-covered, the effect of $l$ being discrete turned out to be quite negligible.

 While the oscillation cancellation described above is indeed very effective in damping the oscillation and revealing the generalized sum, it turns out that fitting the partial sums (at large $l$) as a power series in $1/l$ multiplied by a superposition of $\cos\left[2\pi l/\tilde{\lambda}_l\right]$ and $\sin\left[2\pi l/\tilde{\lambda}_l\right]$  -- also globally multiplied by $l^2$ in the case of fluxes -- is significantly more numerically efficient, and this is the method we use in practice (typically with $\sim10^2$ orders for each of the $\cos$ and $\sin$ terms).
 
Another remark concerns the relationship between the oscillatory behavior one finds in the BH interior and what occurs in the BH exterior, where the mode contributions \emph{decay exponentially} in $l$ at fixed $\omega$ (due to the potential barrier). The mode contributions we consider here represent the analytic extension of those in the BH exterior to the BH interior. As it appears, the (negative) real exponent of $l$ in the BH exterior is analytically extended to the BH interior as a purely imaginary exponent, thereby converting the exponential decay to oscillations in $l$. 

\subsubsection{Regularization of the $\omega$ integral}\label{subsec:regularization_omega_ae}

After performing the sum over $l$ per $\omega$ via the oscillation cancellation procedure described above (or alternatively via a fit, as described), one may construct the \emph{basic integrand function }in $\omega$, denoted $E^{\text{ae}}(\omega)$ [the analytic-extension analog of Eq.~\eqref{eq:E_basic_def}], also denoted by  $T_{yy}^{\text{ae}}(\omega)$ for the fluxes. (Dependence on the spacetime point is implied in this notation, and sometimes added explicitly to the parentheses.)
As in the standard variant, the regularization of this analytic-extension basic integrand includes subtracting the PMR counterterms (see Sec.~\ref{subsec:Counterterms}). However, while in the standard
variant this leaves a converging integrand, here we again remain with
oscillations which interfere with the convergence of the $\omega$-integral (at large $\omega$). These oscillations differ from the single oscillation encountered in the $l$-sum by introducing two major complications: (\emph{i}) the oscillatory
behavior is composed of an entire spectrum (which is reminiscent of the spectrum of oscillations one encounters outside the BH, see Ref.~\cite{AAt:2015}), and (\emph{ii})
most crucially, these oscillations are accompanied by \emph{exponential} growth! (This feature has no exterior counterpart.)
Hence, while the oscillatory behavior in $l$ was damped via a simple oscillation cancellation procedure (merely averaging points roughly half-a-wavelength away), the large-$\omega$ behavior of the basic integrand requires a modified oscillation cancellation procedure -- in particular, one that accounts for the exponential growth (this procedure will be described below).

The oscillations we find, which we number by an index $i$, are of the general form
\begin{equation}
\propto e^{\eta_i\omega}e^{i\omega\Phi_i}\,,\label{eq:osc_form}
\end{equation}
where $\eta_i>0$ and $\Phi_i\in\mathbb{R}$.
$\eta_i$ denotes the exponential growth parameter. We shall refer to $\Phi_i$ as the ``$\omega$-frequency",  and denote its corresponding ``$\omega$-wavelength" by $\tilde{\lambda}_i=2\pi/\Phi_i$. 
We order the oscillations by their $\omega$-frequency -- from the lowest $\omega$-frequency (longer $\omega$-wavelength) towards higher $\omega$-frequencies (shorter $\omega$-wavelengths).

Notably, these oscillations (including their parameters $\eta_i$ and $\Phi_i$ which are discussed below) are shared by both the field square and the fluxes. The full leading order large-$\omega$ asymptotic behavior, however, is numerically found to include multiplication of the above oscillatory (and exponentially-growing) form by $\omega^{1/2}$ for the field square and $\omega^{5/2}$ for the fluxes.

Generally speaking, the spectrum of oscillations in $\omega$ encountered in the BH interior is induced (through the concept of analytic continuation) by the oscillations in $\omega$ existing outside the BH, which are in turn related to a family of non-radial null geodesics connecting the points $x$ and $x'$ associated with the separation in the $t$ direction.
In particular, each such connecting null geodesic determines a certain $\Delta t\equiv t’-t$, which in turn is the  $\omega$-frequency (see Ref.~\cite{AAt:2015} for a detailed treatment of this issue in the Schwarzschild case, but this behavior also carries over to any stationary BH, and in particular to the Kerr case under consideration -- see Ref.~\cite{LeviEilonOriMeentKerr:2017}).

Notably, these (real) $\omega$-frequencies, comprising the spectrum of oscillations outside the BH, are $r$ dependent.
By the very nature of the analytic extension method, the ($r$-dependent) $\omega$-frequencies encountered inside the BH are expected to be related to their external counterparts by analytic continuation from $r>r_+$ to $r<r_+$.

Exploring this process of analytic continuation, one finds that all ($i\geq 1$) $\omega$-frequencies – which are real at $r>r_+$ – acquire a universal \emph{imaginary} part $-\pi/\kappa_+$.
In addition, these $\omega$-frequencies at $r<r_+$ also have a real part, which –- like for their $r>r_+$ counterparts –- does depend on $i$ and $r$. As in the BH exterior, this real part of the $\omega$-frequencies may be obtained by numerically analysing certain connecting null geodesics. However, this analysis is beyond the scope of this paper.

The aforementioned imaginary part of the $\omega$-frequencies inside the BH  gives rise to a universal (i.e., at any radius $r\in (r_-,r_+)$) exponential growth parameter:
\begin{equation}
\eta_i =\frac{\pi}{\kappa_{+}}\,,\,\,\,\,\,\text{for every  }\, i\geq1,\label{eq:eta_short}
\end{equation}
whereas the real part of the $\omega$-frequencies gives rise to the $\omega$-frequencies $\Phi_i$ appearing in Eq.~\eqref{eq:osc_form}.  (As an illustration, for $a/M=0.8$ the first three $\Phi_i$ values are $36.8 M$, $68.0 M$ and $99.1 M$.) Both these aforementioned parameters, $\eta_i$ and $\Phi_i$ for the $i\geq1$ spectrum of oscillations, match what we find numerically.

In addition to this $i\geq1$ spectrum of oscillations, we also find another (single) oscillation with a different exponential factor and a significantly longer $\omega$-wavelength, hence we denote it by $i=0$. 
The origin of this longer oscillation is not entirely clear to us. 
Nevertheless, a simple analysis, based on a somewhat speculative argument (related to a special type of connecting null geodesics \footnote{Specifically, this is a polar (i.e. $\theta=0$) complex null geodesic that connects the $r$ value of the evaluation point with the complex point $r=ia$ — which is the location of the pole in the effective potential $V_{\omega lm}(r)$ give in Eq. \eqref{eq:KerrPot}. The (complex) t-difference $\Delta t$ between these two edge points of the geodesic can be easily computed analytically; and when translated to an $\omega$-frequency (and hence to corresponding $\eta$ and $\Phi$  parameters according to the connection that was briefly described above), one obtains the values given in Eqs. (\ref{eq:eta_long},\ref{eq:Phi_long}).})
suggests that the parameters characterizing the oscillation are given by \footnote{Note, however, that in order to correctly perform the oscillation cancellation, analytical determination of the oscillation parameters $\eta$ and $\Phi$ is not necessary (these parameters can be determined numerically).}:

\begin{equation}
\eta_0 =\frac{\pi}{\kappa_{-}}-\left[2a-\frac{1}{\kappa_{+}}\arctan\left(\frac{a}{r_{+}}\right)+\frac{1}{\kappa_{-}}\arctan\left(\frac{a}{r_{-}}\right)\right]\,,\\\label{eq:eta_long}
\end{equation}

\begin{equation}
\Phi_0\left(r\right)= \left[\frac{1}{2\kappa_{+}}\ln\left(a^{2}+r_{+}^{2}\right)-\frac{1}{2\kappa_{-}}\ln\left(a^{2}+r_{-}^{2}\right)-4M\ln\left(r_{+}-r_{-}\right)\right]-2r_{*}\left(r\right)\,,\\\label{eq:Phi_long}
\end{equation}
where $r_*$ is as given in Eq.~\eqref{eq:rstarKerr}. 

These expressions for $\eta_0$ and $\Phi_0$ seem to match the oscillation we find numerically.

The oscillation cancellation procedure, which accommodates an exponentially growing amplitude, constitutes of taking averages with suitable weights (which sum to 1).
In our case, the suitable self cancellation
operator for a function $f\left(x\right)$ that behaves like $\propto e^{\eta x}e^{(2\pi x)i/\tilde{\lambda}}$  is
\begin{equation}
O_{\Delta x}^{\text{exp}}\left[f\left(x\right)\right]\equiv\frac{f\left(x\right)+\exp\left(-\eta \Delta x\right) f\left(x+\Delta x\right)}{1+\exp\left(-\eta \Delta x\right)},\label{eq:osc_canc_exp}
\end{equation}
where $\Delta x$ is a chosen increment. 

We typically take this operator with $\Delta x=\tilde{\lambda}/2$.  The weights were chosen such that with that $\Delta x$, a pure exponentially-growing oscillation $e^{\eta x}e^{(2\pi x)i/\tilde{\lambda}}$ is fully cancelled in one application of the operator. If $\Delta x$ is not exactly (i.e. slightly deviating from) $\tilde{\lambda}/2$,  the operator still acts to damp the oscillation and yields the same result (even if less efficiently, in the sense that more applications of the operator are needed, hence also a larger range of modes as each repetition shortens the original list by $\Delta x$). Also, note that in our case, the function is not purely of the form $e^{\eta x}e^{(2\pi x)i/\tilde{\lambda}}$  but multiplied by certain powers of the variable $x$. In this case, as discussed for the self cancellation operator of Eq.~\eqref{eq:osc_canc_Delta_x}, applying the operator $O_{\tilde{\lambda}/2}^{\text{exp}}$  does not fully cancel the growing oscillation (but still acts to significantly damp it at each application).
The justification for using this operator is similar to that of the standard oscillation cancellation; see the very brief discussion around  Eq.~\eqref{eq:osc_canc_Delta_x}.  That is, the procedure is constructed to be justified for the accumulation function,
building on the concept of a generalized integral, but it can be translated into a similar procedure for the integrand function itself.\footnote{The procedure for the integrand function (rather than accumulation function) involves various subtleties. In particular, in order to preserve the value of the integral, we attach a zero vector as long as the original $\omega$ vector to the beginning of the original integrand, prior to any application of  $O_{\tilde{\lambda}/2}^{\text{exp}}$. That is, we first double the original $\omega$ range -- the first copy is set to be identically zero and the second copy is the original integrand. The oscillation cancellation procedure is then applied on this lengthened integrand function. Notably, although the processed integrand function depends on the specifics of the oscillation cancellation procedure applied (i.e. the number of operations performed with a certain $\Delta x$), the resulting integral remains invariant.\label{fn:osc_int}}

Comparing Eq.~\eqref{eq:eta_long} with Eq.~\eqref{eq:eta_short}, one finds that the shorter oscillations $i \geq 1$ are accompanied by a stronger exponent compared to that of the long $i=0$ oscillation.
Hence, we typically first damp the short oscillations, until the long oscillation is exposed and damped as well.
Applying a multiple-damping procedure (by repeated application of $O^{\text{exp}}_{\Delta x}$ with suitable $\Delta x$ at each repetition) effectively damps the exponentially growing oscillations.
In the cases we looked at, it has been used to ``kill'' more than a hundred orders of magnitude in the $\omega$-integrand.
Clearly, achieving such fantastic damping is only possible with extremely accurate
data, which is indeed what we have
(see Sec.~\ref{subsec:Computation-of-raw-data}).

Finally, after the oscillations have been sufficiently damped, the ``regularized",
non-oscillatory piece of the integrand is exposed and integrated over.
Then, following a subtraction of the finite PMR counterterm
(see Sec.~\ref{subsec:Counterterms}), the computation of the renormalized quantity of interest is finally complete.

\subsection{Summary of regularization using the analytic extension variant\label{subsec:ae_summary}}

In the standard variant, Eqs.~\eqref{eq:E_basic_def} and \eqref{eq:P_ren_PMR} summarize the regularization procedure and the components involved. Within the \emph{analytic extension} variant, the procedure that yields $P_{\text{ren}}$ (where $P$ is the quantity of interest, either the field square or the fluxes), may be written schematically as
 
\begin{equation}
P_{\text{ren}}\left(x\right)=\int_{0}^{\infty}\left(O_{[\omega]}\left[E^{\text{ae}}\left(\omega,x\right)-E^{\text{sing}}\left(\omega,x\right)\right]\right)\text{d}\omega-e\left(x\right)\,,\label{eq:P_ren_ae}
\end{equation}
where  $O_{[\omega]}$ is an operator damping the oscillations in $\omega$ (described below), 
and $E^\text{ae}(\omega,x)$ is the analytic-extension basic integrand, defined as
\begin{equation}
E^{\text{ae}}\left(\omega,x\right)\equiv\sum_{l=0}^{\infty}\left(O_{[l]}\left[E^{\text{ae}}_{\omega l}\left(x\right)\right]\right)\,.\label{eq:E_ae_def}
\end{equation}
Here,  $O_{[l]}$ is an operator damping the oscillations in $l$ (described below), and the functions $E_{\omega l}^\text{ae}$ are the bare mode contributions of the analytic extension method, given for the field square in Eqs.~\eqref{eq:Eana} and \eqref{eq:Edif}, and replaced by  $T_{yy(\omega l)}^\text{ae}$ of Eqs.~\eqref{eq:Tana} and \eqref{eq:Tdif} for the fluxes.

The $O_{[l]}$ operator mentioned in the above recipe consists of a multiple application of the operator $O_{\Delta x}$ of Eq.~\eqref{eq:osc_canc_Delta_x} with the increment $\Delta x$ taken to be as close as possible to half the wavelength $\tilde{\lambda}_l/2$, as described in Sec.~\ref{subsec:regularization_l_ae}. The $O_{[\omega]}$ operator is more involved, as it damps the entire complex spectrum of oscillations in $\omega$ described above.
It hence consists of a multiple application of the operator $O^{\text{exp}}_{\Delta x}$ of Eq.~\eqref{eq:osc_canc_exp} for each $i$-oscillation, with a suitable increment $\Delta x=\tilde{\lambda}_i/2$ and an exponential growth parameter $\eta_i$, as described in Sec.~\ref{subsec:regularization_omega_ae}.

A comparison of Eqs.~\eqref{eq:P_ren_ae} and \eqref{eq:E_ae_def} with their standard variant counterparts given in Eqs.~\eqref{eq:E_basic_def} and \eqref{eq:P_ren_PMR} reveals the differences, as well as similarities, between the two methods.
Both methods include the subtraction of $t$-splitting PMR counterterms, $E^\text{sing}(\omega,x)$ and the finite counterterm $e(x)$, given in Sec.~\ref{subsec:Counterterms}. In this sense, both are variants of the PMR $t$-splitting method.
However, the bare mode contribution in the two variants is different from the outset, leading to distinct asymptotic behaviors at large $l$ and $\omega$, which naturally influence the convergence of the mode sums. Therefore, different treatments are required when performing the integration and summation.
While in the standard variant, the $l$ sum is performed via the subtraction of an ID $E^\text{div}_{\omega l}$, in the analytic extension variant this step is replaced by oscillation cancellation, as described in Sec.~\ref{subsec:regularization_l_ae}. In addition, while in the standard variant the resulting $\omega$-integrand is convergent following the counterterm subtraction, in the analytic extension variant one still has to deal with exponentially growing oscillations, which is done through a specially tailored procedure of oscillation cancellation, as described in Sec.~\ref{subsec:regularization_omega_ae}.

\subsection{Numerical results and comparison with the standard variant\label{subsec:results}}

The expressions provided in Sec.~\ref{subsec:Expressions-in-coincidence},
given in terms of the numerically-computable ingredients $\psi_{\omega l}^{\text{int}}$,
$\rho_{\omega l}^{\text{up}}$ and $S_{\omega l}\left(0\right)$ (see Sec.~\ref{subsec:Computation-of-raw-data}),
may be used to compute $\left\langle \Phi^{2}\right\rangle _{\text{ren}}^{U}$
and $\left\langle T_{yy}\right\rangle _{\text{ren}}^{U}$ following the regularization steps of the analytic extension variant outlined in Sec.~\ref{subsec:The-renormalization-procedure}. Notably, this includes performing
multiple oscillation cancellations which challenge the numerical implementation of the method, requiring very high accuracy and a wide range of modes (since, as mentioned, each application of an oscillation cancellation operator shortens the original range).
The results of these computations may be subsequently compared with
the standard-variant results. 

We focused on  $\left\langle T_{uu}\right\rangle _{\text{ren}}^{U}$ and  $\left\langle \Phi^{2}\right\rangle _{\text{ren}}^{U}$ at two selected $r$ values, $1.4M$ and $1.5M$, inside a Kerr BH of spin parameter $a/M=0.8$ (in which $r_-=0.4M$ and $r_+=1.6M$), at the pole. (It is worth noting that these quantities are shown as a function of $r$ in Figs.~\ref{Fig:a0p8_general},~\ref{Fig:a0p8_general-phi} in the main text.)
For $\left\langle T_{uu}\right\rangle _{\text{ren}}^{U}$ at $r=1.5M$, we found a striking agreement with the standard-variant result of $-1.019578781\times 10^{-7}\hbar M^{-4}$ up to a relative difference of $6\times10^{-8}$. 
At $r=1.4M$ the accuracy dropped mainly due to longer $\omega$-wavelengths, particularly the $i=0$ one, 
which necessitates more modes for effective damping (while the range of modes prepared was the same for both $r$ values). Then, for $\left\langle T_{uu}\right\rangle _{\text{ren}}^{U}$   at $r=1.4M$, we found a relative difference of $2\times 10^{-5}$ from the standard-variant result of $-5.61533442\times 10^{-7}\hbar M^{-4}$. 
Similar computations of  $\left\langle \Phi^{2}\right\rangle _{\text{ren}}^{U}$ at these $r$  values have also shown nice agreement with the standard-variant $t$-splitting results.
The agreement exposed is very
impressive, given the many orders of magnitude that needed to be damped in order for the physical results
to be extracted from the bare mode contributions.

\begin{figure}
    \centering
    \includegraphics[width=1\linewidth]{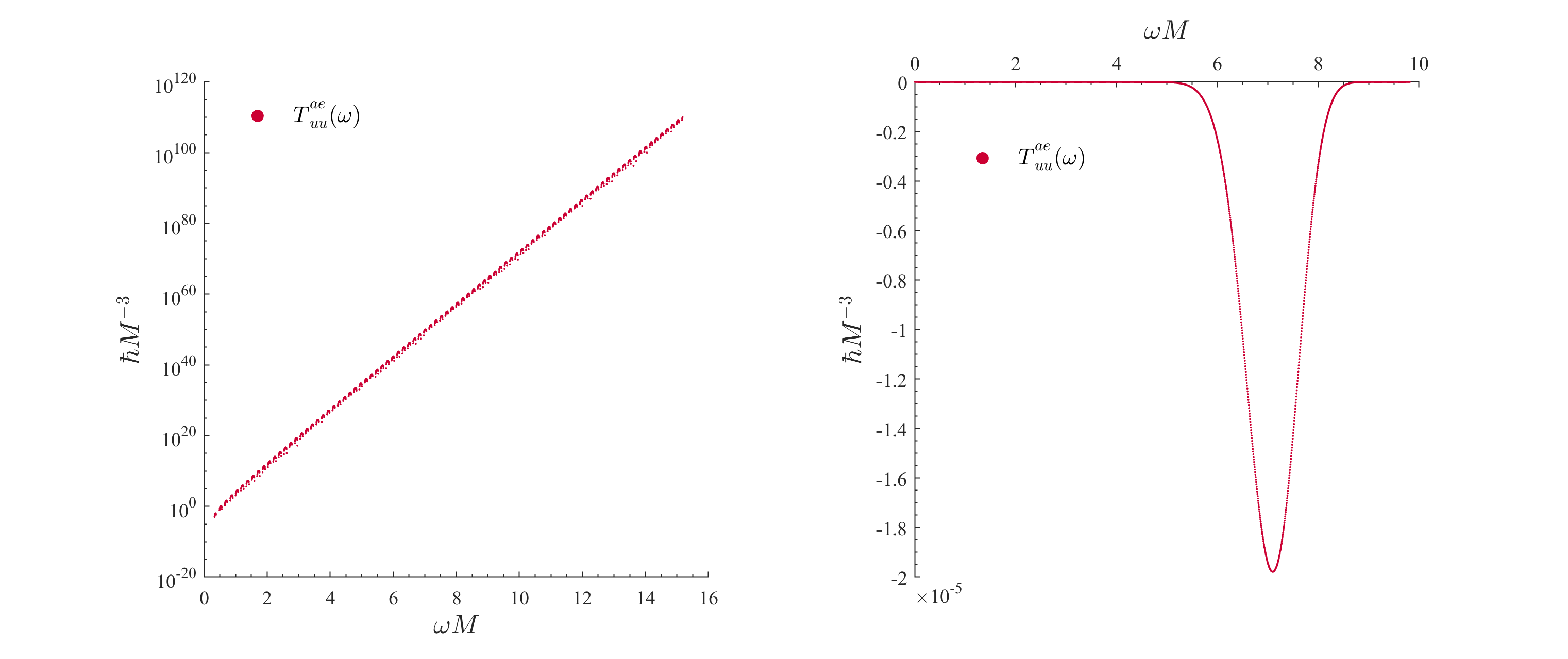}
    \caption{The analytic-extension integrand $T_{uu}^{\text{ae}}(\omega)$ at $r=1.5M$ within a BH of spin $a/M=0.8$, before and after its regularization including the required oscillation cancellation (as described in the main text). \emph{Left:} the basic integrand on a logarithmic scale, where the dominant exponent with the growth parameter given in Eq.~\eqref{eq:eta_short} is clearly visible alongside oscillations. \emph{Right:} the integrand exposed after regularization, to be integrated in the subsequent step. While the details of the applied operations affect the appearance of the resulting integrand shown in this panel, the final integral value remains invariant (refer to footnote~\ref{fn:osc_int}). 
The comparison between the two panels demonstrates the success of the damping procedure, which is vital in order to reveal the `real' physical content of the modes.}
    \label{fig:analytic_ext}
\end{figure}

Fig.~\ref{fig:analytic_ext} illustrates the dramatic decrease in orders of magnitude by the procedure of oscillation cancellation described above, focusing on the integrand $T_{uu}^{\text{ae}}(\omega)$ at $r=1.5M$. The integrand is depicted before and after the oscillation cancellation procedure, showing a reduction by roughly $\sim115$ order of magnitude. 
(Besides this remarkable reduction in the $\omega$-integrand, a decrease of $10-40$ orders of magnitude was already achieved at an earlier stage, when constructing this basic integrand per $\omega$ from the corresponding series in $l$ via the associated oscillation cancellation procedure.)

This challenge of damping such a large number of orders of magnitude is unique to the analytic extension. Yet, the two methods yield the same results in the cases we examined (up to the small error mentioned above), thereby providing the sought-after verification for the standard method described in the main text.

Following the mentioned drop in accuracy from $r=1.5M$ to $r=1.4M$, we did not attempt to decrease $r$ further. The technical difficulty increases with further decreasing $r$, mainly due to the dependence of the oscillation $\omega$-wavelengths on $r$ [along with
the presence of the ($r$-independent) exponential growth which drastically increases the numerical requirements of the entire procedure]; as the $\omega$-wavelengths become
longer, a larger range of highly accurate modes is required for effective damping
of the oscillation. 

We see that the analytic extension variant is far from an effective method of computation, due to the presence of multiple oscillations and exponential growth which put a strong demand on the numerical data required. However, we have demonstrated that it is indeed feasible to use this method to reproduce and verify our standard $t$-splitting results in a few chosen cases. 
\\

\

\end{document}